\documentclass[a4paper,11pt]{article}
\pdfoutput=1 

\usepackage{jheppub} 

\usepackage[T1]{fontenc} 
\usepackage{bm,bbm}
\usepackage{graphics}
\usepackage{subcaption}
\usepackage{amssymb}
\usepackage{hyperref}
\usepackage{slashed}
\usepackage{mathrsfs}
\usepackage{esint}
\usepackage[normalem]{ulem}
\usepackage{longtable}
\usepackage{cancel}

\newcommand{\msb}{{\overline{\mathrm{MS}}}}

\newcommand{\nn}{\nonumber}

\newcommand{\as}{\alpha_s}
\newcommand{\IM}{\mbox{\rm Im}}

\newcommand{\sfrac}[2]{\mbox{$\frac{#1}{#2}$}}

\title{\boldmath
Reconciling the Contour-Improved and Fixed-Order Approaches for $\tau$ Hadronic Spectral 
	Moments I: \\ Renormalon-Free Gluon Condensate Scheme
}

\preprint{
\begin{flushright}
UWThPh-2022-2\\
\end{flushright}
}

\author[a]{Miguel A. Benitez-Rathgeb}
\author[a,b]{Diogo Boito}
\author[a,c]{Andr\'e H. Hoang}
\author[a,d]{Matthias Jamin}

\affiliation[a]{University of Vienna, Faculty of Physics,\\Boltzmanngasse 5, A-1090 Wien, Austria}
\affiliation[b]{Instituto de F\'isica de S\~ao Carlos, Universidade de S\~ao Paulo, \\ CP 369, 13560-970, S\~ao Carlos, SP, Brazil}
\affiliation[c]{Erwin Schr\"odinger International Institute for Mathematics and Physics,\\
University of Vienna, Boltzmanngasse 9, A-1090 Wien, Austria}
\affiliation[d]{Department of Addictive Behaviour, Central Institute of Mental Health, Medical Faculty Mannheim, Heidelberg University, Mannheim, Germany}

\emailAdd{miguel.angel.benitez-rathgeb@univie.ac.at}
\emailAdd{boito@ifsc.usp.br}
\emailAdd{andre.hoang@univie.ac.at}
\emailAdd{matthias.jamin@gmail.com}

\abstract{We propose a simple and easy-to-implement scheme for a renormalon-free gluon condensate (GC) matrix element, which is analogous to implementations of short-distance heavy-quark mass renormalization schemes existing in the literature already for a long time.
Because the scheme is based on a perturbative subtraction at the level of the matrix element, with a freely adaptable
infrared factorization scale, it can be implemented with little effort for any observable where the GC
is relevant. The scheme depends on the renormalon norm of the GC which has to be supplemented independently.
We apply the scheme to the fixed-order (FOPT) and contour-improved (CIPT)
perturbative expansions of  $\tau$ hadronic spectral function moments. These expansions exhibit a long-standing discrepancy
for moments used in high-precision determinations of the strong coupling in the commonly used GC scheme that is not renormalon-free.
We show that the scheme is capable of resolving the FOPT-CIPT discrepancy problem.
At the same time, the perturbative behaviour of the moments that previously showed
bad convergence properties and for which the non-perturbative corrections from the GC are sizeable, is substantially improved.
The new GC scheme may provide a powerful theoretical tool for future phenomenological applications.
}

\begin{document}
\maketitle
\flushbottom

\section{Introduction}
\label{sec:intro}

The extraction of the strong coupling $\alpha_s$ from inclusive hadronic $\tau$ decay data remains to this date one of the most precise determinations of the QCD coupling from experiment. It also provides one of the most stringent tests of asymptotic freedom, since it is obtained at relatively low energies.
The standard procedure in this analysis consists in the use of Finite Energy Sum Rules (FESRs) to relate weighted integrals over the
experimentally accessible spectral functions, running from threshold to $m_\tau^2$, to integrals over the theory prediction which are
performed in a closed contour in the complex plane, often chosen to be a circle of radius $m_\tau^2$~\cite{Braaten:1991qm}. These weighted integrals are also called $\tau$ hadronic spectral function moments. In the theory
integral, a prescription for setting the renormalization scale $\mu$ must be adopted. The two most widely employed prescriptions are Fixed
Order Perturbation Theory (FOPT)  and Contour Improved Perturbation Theory (CIPT). The FOPT approach consists of setting $\mu=m_\tau$ and of
performing a strict fixed-order expansion in powers of $\alpha_s(m_\tau)$ (see e.g. Ref.~\cite{Beneke:2008ad}).  The latter approach allows to treat the strong coupling as a constant for the contour integration. In the CIPT approach the renormalization
scale runs along the contour and one resums an infinite set of logarithmic contributions related to the QCD $\beta$-function~\cite{LeDiberder:1992jjr,Pivovarov:1991rh}. As a consequence, after the contour integration is carried out, the individual terms in the CIPT series do not exhibit explicit powers of the strong coupling,
and it is not possible to transition between the terms of the CIPT and FOPT series through a change of renormalization scale for the strong coupling. For a given value of $\alpha_s$ and for moments such as the so-called kinematic moment, which corresponds to the $\tau$ hadronic inclusive decay width (see Eq.~(\ref{eq:Wtau})),
where the gluon condensate (GC) operator product expansion (OPE) correction and the asymptotic character of the series related to it are suppressed,\footnote{The kinematic moment is the most prominent member of the class of spectral function moments which in this work we refer to as `gluon condensate suppressed moments' , see Sec.~\ref{sec:renormalon}.} the CIPT approach leads to systematically smaller values for the truncated perturbative series. As a result $\alpha_s$ determinations based on CIPT yield systematically larger values than being based on FOPT~\cite{Boito:2012cr,Davier:2013sfa,Boito:2014sta,Boito:2020xli,Pich:2016bdg,Ayala:2021mwc}.

For a long time, the discrepancy between the results obtained with the two prescriptions has been one of the dominant uncertainties in
the final value of $\alpha_s$ from $\tau$ spectral functions quoted  by the Particle Data Group~\cite{ParticleDataGroup:2020ssz,Salam:2017qdl}. This situation has not improved with the calculation of the
$\mathcal{O}(\alpha_s^4)$ (five-loop) contribution to the perturbative expansion~\cite{Baikov:2008jh,Herzog:2017dtz}.
Until recently, attempts to address the discrepancy between FOPT and CIPT focussed mostly on forecasting the behaviour of the series at high orders in the context of concrete models or approximations of the Borel function of the underlying Adler function~\cite{Beneke:2008ad,Boito:2018rwt,Caprini:2019kwp}, or on the suggestions of modified versions of the perturbative expansion~\cite{Caprini:2009vf,Caprini:2011ya,Cvetic:2010ut}.
A possible connection of the FOPT-CIPT discrepancy with IR renormalons was also hinted upon in Ref.~\cite{Beneke:2012vb}, but this aspect was not further explored. The fundamental reason for the observed discrepancy was not fully addressed in these previous works.\footnote{ In Refs.~\cite{Caprini:2009vf,Caprini:2011ya} `new/improved contour-improved' and `new/improved fixed-order' moment series were defined from the Borel sums over series of functional approximations to the Borel function of the Adler function. These moment series and their properties are fundamentally different from the actual CIPT and FOPT series expansions. In Ref.~\cite{Cvetic:2010ut} a `modified contour improved expansion' based on derivatives of the strong coupling was defined, which is equivalent to the actual CIPT expansion in the large-$\beta_0$ approximation and which agrees with the actual CIPT expansion within perturbative uncertainties.}

In  Ref.~\cite{Hoang:2020mkw} (see also the review~\cite{Hoang:2021nlz}) it was shown at an analytic level that a discrepancy in the behavior of the FOPT and CIPT spectral function moment series can arise from a different infrared (IR) sensitivity inherent to the two expansion methods. To be concrete, it was demonstrated how the asymptotic (i.e.\ factorially non-convergent) character of the Adler function's perturbation series due to IR renormalons~\cite{Mueller:1984vh,Beneke:1998ui} can lead to a systematic and computable disparity in the value that CIPT and FOPT moment series approach at intermediate orders, i.e.\ before the divergent behavior of the asymptotic series sets in at higher orders.\footnote{ We stress that the results of Ref.~\cite{Hoang:2020mkw} and the asymptotic separation do not apply to the `new/improved contour-improved' and `new/improved fixed-order' expansions suggested in Refs.~\cite{Caprini:2009vf,Caprini:2011ya}.}
This disparity, called the {\it asymptotic separation}, can be computed analytically for any spectral function moment, if the singular and non-analytic structures due to IR renormalons in the Adler function's Borel function are known. It was also shown that the asymptotic separation does not arise for UV renormalons. In particular, it was found that only the asymptotic separation caused by the GC renormalon can be sufficiently sizeable to have practical relevance and that it even has the right sign to explain the discrepancy. Under the assumption that the 5-loop QCD corrections to the Adler function already have a sizeable contribution from the asymptotic behavior associated to the GC renormalon and from concrete numerical studies, it was therefore suggested in Ref.~\cite{Hoang:2020mkw} that the asymptotic separation could explain the difference between the CIPT and FOPT spectral function moment series that is observed at the 5-loop level.  From the properties of the asymptotic separation, and based on detailed studies of the FOPT and CIPT moment series, it was furthermore concluded in Ref.~\cite{Hoang:2020mkw} that the CIPT expansion is not consistent with the standard way\footnote{What we refer to as the standard treatment of the OPE corrections is explained in detail in Sec.~\ref{sec:theorysetup}.}  the OPE corrections are accounted for in spectral function moment analyses.

In this work we explore from  a practical perspective the suggestion made in Refs.~\cite{Hoang:2020mkw,Hoang:2021nlz}, that the discrepancy between the CIPT and FOPT spectral function moment perturbation series is of IR origin and mainly caused by the IR renormalon associated to the GC. Even though we use analytic results for the asymptotic separation derived in~\cite{Hoang:2020mkw,Hoang:2021nlz} for comparison in our numerical analyses, our  concrete considerations do not rely  on the concept of the asymptotic separation.

To be definite, we start (a) from the  proposition that the GC renormalon gives a sizable contribution to the Adler function corrections already at lower orders  such that its normalization can be inferred at least approximately from the known 5-loop corrections and (b) from the fact that the OPE corrections do not only provide non-perturbative corrections but, at the same time,  also compensate order-by-order in perturbation theory for the asymptotically increasing behavior of the QCD corrections~\cite{Gross:1974jv,tHooft:1977xjm,David:1983gz,Mueller:1984vh,Beneke:1998ui,Grozin:2004ez}. The proposition (a), which has been advocated based on quantitative plausibility studies in Refs.~\cite{Beneke:2008ad,Beneke:2012vb} before, is a natural one, because the GC represents the parametrically dominant quartic IR sensitivity of the Adler function and constitutes the leading term in its OPE. Assuming that the normalization of the GC renormalon is not unnaturally suppressed or enhanced and that there are no fine-tuned cancellations among different renormalons, it should have a sizeable effect on the series terms at the 5-loop level and information on its size should be attainable from them.\footnote{This means that the GC renormalon contributions is sufficiently sizeable such that its asymptotic behavior is already visible in the 5-loop corrections, but it does not mean that the GC renormalon dominate these coefficients numerically.} Admittedly, this cannot be proven from first principles.  The situation, however,  is somewhat analogous to observables sensitive to heavy quark masses, where the dominance of the so-called pole mass renormalon (associated to linear IR sensitivity) cannot be proven from first principles either, but is in excellent agreement with the behavior of the computed QCD corrections and is a generally accepted practical paradigm (see e.g.\ the recent reviews~\cite{Hoang:2020iah,Beneke:2021lkq}).
Property (b) is tied to the common approach that the coefficients in QCD perturbation theory are calculated in the limit of vanishing IR cutoff. While IR singularities cancel in IR-safe observables (typically between real and virtual corrections), sub-leading (power) IR sensitivities remain and cause fixed-sign divergent structures in the perturbation series at large orders.

Here we explore the idea that it should be possible to reconcile the CIPT-FOPT discrepancy problem by removing the leading quartic IR sensitivity, which is related to the GC renormalon, from the perturbative coefficients of the underlying Adler function. The final aim we have in mind is that, upon this removal, the discrepancy problem simply does not arise any more, and that this source of theoretical uncertainty in the strong coupling determinations from hadronic $\tau$ decay spectral function moments is eliminated.

In the context of heavy quark physics the standard approach to eliminate the linear IR sensitivity in the pole mass renormalization scheme is to switch to so-called short-distance quark-mass schemes, which effectively reimplement a finite IR cut for the perturbative calculation. If the short-distance scheme is chosen in an appropriate way, the perturbative behavior of heavy-quark mass dependent observables can then be improved substantially. In fact, recent progress on charm and bottom quark mass determinations with a precision of around $10$-$20$~MeV has only been possible adopting short-distance quark mass renormalization schemes~\cite{ParticleDataGroup:2020ssz}.

In this work, which consists of two parts, we go along a similar way. In Part I, which is the present work, we set up a novel but also very simple renormalon-free short-distance scheme for the GC matrix element, which in turn eliminates the impact of the GC renormalon in the perturbation series for  the Adler function. While previous similar suggestions to define a renormalon-free GC  scheme (or even a renormalon-free OPE scheme)~\cite{Caprini:2009vf,Lee:2010hd,Caprini:2011ya,Bali:2014sja,Ayala:2020pxq,Hayashi:2021vdq} implemented the scheme through modifications or prescriptions imposed directly to the perturbation series, our scheme starts from a redefinition of the GC matrix element itself, in close analogy to the change from the pole mass scheme to short-distance mass scheme through a redefinition of the heavy quark mass parameter. In this scheme the norm of the GC renormalon appears as an explicit parameter that has to be supplemented independently.\footnote{The perturbative subtractions involved in the renormalon-free GC scheme we propose are analogous to those involved in the renormalon-subtracted (RS) quark-mass scheme, which is a short-distance quark-mass scheme that was devised previously in Ref.~\cite{Pineda:2001zq}.}
Applying the  renormalon-free GC scheme to spectral function moments we corroborate the statements and implications of Refs.~\cite{Hoang:2020mkw,Hoang:2021nlz}. In particular we show that the discrepancy between the FOPT and CIPT expansion series for GC suppressed spectral function moments can be indeed removed, such that the CIPT expansion remains as a viable expansion method. At the same time, the previous bad perturbative behavior of the FOPT and CIPT expansion series for GC enhanced spectral function moments, can be amended substantially. These two improvements arise systematically in the large-$\beta_0$ approximation where the all-order perturbative series and the GC norm are known exactly. More importantly, the two improvements also arise systematically in full QCD at ${\cal O}(\alpha_s^4)$ and ${\cal O}(\alpha_s^5)$\footnote{It is customary in state-of-the-art hadronic $\tau$ decay spectral function moment analyses to include estimates for the ${\cal O}(\alpha_s^5)$ corrections.}, if we assume a value for the GC renormalon norm determined from a renormalon model of the Adler function suggested in Ref.~\cite{Beneke:2008ad} that is in line with proposition (a). The consistency of the improvements we find in the renormalon-free GC scheme for different kinds of spectral function moments makes it highly unlikely that the true GC renormalon norm is substantially different and that the improvements are based on a pure coincidence that may not persist at higher orders. In the context of proposition (a), we therefore believe that the FOPT-CIPT discrepancy problem, that affected strong coupling determination from $\tau$ decay spectral function moments for a long time, can be considered as resolved.

What remains to be done in this context is to obtain an estimate of the GC renormalon norm with a reliable theoretical uncertainty and to demonstrate the dependability of the renormalon-free GC scheme in concrete strong coupling measurements in the face of experimental data. This is addressed in Part II of this work, which is going to be published as a follow-up paper.

Some of the elements of our renormalon-free GC scheme have recently been briefly outlined  in the large-$\beta_0$ approximation in Ref.~\cite{Benitez-Rathgeb:2021gvw}. In the current paper we discuss the case of full QCD which is more involved, provide many more details and discuss numerous conceptual and practical aspects. We note that, in this paper, for the formulation of all analytic expressions concerning the renormalon calculus, without any loss of generality,
we use the strong coupling in the $C$-scheme~\cite{Boito:2016pwf} (for the concrete value $C=0$).  This is because in the $C$-scheme the analytic expressions for the Borel functions, that encode the non-analytic renormalon structures, and the subtraction scheme can be written down in closed form to all orders similar to the large-$\beta_0$ approximation (in contrast to the common $\overline{\rm MS}$ scheme for the strong coupling, where the expressions involve infinite sums and have to be truncated). Our practical (finite order) numerical studies are, however, still carried out in the usual $\overline{\rm MS}$ scheme and are obtained from the $C$-scheme expressions by a finite-order re-expansion. For the numerical evaluations of the strong coupling in the complex plane we used the quasi-exact routine from the REvolver library package~\cite{Hoang:2021fhn}.

The content of this paper is as follows:
In Sec.~\ref{sec:Notation} we set up our notations and naming conventions, review the $C$-scheme for the strong coupling and provide a brief introduction on the renormalon calculus, the GC and the association of OPE corrections with IR renormalons. This section forms the conceptual background for the rest of the paper.
In Sec.~\ref{sec:subtractedPT} we comment on previous suggestions in the literature to set up renormalon-free OPE schemes and then set up our observable-independent renormalon-free GC scheme. We show how the scheme is applied to the Adler function, explain its properties and address the relation of the GC in the new scheme to schemes devised prior to this work. The renormalon-free GC scheme is then tested on FOPT and CIPT spectral function moment expansions using an all-order toy model for the Adler function which consists of a pure ${\cal O}(\Lambda_{\rm QCD}^4)$ renormalon series,
corroborating the findings of Ref.~\cite{Hoang:2020mkw} and its implications and
verifying the effectiveness of the new scheme in eliminating the FOPT-CIPT discrepancy caused by the GC renormalon for GC condensate suppressed moments.
In Sec.~\ref{sec:largeb0} the renormalon-free GC scheme is then applied to the FOPT and CIPT spectral function moment expansions based on the Adler function in the large-$\beta_0$ approximation where exact all-order results can be obtained. Here we demonstrate the effectiveness of the new scheme in eliminating the FOPT-CIPT discrepancy
in a more realistic environment where also other IR and UV renormalons are present. We also show that the new scheme at the same time substantially improves the previously bad perturbative behavior of GC enhanced moments.
In Sec.~\ref{sec:MultiRenormalonModel} we apply the new GC scheme to the FOPT and CIPT spectral function moment expansions in full QCD using corrections up to ${\cal O}(\alpha_s^5)$ and adopting an Adler function renormalon model based on Ref.~\cite{Beneke:2008ad} that provides a concrete value for the GC renormalon norm and an estimate for the orders beyond.  The results demonstrate the practical effectiveness of the new scheme.
A summary and a conclusion are given in Sec.~\ref{sec:conclusions}.
A number of relevant formulae that did not find their way into the main body of the paper are given in App.~\ref{app:modelsC}.

\section{Notation, Strong Coupling and Renormalon Calculus}
\label{sec:Notation}

\subsection{Theoretical Setup}
\label{sec:theorysetup}

The theoretical description of hadronic $\tau$ decay spectral functions receives
contributions from all basic two-point correlation functions: vector/axialvector
and scalar/pseudoscalar. The latter scalar/pseudoscalar contributions only
arise suppressed by at least two powers of light-quark masses, and hence
are very small. Furthermore, because of chiral symmetry, in the chiral limit,
for the dominant vector/axialvector contribution, the purely perturbative parts
are identical. Therefore, in this work, it will be sufficient to concentrate
our discussion and analysis on the vector correlation function, and its
corresponding contribution to hadronic tau decay spectral function moments.

In momentum space, the vector correlation function $\Pi_{\mu\nu}(p)$ is
defined as
\begin{equation}
\label{Pimunu}
\Pi_{\mu\nu}(p) \,\equiv\,  i\!\int \! {\rm d}x \, e^{ipx} \,
\langle\Omega|\,T\{ j_\mu(x)\,j_\nu(0)^\dagger\}|\Omega\rangle\,,
\end{equation}
where $|\Omega\rangle$ denotes the physical QCD vacuum state. For hadronic $\tau$
decays, only two flavor non-diagonal currents with light quarks contribute,
\begin{equation}
\label{jvec}
j_\mu(x) \,=\; :\!\bar u(x)\gamma_\mu d(x)\!: \,,
\end{equation}
and the current where the down quark is replaced by a strange quark.
The correlator $\Pi_{\mu\nu}(p)$ admits the Lorentz decomposition
\begin{equation}
\label{PimunuDecomp}
\Pi_{\mu\nu}(p) \,=\, (p_\mu p_\nu - g_{\mu\nu}p^2)\,\Pi^{(1+0)}(p^2) +
g_{\mu\nu}\,p^2\,\Pi^{(0)}(p^2) \,,
\end{equation}
where the superscripts denote the components corresponding to angular momentum
$J=1$ (transversal) and $J=0$ (longitudinal) in the hadronic rest frame.
The correlator $\Pi^{(0)}(s)$ with $s\equiv p^2$ is related to the
scalar/pseudoscalar correlators, and vanishes in the limit of massless quarks.
For simplicity, in the following, we will write $\Pi(s)\equiv\Pi^{(1+0)}(s)$,
and we frequently refer to it as the vacuum polarization function.

$\Pi(s)$ itself is not a physical quantity in the
sense that it contains a renormalization scale and scheme dependent subtraction
constant.  This subtraction constant can either be removed by taking the
imaginary part, which corresponds to the spectral function $\rho(s)$, or by
taking a derivative with respect to $s$, which leads to the (reduced) Adler function
$D(s)$:
\begin{equation}
\label{rhoD}
\rho(s) \,\equiv\, \frac{1}{\pi}\,\IM\,\Pi(s+i 0) \,, \quad
\frac{1}{4\pi^2}\big[1+D(s)\big] \,\equiv\, -\,s\,
\frac{\rm d}{{\rm d}s}\,\Pi(s) \,.
\end{equation}

The Adler function $D(s)$ for a general complex $s$ satisfies a homogeneous RGE, and hence the logarithms, which appear at every order in perturbation theory, can be summed adopting $\mu^2=-s$ as the renormalization scale. It is therefore possible to formulate the perturbative series for the partonic Adler function in the chiral limit, $\hat D(s)$, in the form
\begin{equation}
\label{eq:Dresum}
\hat D(s) \,=\, \sum\limits_{n=1}^\infty \,
c_{n,1} \left(\frac{\alpha_s(-s)}{\pi}\right)^n \,,
\end{equation}
with complex-valued powers of the strong coupling and real-valued coefficients $c_{n,1}$.
When we refer to the Adler function along the Euclidean axis, i.e.\ for $s=-|s|$, we adopt the definition $Q^2\equiv-s=|s|$, such that the perturbation series for the Euclidean Adler function adopts the form
$\hat D(-Q^2)=\sum_{n=1}^\infty c_{n,1}\,(\frac{\alpha_s(Q^2)}{\pi})^n$.
The coefficients $c_{n,1}$ are known analytically up to order $\as^4$
\cite{Gorishnii:1990vf,Surguladze:1990tg,Baikov:2008jh}.
For $n_f=3$ dynamical flavors, which is the flavor number scheme we adopt throughout this work, the coefficients $c_{n,1}$ have the following form in the $\msb$-scheme  for the strong coupling:
\begin{eqnarray}
\label{cn1}
c_{1,1} &=& 1 \,, \quad
c_{2,1} \,=\, \sfrac{299}{24} - 9\zeta_3 \,=\, 1.640 \,, \nn \\
\vbox{\vskip 6mm}
c_{3,1} &=& \sfrac{58057}{288} - \sfrac{779}{4}\zeta_3 + \sfrac{75}{2} \zeta_5
\,=\, 6.371 \,, \\
\vbox{\vskip 6mm}
c_{4,1} &=& \sfrac{78631453}{20736} - \sfrac{1704247}{432}\zeta_3 +
\sfrac{4185}{8}\zeta_3^2 + \sfrac{34165}{96}\zeta_5 - \sfrac{1995}{16}\zeta_7
\,=\, 49.076 \,. \nn
\end{eqnarray}

As is customary in $\tau$ decay analyses we will in fact work at $\mathcal{O}(\alpha_s^5)$ employing an estimate for $c_{5,1}$. There are several independent estimates of this 6-loop coefficient, all of them in good agreement, and we adopt
\begin{equation}
\label{eq:c51MSb}
  c_{5,1} \,=\, 280\pm 140 \,,
\end{equation}
which  covers all the estimates available in the recent literature~\cite{Boito:2018rwt,Baikov:2008jh,Beneke:2008ad,Caprini:2019kwp,Jamin:2021qxb}. We use the central value for the numerical studies in Sec.~\ref{sec:MultiRenormalonModel}. The impact of the uncertainties are addressed in our follow-up paper.

Restricting the analysis to the $ud$ quark current, for which the mass corrections can safely be neglected, moments of the experimentally accessible $\tau$ hadronic spectral functions can be parametrized in the following general form
\begin{equation}
\label{eq:momdef}
R_{V/A}^{(W)}(s_0) \, =\, \frac{N_c}{2} \,S_{\rm ew}\,|V_{ud}|^2 \Big[\,
\delta^{\rm tree}_{W} + \delta^{(0)}_{W}(s_0)  +
\sum_{d\geq 4}\delta^{(d)}_{W,V/A}(s_0) +\delta_{W,V/A}^{(\rm DV)}(s_0)\Big] \,,
\end{equation}
where $N_c=3$ and the factor $S_{\rm ew}$ represents factorizable electroweak corrections.
The term $\delta^{\rm tree}_{W}$ is the tree level contribution while  $\delta^{(0)}_{W}(s_0)$ contains the higher-order perturbative QCD corrections in the chiral limit. The power-suppressed non-perturbative corrections from the OPE condensates of dimension $d$ are encoded in the terms
$\delta^{(d)}_{W,V/A}(s_0)$ and, finally, $\delta_{W,V/A}^{(\rm DV)}(s_0)$ is the  duality violation  (DV) contribution. The latter term is not essential for the discussion of the present paper and is  going to be ignored in the remainder.

The experimental counterpart to Eq.~(\ref{eq:momdef}) can be obtained from a weighted integral over the spectral functions $\rho_{V/A}(s)$ as
\begin{equation}
\label{eq:Rexperimental}
R_{V/A}^{(W)}(s_0) \, =\, 12 \pi^2 \,S_{\rm ew}\,|V_{ud}|^2\,\int\limits_0^{s_0} \frac{{\rm d}s}{s_0}\left[\,w({\textstyle \frac{s}{s_0}})  \rho_{V/A}^{(1+0)}(s) -w_L({\textstyle \frac{s}{s_0}})\rho_{V/A}^{(0)}(s) \right]\,,
\end{equation}
where the longitudinal term involving $\rho_{V/A}^{(0)}$ is shown for completeness, but does not play any role in the present analysis, where we consider the massless quark limit.
The QCD corrections in $\delta^{(0)}_{W}(s_0)$ and the OPE corrections in $\delta^{(d)}_{W}(s_0)$ can be obtained from the invariant mass contour integral over the Adler function as ($x\equiv s/s_0$)
\begin{align}
\label{eq:deltadef}
\delta^{(0)}_{W}(s_0)  +
 \sum_{d\geq 4}\delta^{(d)}_{W}(s_0) =
 \frac{1}{2\pi i}\,
\ointctrclockwise\limits_{|s|=s_0} \!\! \frac{{\rm d}s}{s}\,W({\textstyle \frac{s}{s_0}})\, D(s)
\, =\,
\frac{1}{2\pi i}\, \ointctrclockwise\limits_{|x|=1}  \!\!\frac{{\rm d}x}{x}\,W(x)\, D(x s_0)\,,
\end{align}
where the contour  integral starts/ends at a positive real-valued $s_0\pm i 0$ (or $1\pm i 0$) and is a counterclockwise path in the complex $s$-plane (or $x$-plane) around the origin~\cite{Braaten:1991qm,LeDiberder:1992jjr}. Commonly a circular path with radius $s_0$ (or $1$) is adopted, but it may be deformed as long as the path encloses the Landau pole of the strong coupling and remains in the perturbative regime.
The terms   $w(x)$ and $W(x)$ are analytic weight functions which are taken to be polynomials in practical applications. Due to analyticity they satisfy the relation
\begin{eqnarray}
\label{eq:Wrelation}
W(x)&=& 2\int_{x}^1 {\rm d}z \,\,w(z)\,,
\end{eqnarray}
so that for all physically relevant spectral function moments we have $W(1)=0$.
Spectral function moments for which $w(x)$ ($W(x)$) vanish linearly (quadratically) at $x=1$ are referred to as 'pinched' and those for which $w(x)$ ($W(x)$) vanish quadratically (cubically) at $x=1$  are called `doubly-pinched'.
Since we mostly consider the spectral function moments from the perspective of the contour integrations over the Adler function, as shown in the second line of Eq.~(\ref{eq:deltadef}), we refer to them specifying the weight functions $W(x)$.
For the so-called kinematic weight functions
\begin{eqnarray}
\label{eq:Wtau}
w_\tau(x) &=&  (1-x)^2(1+2x) = 1-3x^2+2x^3 \\
W_\tau(x)&=& (1-x)^3(1+x)=1-2x+2x^3-x^4
\end{eqnarray}
and $s_0=m_\tau^2$ the moments $R_{V/A}^{(W_\tau)}(m_\tau^2)\equiv R^\tau_{V/A}$ give the normalized non-strange hadronic $\tau$ decay rates  $R^\tau_{V/A}=\Gamma(\tau^-\to\mbox{(hadrons)}_{V/A;ud}\,\nu_\tau(\gamma))/\Gamma(\tau^-\to e^-\bar{\nu}_e\nu_\tau(\gamma))$. The total non-strange hadronic $\tau$ decay is then $R^\tau_{V+A}=R^\tau_V+R^\tau_A$.

The non-perturbative power corrections $\delta^{(d)}_{W}(s_0)$ are obtained from the higher dimensional OPE corrections~\cite{Shifman:1978bx} to the Adler function $D(s)=\hat D(s)+D^{\rm OPE}(s)$, which for massless quarks can be generally parametrized in the form
\begin{equation}
\label{eq:DOPE}
D^{\rm OPE}(s) \, = \,
\frac{C_{4,0}(\alpha_s(-s))}{s^2} \langle \bar {\cal O}_{4,0} \rangle  +
\sum\limits_{d=6}^\infty \frac{1}{(-s)^{d/2}} \sum_i  C_{d,i}(\alpha_s(-s)) \langle \bar {\cal O}_{d,\gamma_i}\rangle \,
\end{equation}
Here the terms $\langle \bar {\cal O}_{d,\gamma_i}\rangle$ are non-perturbative vacuum matrix elements of light quark and gluon field operators with anomalous dimensions $\gamma_i$, which are called condensates. The functions $C_{d,i}$ are the corresponding Wilson coefficients, which are computable perturbative series in $\alpha_s(-s)$ and describe the short-distance information.  The sum of $i$ arises since for dimension $6$ and higher there are several different condensates that have to be considered.

 We refer to the form of the OPE corrections as given in Eq.~(\ref{eq:DOPE}) as the {\it standard form of the OPE}. This standard form is generally accepted to be valid for any IR factorization scheme to separate low-energy non-perturbative effects from the high-scale perturbative corrections in $\delta^{(0)}_{W}(s_0)$. The perturbative series for the partonic Adler function in Eqs.~(\ref{eq:Dresum}) and (\ref{cn1}) has been obtained using dimensional regularization and the $\overline{\rm MS}$ renormalization scheme for the strong coupling which entails that the limit of zero IR cutoff is applied to the computation of the  coefficients $c_{n,1}$. We refer to this IR factorization scheme as the {\it $\msb$-scheme for the OPE}, and we indicate this scheme by a bar over each of the operator matrix elements. We stress that the wording should not be confused with the $\msb$ renormalization scheme for the strong coupling. The scheme for the strong coupling (such as the $\msb$- or the $C$-scheme~\cite{Boito:2016pwf} which we employ in this paper) can be chosen independently from the choice of IR factorization scheme.

The CIPT method to determine the perturbation series $\delta^{(0)}_{W}(s_0)$ is based on the perturbative series for the Adler function given in Eq.~(\ref{eq:Dresum}), and one carries out the contour integration of Eq.~(\ref{eq:deltadef}) over powers of the complex-valued strong coupling $\alpha_s(-s)$. The CIPT series arises from truncating the sum in Eq.~(\ref{eq:Dresum}).
For the FOPT method to determine $\delta^{(0)}_{W}(s_0)$ one expands the series~(\ref{eq:deltadef}) in powers of $\alpha_s(s_0)$. The complex phases then appear as powers of $\ln(-s/s_0)$ in the coefficients of the power series in $\alpha_s(s_0)$. The contour integration of Eq.~(\ref{eq:deltadef}) is then carried out over the polynomials of $\ln(-s/s_0)$. For the FOPT moments, the series arises from truncating the sum in powers of $\alpha_s(s_0)$. The CIPT method differs from  FOPT in that it resums the powers of $\ln(-s/s_0)$ to all orders along the contour integration~\cite{LeDiberder:1992jjr,Pivovarov:1991rh}. Furthermore, while the FOPT series for $\delta^{(0)}_{W}(s_0)$ is an explicit power series in the strong coupling $\alpha_s$ at a definite renormalization scale, the CIPT series is not, since the renormalization scale of the strong coupling is the integration variable. This is the reason why
it is not possible to  obtain the CIPT series through a change of renormalization scheme from the FOPT series.

\subsection[Strong coupling in the   \texorpdfstring{$C$}{C}-scheme]{Strong coupling in the  \boldmath \texorpdfstring{$C$}{C}-scheme}
\label{sec:coupling}

For the definition of our renormalon-free GC scheme, given in Sec.~\ref{sec:subtractedPT}, we use for the strong coupling the $C$-scheme~\cite{Boito:2016pwf}, for the particular value $C=0$, for which the exact $\beta$-function can be written down in closed form. This will allow us to write down closed and exact all-order expressions for the renormalon-free GC scheme. While in the $\msb$-scheme the renormalization group equation of the strong coupling has the form ($\beta_0=11-2\,n_f/3$, $\beta_1=102-38\,n_f/3$)
\begin{equation}
\label{eq:betaMSbar}
\frac{d\alpha_s(Q^2)}{d\ln Q}\, = \, \beta(\alpha_s(Q^2)) \, \equiv \,
-2\,\alpha_s(Q^2)\,\sum\limits_{n=0}^\infty \beta_n\Big(\frac{\alpha_s(Q^2)}{4\pi}\Big)^{n+1}
\end{equation}
where the coefficients $\beta_n$ are independent, the $C$-scheme evolution has the closed all-order form
\begin{equation}
\label{eq:Cscheme}
\frac{d\bar\alpha_s(Q^2)}{d\ln Q}\, = \, \bar \beta(\bar \alpha_s(Q^2))\, \equiv \, -2\,\bar \alpha_s(Q^2)\,\frac{\beta_0 \,\bar\alpha_s(Q^2)}{4\pi-\frac{\beta_1}{\beta_0}\bar\alpha_s(Q^2)}\,,
\end{equation}
where only the universal coefficients $\beta_0$ and $\beta_1$ enter.
The relation between the strong coupling in the  $\msb$ and in the $C$-scheme (for $C=0$) is then given by the equality
\begin{equation}
\label{eq:MSbarC}
\frac{\pi}{\bar\alpha_s(Q^2)} + \frac{\beta_1}{4\beta_0}
\ln(\bar\alpha_s(Q^2)) \,=\,
\frac{\pi}{\alpha_s(Q^2)} + \frac{\beta_1}{4\beta_0}
\ln(\alpha_s(Q^2)) + \frac{\beta_0}{2}\,\int\limits_0^{\alpha_s(Q^2)}
\!\!\! d\tilde\alpha\,\left[\frac{1}{\beta(\tilde\alpha)}+\frac{2\pi}{\beta_0\tilde \alpha^2}-\frac{\beta_1}{2\beta_0^2\tilde\alpha}\right]\,,
\end{equation}
which ensures that the perturbative relation of the strong coupling in both schemes starts at ${\cal O}(\alpha_s^3)$ and that their QCD scales are equivalent~\cite{Celmaster:1979km}.

As already displayed in Eqs.~(\ref{eq:betaMSbar})-(\ref{eq:MSbarC}) we use unbarred quantities to indicate the $\msb$-scheme and barred quantities for the $C$-scheme (with $C=0$). In the following we will call this scheme simply `the $C$-scheme' for brevity. Note that from the perspective of the $C$-scheme, the information on the perturbative coefficients $\beta_{n>1}$ contained in the $\msb$-scheme $\beta$-function is fully encoded  in the perturbative relation between both couplings and not in strong coupling evolution. Since  the analytic expressions of the renormalon calculus only depend on the form of the $\beta$-function, this ensures that the resulting  expressions are particularly simple in the $C$-scheme. When converting between $\msb$- and $C$-scheme couplings we assume that the truncated 5-loop $\msb$ $\beta$-function with the known coefficients $\beta_{0,1,2,3,4}$ is exact. The resulting perturbative relation between both couplings for $n_f=3$ is given in Eq.~(\ref{eq:asCtoMSbar})  for convenience of the reader for the orders relevant for most of our numerical discussions. The value of $\bar\alpha_s(Q^2)$ is obtained from the $\overline{\rm MS}$ coupling, $\alpha_s(Q^2)$, from the numerical solution of Eq.~(\ref{eq:MSbarC}).

Regarding the coefficients of the Adler function in the $C$-scheme at $C=0$, the first two perturbative coefficients remain unchanged: $\bar c_{1,1}=c_{1,1}$ and $\bar c_{2,1}=c_{2,1}$. The higher coefficients turn out different from the $\msb$ scheme. The next two coefficients $\bar c_{3,1}$ and $\bar c_{4,1}$ are found to be:
\begin{eqnarray}
\label{cb3cb4}
\bar c_{3,1} &=& \sfrac{262955}{1296} - \sfrac{779}{4}\zeta_3 +
\sfrac{75}{2} \zeta_5 \,=\, 7.682 \,, \nn \\
\vbox{\vskip 6mm}
\bar c_{4,1} &=& \sfrac{357259199}{93312} - \sfrac{1713103}{432}\zeta_3 +
\sfrac{4185}{8}\zeta_3^2 + \sfrac{34165}{96}\zeta_5 -
\sfrac{1995}{16}\zeta_7 \,=\, 61.060 \,.
\end{eqnarray}
For $\bar c_{5,1}$, the first unknown coefficient of the Adler function expansion, the result  is
\begin{eqnarray}
\label{eq:barc51}
\bar c_{5,1} &=&c_{5,1}+\sfrac{73019059337}{80621568}-\sfrac{273360539 }{373248}\zeta_3-\sfrac{445}{8}\zeta_3^2+\sfrac{2522915 }{20736}\zeta_5-\sfrac{89}{1536}\pi^4 \nn \\
\vbox{\vskip 6mm}
 &=& c_{5,1} + 65.477 = 345.4774\pm 140,
\end{eqnarray}
where in the numerical value we used the estimate of $c_{5,1}$ given in Eq.~(\ref{eq:c51MSb}).
We remind the reader that we use the $C$-scheme for setting up our renormalon-free GC scheme, but we convert back to the common $\msb$-scheme for the strong coupling for all our numerical studies.

For convenience of notation in the following sections we also define the following abbreviations involving the $C$-scheme strong coupling and coefficients:
\begin{align}
\label{eq:adef}
& \bar a(-s)\,\equiv\, \frac{\beta_0\,\bar \alpha_s(-s)}{4\pi}
\,,
\nonumber \\
& \bar a_Q\, \equiv\, \frac{\beta_0\,\bar \alpha_s(Q^2)}{4\pi} \,,
\nonumber \\
& \bar c_n\, \equiv\, \frac{4^n\,\bar c_{n,1}}{\beta_0^n}\,,
\end{align}
which leads to the following form of the renormalization group equation for $\bar a_Q$
\begin{equation}
\frac{d\,\bar a_Q}{d\ln Q^2} \,=\,
\frac{-\,\bar a^2_Q}{[\,1-2\hat b_1\bar a_Q]} \,,
\end{equation}
where we have defined
\begin{equation}
\label{eq:b1hatdef}
\hat b_1 \, \equiv \, \frac{\beta_1}{2\beta_0^2}\,.
\end{equation}
For $n_f=3$ quark flavors we have $\hat b_1=32/81=0.395$.

\subsection{Renormalon Calculus and Standard OPE Corrections}
\label{sec:renormalon}

According to the standard approach to the OPE and in the chiral limit, each term in the series of the OPE corrections indicated in Eq.~(\ref{eq:DOPE}) arises from a particular type of sensitivity of the spectral function moments and the underlying Adler function to infrared (IR) effects contained in their perturbation series. Each type of IR sensitivity can be associated to a certain QCD low-energy operator, containing products of light-quark and gluon field operators that match the dimension and the tensorial structure of the infrared momentum dependence~\cite{Shifman:1978bx}.
In the common $\overline{\rm MS}$ approach for performing perturbative calculations in QCD, where one considers the limit of a vanishing IR cutoff, each type of IR sensitivity generates a characteristic equal-sign divergent pattern of large-order behavior in the perturbation series. These patterns, referred to as IR renormalons, are generated from the logarithmic sensitivity of the perturbative strong coupling to vanishing momenta and thus depend on the perturbative coefficients of the QCD $\beta$-function~\cite{Gross:1974jv,tHooft:1977xjm}.
These patterns of divergence are a general characteristic of QCD perturbation series in the limit of a vanishing IR cutoff and cause them to be asymptotic. On the conceptual side, these IR-sensitive contributions in the coefficients of the perturbation series arise from momentum regions below the domain of validity of perturbation theory. They also imply that the convergence radius of the perturbation series (in any perturbative scheme for the strong coupling) is zero. These asymptotic patterns can be parametrized conveniently using the renormalon calculus \cite{David:1983gz,Mueller:1984vh,Beneke:1998ui}.

Starting e.g.\ from the perturbation series for the Euclidean Adler function
\begin{equation}
\label{eq:Adlerseriesgeneric}
\hat D(-Q^2) \, =  \, \sum_{\ell=1}^\infty \,
\bar c_{\ell} \,\bar a^\ell_Q\,,
\end{equation}
where we use the C-scheme for the strong coupling,
one defines the Borel function with respect to the expansion in powers of $\bar\alpha_s(Q^2)$ as
\begin{equation}
\label{eq:BTaylor}
\bigg[\, B[\hat D(-Q^2)](u) \,\bigg]_{\rm Taylor}\, =\,
\sum\limits_{\ell=1}^\infty\,
\frac{u^{\ell-1}}{\Gamma(\ell)}\bar c_\ell\,.
\end{equation}
Due to the $\ell$-factorial suppression, the Borel function has a finite radius of convergence and one can analytically continue it to cover the entire complex $u$-plane except for cuts (or poles) located at some distance to the origin. Each type of IR renormalon is associated to
cuts with non-analytic properties of a particular kind.
The original series can then term-by-term be recovered through the inverse Borel transform
\begin{equation}
\label{eq:invBorelD}
\hat D(-Q^2) = \int_0^\infty \!\! {\rm d} u \,\, \bigg[\,
B[\hat D(-Q^2)](u)\,\bigg]_{\rm Taylor}\,e^{-\frac{u}{\bar a_{Q}}}\,,
\end{equation}
using the relation
\begin{equation}
\label{eq:invBorelDv2}
\int_0^\infty \!\! {\rm d} u \,u^{\ell-1}\,e^{-\frac{u}{\bar a_{Q}}} \, = \, \Gamma(\ell)\,\bar a_Q^\ell
\,.
\end{equation}
The Borel calculus and the inverse Borel transform are frequently used in the literature to discuss an all order value of the  perturbation series, called the Borel sum.
However,  at this point {\it we are only concerned with the perturbation series and the parametrization of the contributions of the perturbative coefficients that arise from the IR renormalons}. This is signified by the subscript ``Taylor".

A dimension $d$ term in the Euclidean Adler function's OPE series in Eq.~(\ref{eq:DOPE}), which has a Wilson coefficient that starts with unity at tree-level, assuming no operator mixing and  the renormalization-group-invariant (RGI) convention, can be cast into the generic form {
\begin{eqnarray}
\label{eq:AdlerOPE}
\delta D^{\rm OPE}_{d,\gamma}(-Q^2) & = &
\frac{1}{Q^d}\,(\bar a_Q)^{-\alpha} \,\Big[ 1 + \bar c_{d,\gamma}^{(1)}\, \bar a_Q
+ \ldots  \Big]\, \langle \bar{\cal O}_{d,\gamma}\rangle\,,
\end{eqnarray}
where  $\alpha=-2 \gamma^{(1)}/\beta_0$.  Here $\gamma^{(1)}$ is the leading-logarithmic (one-loop) anomalous dimension of the low-energy QCD operator $\bar{\cal O}_{d,\gamma}$, defined by
$\frac{d}{d\ln(\mu)}\bar{\cal O}_{d,\gamma}=-\gamma[\bar \alpha_s(\mu)]\bar{\cal O}_{d,\gamma}$ with $\gamma[\bar \alpha_s(\mu)] = \gamma^{(1)}(\frac{\bar \alpha_s(\mu)}{\pi})+\ldots$, and the coefficient $\bar c^{(1)}_{d,\gamma}$ arises from the one-loop correction to the Wilson coefficient and the two-loop anomalous dimension.
In the $C$-scheme the associated term in the Adler function's Borel function (with respect to the expansion in powers of $\bar\alpha_s(Q^2)$) can be written down in closed form and reads
\begin{eqnarray}
\label{eq:AdlerOPEBorel}
B_{d,\gamma}(u) & = &
\Big[ 1 + \bar c_{d,\gamma}^{(1)}\, \bar a_Q
+ \ldots  \Big]\, \frac{N_{d,\gamma}}{(\frac{d}{2}-u)^{1+d \hat b_1+\alpha}}\,.
\end{eqnarray}
The term $N_{d,\gamma}$ is the normalization of the corresponding IR renormalon. Together with the non-analytic behavior of the term $1/(d/2-u)^{1+d \hat b_1+\alpha}$, which has a cut along the positive real $u$-axis for $u>d/2$, the Borel function term $B_{d,\gamma}(u)$ fully quantifies the IR renormalon contribution in the coefficients of the perturbation series that is associated to the OPE correction term $\delta D^{\rm OPE}_{d,\gamma}(-Q^2)$:
\begin{equation}
\label{eq:AdlerseriesOPEterm}
\delta  \hat D_{d,\gamma}(-Q^2) \, =  \,N_{d,\gamma}\, \Big[ 1 + \bar c_{d,\gamma}^{(1)}\, \bar a_Q
+ \ldots  \Big]\,
\sum_{\ell=1}^\infty \,
r_{\ell}^{(d,\alpha)} \,\bar a^\ell_Q\,,
\end{equation}
where the perturbative coefficients associated with each renormalon singularity are
\begin{equation}\label{eq:rncoeff}
  r_{\ell}^{(d,\alpha)} =\left(\frac{2}{d}\right)^{\ell+d \hat b_1+\alpha}\,\frac{\Gamma(\ell+d\, \hat b_1+\alpha)}{\Gamma(1+d\, \hat b_1+\alpha)}\, .
\end{equation}
 Because of the exact and compact form of the $C$-scheme $\beta$-function given in Eq.~(\ref{eq:Cscheme})
there aren't any $1/\ell$-suppressed subleading asymptotic contributions in Eqs.~(\ref{eq:AdlerOPEBorel}) and (\ref{eq:rncoeff}), which means that the expression for $r_{\ell}^{(d,\alpha)}$ is exact for any $\ell$. This makes the $C$-scheme particularly convenient for the renormalon calculus.

The order-by-order pattern encoded in the coefficients $ r_{\ell}^{(d,\alpha)}$ is associated with a fixed-sign divergence which is characteristic for IR renormalons. There are also sign-alternating divergent patterns in the coefficients of the perturbation series related to various types of UV sensitivity, referred to as UV renormalons. These UV renormalons can also be parametrized by non-analytic terms in the Adler function's Borel function and have the generic form $1/(k+u)^{1+d \hat b_1+\alpha}$. They are quite similar to Eq.~(\ref{eq:AdlerOPEBorel}) but have cuts along the negative real $u$-axis for $u<-k$ with $k=1,2,\ldots$. Since we do not need to know more about UV renormalons for the purpose of this paper, we do not go into more detail at this point and refer to Ref.~\cite{Beneke:1998ui}.

It is possible to use integration-by-parts in the inverse Borel transform integral of Eq.~(\ref{eq:invBorelD}) to turn the powers of $\bar a_Q$ in Borel function $B_{d,\gamma}(u)$ shown in Eq.~(\ref{eq:AdlerOPEBorel}) into a sum of terms $1/(d/2-u)^{1+d \hat b_1+\alpha-\ell}$, with $\ell=1,2,\dots$, plus regular polynomials in $u$. This leads to a form of the Borel function that is independent of $\bar a_Q$.
The resulting sum of the non-analytic terms $1/(d/2-u)^{1+d \hat b_1+\alpha-\ell}$  with $\ell=0,1,2,\dots$ yields the parametrization of the Borel function associated to a particular renormalon which is most frequently used in the literature~\cite{Beneke:1998ui}.
However, we use the form given in Eq.~(\ref{eq:AdlerOPEBorel}) since it allows us to write down very simple closed expressions that are very similar to those in the large-$\beta_0$ approximation.

The full Borel function is the sum of the functions $B_{d,\gamma}(u)$, matching the terms in the OPE corrections for different $d\ge 4$ and $\gamma$, of non-analytic UV renormalon terms plus regular functions of $u$ which are well-defined and analytic in the entire complex $u$ plane. Generically, at large orders  the renormalon contributions for smaller values of $d/2$ or $k$ dominate over renormalons for larger values of $d/2$ or $k$, but at lower and intermediate orders the size of $N_{d,\gamma}$ and the corresponding norms for UV renormalons play an additional important role for the relative size of the renormalon contributions in the perturbative coefficients.
As a matter of principle, the values of the normalization factors, which quantify the relative weight of the different non-analytic terms, could only be determined exactly if the perturbation series were known to all orders. In practice, in full QCD, they are known at best approximately for renormalons with small values of $d$ and $k$.
For the Adler function, previous studies in full QCD~\cite{Beneke:2008ad,Beneke:2012vb,Boito:2018rwt} in the context of the available  QCD corrections and proposition (a) have shown that, even though the UV renormalons for  $k=1$ are formally dominant with respect to the IR renormalons $d/2=2,3,\ldots$, the respective norms for the UV renormalons are sufficiently suppressed such that in the $\msb$- and the $C$-scheme for the strong coupling the contributions and the sign-alternating effects of the UV renormalons are subleading with respect to the IR renormalons for intermediate orders.

For massless quarks, the leading dimension $d=4$ OPE correction to the Adler function consists of a single term  and is related to the well-known GC matrix element. It can be defined to be renormalization scale invariant, so that its anomalous dimension vanishes, $\gamma=0$. For massless quarks the renormalization scale invariant GC in the $\overline{\rm MS}$ scheme can be conveniently defined as~\cite{Pich:1999hc}
\begin{equation}
\label{eq:GCdefinition}
  \langle \bar{\cal O}_{4,0}\rangle \,=\,\frac{2\pi^2}{3} \,\langle\Omega|
\tilde{\beta}(\alpha_s)
\,G^{\mu\nu}G_{\mu\nu}|\Omega \rangle
  \,\equiv\, \frac{2\pi^2}{3}\,\langle \bar G^2\rangle\,,
\end{equation}
where
\begin{equation}
\tilde{\beta}(\alpha_s)\,\equiv\,- \frac{2\,\beta(\alpha_s)}{\beta_0\,\alpha_s}
\, = \, \Big(\frac{\alpha_s}{\pi}\Big) + \frac{\beta_1}{4\beta_0}\Big(\frac{\alpha_s}{\pi}\Big)^2+\ldots\,.
\end{equation}
Analogous relations also hold for any other scheme to define the strong coupling including the $C$-scheme.
Its one-loop Wilson coefficient correction is known and with the results of Ref.~\cite{Surguladze:1990sp} its contribution to the Euclidean Adler function's OPE series reads
\begin{eqnarray}
\label{eq:AdlerOPEGC}
\delta  D^{\rm OPE}_{4,0}(-Q^2) & = &
\frac{1}{Q^4} \frac{2\pi^2}{3}\,\Big[ 1 + \bar c_{4,0}^{(1)} \, \bar a_Q \Big]\, \langle \bar G^2\rangle \,,
\end{eqnarray}
with
\begin{eqnarray}
\label{eq:AdlerOPEc1}
\bar c_{4,0}^{(1)} = \frac{4}{\beta_0}\left(\frac{C_A}{2}-\frac{C_F}{4}-\frac{\beta_1}{4\beta_0}    \right),
\end{eqnarray}
where $C_A=3$, $C_F=4/3$. For $n_f=3$ we have $\bar c_{4,0}^{(1)}=-22/81$.
The term in the Euclidean Adler function's Borel function (with respect to the expansion in powers of $\bar\alpha_s(Q^2)$) that corresponds to the GC OPE correction has the form
\begin{eqnarray}
\label{eq:AdlerBorelGC}
B_{4,0}(u) & = &
\Big[ 1 + \bar c_{4,0}^{(1)}\, \bar a_Q  \Big]\, \frac{N_{4,0}}{(2-u)^{1+4 \hat b_1}}\,.
\end{eqnarray}
For the purpose of this work we adopt the exact form of Eqs.~(\ref{eq:AdlerOPEGC}) and (\ref{eq:AdlerBorelGC}) in the $C$-scheme, i.e.\  in the Wilson coefficient we truncate all terms at ${\cal O}(\alpha_s^2)$ and beyond. When switching to the $\msb$-scheme, however, we keep all resulting higher-order terms that are generated by the term $\bar c_{4,0}^{(1)}\, \bar a_Q$.

In this work we refer to the non-analytic structure of $B_{4,0}(u)$, its associated asymptotic power series of Eq.~(\ref{eq:AdlerseriesOPEterm}) and the corresponding OPE correction $\delta  D^{\rm OPE}_{4,0}(-Q^2)$ collectively as the `GC renormalon'. Our notation also applies when a general complex-valued momentum transfer $s$ is considered instead of $Q^2$.
A very important phenomenological aspect of the GC renormalon is that for spectral function moments with polynomial weight functions $W(x)$ that do not contain a quadratic term $x^2$ (corresponding to the absence of a linear term $x$ in $w(x)$), the GC renormalon is strongly suppressed. For the GC OPE correction $\delta^{(4)}_{W}(s_0)$ this suppression can be easily seen from the form of the GC corrections to the Adler function for complex-valued momentum transfer $s$.
\begin{eqnarray}
\label{eq:AdlerOPEGCv2}
\delta  D^{\rm OPE}_{4,0}(s) & = &
\frac{1}{s^2} \frac{2\pi^2}{3}\,\Big[ 1 + \bar c_{4,0}^{(1)} \, \bar a(-s) \Big]\, \langle \bar G^2\rangle \,.
\end{eqnarray}
Accounting only for the tree-level Wilson coefficient, i.e.\ neglecting the one-loop correction proportional to $\bar c_{4,0}^{(1)}$, due to the residue theorem $\oint \frac{{\rm d}s}{s} \,\frac{s^m}{s^2}=0$ for an integer $m\neq 2$, the GC OPE correction $\delta^{(4)}_{W}(s_0)$ vanishes identically
for a weight function $W(x)$ that does not contain a quadratic term $x^2$. So for spectral function moments of this kind the GC OPE correction can contribute only through the $s$-dependence of the ${\cal O}(\alpha_s)$ correction of the Wilson coefficient. Since this dependence on $s$ is only logarithmic, the net effect of the GC OPE correction is tiny and negligibly small for practical applications. The total hadronic tau decay rate, which is obtained from the kinematic weight function $W_\tau(x)=1-2x+2x^3-x^4$, belongs to this kind of spectral function moments.
In our work we call moments based on weight functions without a quadratic term {\it GC suppressed (GCS) spectral function moments}. In contrast, for spectral function moments with polynomial weight functions $W(x)$ that contain a quadratic term $x^2$, the numerical size of the GC renormalon is not suppressed. We call these kind of moments {\it GC enhanced (GCE) spectral function moments}.

Due to the one-to-one association of the OPE corrections and the renormalon behavior of the perturbation series, the diverging GC  renormalon corrections in GCS moments are strongly inhibited and the convergence properties of their perturbation series $\delta^{(0)}_{W}(s_0)$ are substantially better behaved at large orders of perturbation theory. It was first pointed out in Ref.~\cite{Beneke:2012vb} that this improved behavior already takes place at the level of the ${\cal O}(\alpha_s^4)$ and ${\cal O}(\alpha_s^5)$ corrections, and that this provides strong support of the proposition that the norm of the GC renormalon  in the Adler function, $N_{4,0}$, is sufficiently sizeable such that the known corrections are quite sensitive to it. In Ref.~\cite{Beneke:2012vb} additional plausibility arguments were provided that  $N_{4,0}$ is sizeable, and it was suggested for that reason that GCE moments may not be employed for high precision strong coupling determinations.
For this reason, most of the recent extractions of the strong coupling  based on hadronic tau decay data rely exclusively~\cite{Boito:2020xli,Boito:2014sta,Boito:2012cr} (or mostly~\cite{Pich:2016bdg}) on GCS spectral function moments. In these recent strong coupling analyses the resulting higher precision concerning the perturbative uncertainties for the GCS moments has already been used as an integral part of their phenomenological analyses  --- regardless of whether this has been  explicitly stated or not. We come back to this point in Sec.~\ref{sec:MultiRenormalonModel}.

The concept of the asymptotic separation suggested in Refs.~\cite{Hoang:2020mkw,Hoang:2021nlz} is particularly important for the GCS spectral function moments. It
states that the discrepancy problem for the FOPT and CIPT expansions of  $\delta^{(0)}_{W}(s_0)$ can be caused by the GC renormalon if the value $N_{4,0}$ is not strongly suppressed. The quite surprising aspect of the asymptotic separation is that the CIPT-FOPT discrepancy for the GCS moments is not at all negligible in contrast to the negligible size of the GC OPE correction. This apparent contradiction was one aspect of the suggestion made in Refs.~\cite{Hoang:2020mkw,Hoang:2021nlz} that the OPE correction that need to be added to $\delta^{(0)}_{W}(s_0)$ in the CIPT approach cannot be based on the standard form assumed in Eq.(\ref{eq:DOPE}), and in particular on Eq.~(\ref{eq:AdlerOPEGCv2}) for the GC. We will come back to this essential issue in Sec.~\ref{sec:toymodel} from the perspective of our renormalon-free GC scheme.

\section{The Renormalon-Free Gluon Condensate Scheme}
\label{sec:subtractedPT}

In the context of the common $\msb$-scheme for the OPE where the limit of a vanishing IR cutoff is taken, the terms in the OPE series do not only provide non-perturbative corrections but, at the same time, also compensate, order-by-order in perturbation theory, for the asymptotically diverging behavior of the QCD corrections associated to the corresponding IR renormalon shown in Eq.~(\ref{eq:AdlerseriesOPEterm}). For the spectral function moments, this mutual cancellation of corrections with IR origin between the perturbation series $\delta^{(0)}_{W}(s_0)$ and the terms in the OPE series implies that the values of the condensate matrix elements $\langle \bar{\cal O}_{d,\gamma}\rangle$ are in general order-dependent and eventually divergent with increasing order of the perturbative calculations (if the series approximation is not truncated at the minimal series term). This $\msb$-scheme OPE approach to set up the perturbative calculations and the OPE corrections, however, only constitutes a particular choice of scheme.

There are two main approaches in the literature to devise alternative OPE schemes to deal with the asymptotically diverging character of the perturbation series associated to IR renormalons. The first is related to adopting a Borel sum of the perturbation series (based on a physically sensible definition that respects properties such as renormalization scale or scheme invariance) as the definition of its true value. In one variant of this approach, approximations to the Borel function of the original series are constructed from the known series coefficients such that the Borel sum can be calculated~\cite{Caprini:2009vf,Caprini:2011ya,Takaura:2020byt}.\footnote{ In Refs.~\cite{Caprini:2009vf,Caprini:2011ya} and a number of subsequent papers the construction of a renormalon-free OPE scheme, and in particular a renormalon-free GC scheme, is not mentioned as a dedicated aim as the primary focus is on the perturbative series. But their approach effectively represents a realization of such a scheme.} In another variant, the fact that the Borel sum can be approximated accurately employing a truncation procedure for the series at the minimal term is used~\cite{Ayala:2019uaw,Ayala:2020pxq}. This variant requires either estimates or the explicit computation of the perturbation series to very high order. In this approach the resulting renormalon-free OPE condensates are scale-invariant.
The second approach is based on subtractions that remove order-by-order in the perturbative series the IR sensitivity of the coefficients (at least with respect to the dominant IR renormalon) such that the convergence properties of the series are improved. This second approach is in the spirit of the use of heavy quark short-distance mass schemes instead of the pole mass scheme~\cite{Hoang:2020iah,Beneke:2021lkq}. In one variant of the second approach the perturbative subtractions are defined from physical quantities which have the same (dominant) IR renormalon such that the subtractions are intrinsically depending on a subtraction scale~\cite{Hoang:2009yr}. This also implies that the resulting renormalon-free OPE condensates are scale-dependent and obey power-dependent renormalization group equations~\cite{Hoang:2008yj}. In another variant of this approach, the momentum integration of the loop diagrams are reformulated analytically such that the Landau pole singularity of the perturbative strong coupling, which is the computational origin of the IR renormalons, can be regulated at least concerning the dominant IR renormalon~\cite{Hayashi:2020ylq}. In this variant it is possible to define the regulated loop expansion in such a way that the resulting series approaches a Borel sum so that the OPE condensates are scale-invariant.

The two main approaches are not fundamentally different but can be viewed as different technical implementations of the same basic idea.
A mixture of these two main approaches has been suggested in Refs.~\cite{Lee:2002sn,Lee:2010hd}, where the construction of the dominant IR renormalon contribution of the Borel function is combined with corresponding order-by-order subtractions for the remaining contributions. All approaches have in common that only one specific kind of IR renormalon or a certain subset of IR renormalons can be dealt with in a systematic manner. Furthermore, because typically only a small number of loop corrections are available and there is no unambiguous way to identify the contributions of different IR renormalons in the perturbative coefficients, all approaches rely on the assumption that the resulting improvement is a systematic all-order effect and not accidental for the available truncation order. This assumption implies that the IR renormalon that is dealt with has sizeable contributions in the known perturbative series terms.

As far as a renormalon-free scheme for the GC is concerned,  which is the topic of this work, the analyses in Refs.~\cite{Caprini:2009vf,Lee:2010hd,Caprini:2011ya,Bali:2014sja,Ayala:2020pxq,Hayashi:2021vdq} (can be viewed to) provide definite implementations of such a scheme. In Refs.~\cite{Caprini:2009vf,Caprini:2011ya,Abbas:2013usa} $\tau$ hadronic spectral function moments were studied, the analyses of Refs.~\cite{Lee:2010hd,Bali:2014sja,Ayala:2020pxq} considered the average SU(3) gauge theory plaquette, and the Euclidean Adler function was investigated in Ref.~\cite{Hayashi:2021vdq}. All analyses based their GC scheme on the common principle value (PV) prescription for the Borel sum\footnote{We specify this prescription concretely in the subsequent sections.} to define the value of the perturbative series, but no dedicated comparison of the various scheme implementations has been carried out so far. Despite this fact we collectively refer to these schemes for the GC as the {\it Borel sum scheme}.
A common aspect of all these previous studies is that they focused primarily on implementing the Borel sum scheme for the GC through modifications of (or prescriptions imposed on)  the original perturbation series. Since the original perturbation series in the $\overline{\rm MS}$ OPE scheme has contributions from infinitely many IR and UV renormalons in addition to regular convergent contributions, the resulting prescriptions are partly quite involved analytically and make it somewhat difficult to implement the scheme exclusively for the GC, i.e.\ without affecting other renormalons at the same time. This makes these scheme implementations somewhat technical, and discussions concerning the universality and observable independence involved.

In our approach to implement a renormalon-free GC scheme, we do not start from modifications of the original perturbation series, but from the GC OPE correction itself. The scheme is set up from a concrete perturbative scheme change of the GC matrix element using the fact that the associated IR renormalon is universal.  This scheme also follows the well-known approach for heavy quark mass schemes where the pole mass can be replaced by a short-distance mass scheme plus scheme change corrections which subsequently act as order-by-order subtraction terms that are combined with the coefficients of the original perturbation series of the mass-sensitive observable of interest. This approach guarantees by construction that our particular scheme choice for the subtraction series is universal among different observables, and at the same time transparent concerning the relation to other alternative renormalon-free GC schemes that follow the same construction principle.

We devise our renormalon-free GC scheme in two steps, where in the first a scale-dependent renormalon-free GC is defined. In the second step this GC scheme is related to a renormalon-free and scale-invariant GC in the Borel sum scheme. The interesting practical aspect of our scheme is that it is defined by a few simple equalities that can be implemented with very little effort for any other observable where the GC OPE correction plays an important role. Since the subtractions generated by our renormalon-free GC scheme do not modify the intrinsic structure of the original Adler function perturbation series, it is then straightforward to study their impact on the FOPT and CIPT series expansions for the spectral function moments in their original definition as given at the end of Sec.~\ref{sec:theorysetup}. To the best of our knowledge, such an analysis has not been carried out before in the literature.

\subsection{Setting up the Scheme}
\label{sec:schemesetup}

As already describe above, the IR subtraction scheme we construct in this work, only deals with the GC and leaves all other OPE corrections and their associated renormalons strictly unaffected. We construct the scheme in two steps.
We start by imposing that for the Euclidean Adler function $D(-Q^2)$ the order-dependent compensating contribution of the GC related to the series terms in Eq.~(\ref{eq:AdlerseriesOPEterm}) for $d=4$, $\gamma=0$ and $\alpha=0$ is made explicit and that the GC correction in the new scheme still has the form shown in Eq.~(\ref{eq:AdlerOPEGC}). We can then write down the relation between the original order-dependent GC $\langle \bar G^2\rangle^{(n)}$ in the $\overline{\rm MS}$ scheme and our new renormalon-free and order-independent GC $\langle G^2\rangle(R)$:
\begin{equation}
\label{eq:GCIRsubtracted}
\langle \bar G^2\rangle^{(n)}
\, \equiv \,
\langle G^2\rangle(R^2)
\,- \,
R^4 \, \sum\limits_{\ell=1}^n \,
N_g\,
r_{\ell}^{(4,0)} \,\bar a^\ell_R\,,
\end{equation}
 where $N_g$ is the universal GC renormalon norm which is related to the GC renormalon norm of the Adler function as defined in Eq.~(\ref{eq:AdlerBorelGC}) by the relation
\begin{equation}
\label{eq:NgN40}
N_g\, = \, \frac{3}{2\pi^2}\,N_{4,0}
\end{equation}
The coefficients $r_{\ell}^{(4,0)}$ are obtained from Eq.~(\ref{eq:rncoeff}).
We remind the reader that the series on the RHS is a power series in the $C$-scheme strong coupling $\bar\alpha_s(R^2)$. The explicit expression for $r_{\ell}^{(4,0)}$ reads
\begin{equation}\label{eq:rn4zero}
r_{\ell}^{(4,0)}=\Big(\frac{1}{2}\Big)^{\ell+4 \hat b_1}\,\frac{\Gamma(\ell+4 \hat b_1)}{\Gamma(1+4 \hat b_1)}.
\end{equation}
The GC in this renormalon-free scheme is by construction scale-dependent and we refer to this (quadratic) scale generically as $R^2$ since it does not need to be equal to $Q^2$.  From a conceptual point of view $R$ plays the role of an IR factorization scale which may be naturally chosen to be smaller than the relevant dynamical scale of the observable of interest, which is $Q$ for $D(-Q^2)$.
We discuss the role of $R$ in more detail in Sec.~\ref{sec:momentsnewscheme}.
The subtraction series in Eq.~(\ref{eq:GCIRsubtracted}), which encodes the ${\cal O}(\Lambda^4_{\rm QCD})$ GC renormalon, is analogous to the one that has been used previously for the definition of the renormalon-subtracted heavy-quark mass definition~\cite{Pineda:2001zq} in order to remove the ${\cal O}(\Lambda_{\rm QCD})$ pole mass renormalon. In Ref.~\cite{Pineda:2001zq} the subtraction series was given in the $\overline{\rm MS}$ scheme for the strong coupling, where an additional summation of subleading terms in powers of $1/\ell$ is mandatory. In the $C$-scheme these subleading terms are absent~\cite{Boito:2016pwf}.

Since here we are considering the Euclidean Adler function, it is reasonable to consider $R^2$ as well as  $\langle G^2\rangle(R^2)$ as real-valued, but this is not strictly mandatory.
The purpose of this renormalon-free GC is to reshuffle the series on the RHS of Eq.~(\ref{eq:GCIRsubtracted}) back into the perturbative series for the Euclidean Adler function $\hat D(-Q^2)$ so that it can explicitly eliminate the effects of the GC renormalon from the original series in the $\msb$ OPE scheme of Eq.~(\ref{eq:Adlerseriesgeneric}). The resulting subtraction series depends on the norm $N_{4,0}=\frac{2\pi^2}{3}N_g$ and is generated by the inverse Borel transform
\begin{eqnarray}
\label{eq:AdlerBorelGCsub}
\delta \hat D_{4,0}(-Q^2,R^2) & = &
\,-\,\Big[ 1 + \bar c_{4,0}^{(1)}\, \bar a_Q  \Big]\,
\int_0^\infty \!\! {\rm d} u \,\, \bigg[\,
 \frac{R^4}{Q^4} \frac{N_{4,0}}{(2-u)^{1+4 \hat b_1}}
\,\bigg]_{\rm Taylor}\,e^{-\frac{u}{\bar a_{R}}}\,,
\end{eqnarray}
where the series in $\bar a_R$ must still be consistently expanded in $\bar a_Q$ and truncated at the same order as the original unsubtracted series, so that the GC renormalon cancels properly.

To see that this subtraction indeed works, let us consider the sum of the GC renormalon contribution of the original series and the subtraction in Eq.~(\ref{eq:AdlerBorelGCsub}),
\begin{eqnarray}
\label{eq:AdlerBorelGCsubtracted}
\Delta \hat D_{4,0}(-Q^2,R^2) & = & \Big[ 1 + \bar c_{4,0}^{(1)}\, \bar a_Q  \Big]\, N_{4,0}\,
\int_0^\infty \!\! {\rm d} u \,\, \bigg[\,
\frac{e^{-\frac{u}{\bar a_{Q}}}}{(2-u)^{1+4 \hat b_1}}
\,-\,
\frac{R^4}{Q^4} \frac{e^{-\frac{u}{\bar a_{R}}}}{(2-u)^{1+4 \hat b_1}}
\,\bigg]\,.
\end{eqnarray}
It is straightforward to show that the ambiguity due to the cuts cancels in the difference of the two terms and that the net series (consistently expanded in $\bar a_Q$) is convergent.\footnote{The factor $R^4/Q^4$ multiplying the second term in the brackets on the RHS of Eq.~(\ref{eq:AdlerBorelGCsubtracted}) is essential for the cancellation of the ambiguity.} Since the integral is Eq.~(\ref{eq:AdlerBorelGCsubtracted}) is well-defined, we dropped the subscript `Taylor'.
The proof can be carried out for example by deforming the $u$-contour slightly above or below the real axis. In this calculation the net imaginary part vanishes identically, which means that the subtracted series is convergent. This can be easily checked numerically. A different more explicit analytic proof starts from the fact that $\Delta \hat D_{4,0}(-Q^2,Q^2)=0$ and uses the equality
\begin{eqnarray}
\label{eq:subtractedderivative}
\frac{{\rm d}}{{\rm d}\ln R^2}\,\Delta \hat D_{4,0}(-Q^2,R^2)
& = &
\,-\,\Big[ 1 + \bar c_{4,0}^{(1)}\, \bar a_Q  \Big]\, N_{4,0}\,
\int_0^\infty \!\! {\rm d} u \,\frac{{\rm d}}{{\rm d}\ln R^2}\,\bigg[\,
\frac{R^4}{Q^4} \frac{e^{-\frac{u}{\bar a_{R}}}}{(2-u)^{1+4 \hat b_1}}
\,\bigg]\nonumber \\
& = &
\,-\,\Big[ 1 + \bar c_{4,0}^{(1)}\, \bar a_Q  \Big]\, N_{4,0}\,
\frac{R^4}{Q^4}\,\frac{\bar a_R}{1-2\hat b_1 \bar a_R}\,2^{-4\hat b_1}\,.
\end{eqnarray}
The derivative with respect to $R$ of Eq.~(\ref{eq:subtractedderivative}) (and thus of the subtraction series on the RHS in Eq.~(\ref{eq:GCIRsubtracted})) is a convergent series as long as the strong coupling remains in the perturbative regime. This is related to the fact that the ambiguity related to the divergence of the subtraction series in Eq.~(\ref{eq:GCIRsubtracted}) is proportional to the fourth power of the QCD scale and independent of the value of $R$, see also  Refs.~\cite{Hoang:2009yr,Hoang:2008yj,Hoang:2017suc,MasterThesisRegner}.  So taking the derivative removes this ambiguity and thus also the divergent asymptotic character of the series.
We can therefore rewrite $\Delta \hat D_{4,0}(-Q^2,R^2)$ as
\begin{eqnarray}
\label{eq:AdlerBorelGCsubtractedv2}
\Delta \hat D_{4,0}(-Q^2,R^2) & = & \,\Big[ 1 + \bar c_{4,0}^{(1)}\, \bar a_Q  \Big]\,
\frac{N_{4,0}}{Q^4\,2^{4\hat b_1}}\,\int\limits_{\ln(R^2)}^{\ln(Q^2)} \!\!  {\rm d}\ln(\tilde R^2) \,\,\frac{\tilde R^4\,\bar a_{\tilde R}}{1-2\hat b_1 \bar a_{\tilde R}}\,,
\end{eqnarray}
which is an exact all-order expression, that can be written as a convergent power series as long as the strong coupling renormalization scale is in the perturbative regime.

The introduction of the subtraction scale $R$ has the important consequence that the parametric power counting concerning the size of the renormalon-free and $R$-dependent GC $\langle G^2\rangle(R^2)$ is ${\cal O}(R^4)$ rather than the fourth power of the QCD hadronization scale that is commonly assigned to the GC in the $\msb$-scheme.
In phenomenological applications the $R$-dependence of the GC $\langle G^2\rangle(R^2)$ can be controlled
by an $R$-evolution equation~\cite{Hoang:2009yr} that has the form
\begin{eqnarray}
\label{eq:G2Revolution}
\frac{{\rm d}}{{\rm d}\ln R^2}\,\langle G^2\rangle(R^2)
& = &
\frac{N_g}{2^{4\hat b_1}}\,\frac{R^4\,\bar a_R}{1-2\hat b_1 \bar a_R}
\, = \,
-\frac{N_g}{2^{1+4\hat b_1}}\,R^4\, \frac{\bar \beta(\bar\alpha_s(R^2))}{\bar\alpha_s(R^2)}
\,.
\end{eqnarray}
It is an interesting fact that the $R$-dependence of the GC subtraction series defined in Eq.~(\ref{eq:GCIRsubtracted}) is associated to the $\beta$-function in the $C$-scheme and appears to have an association to the definition of the scale-invariant GC given in Eq.~(\ref{eq:GCdefinition}). However, the form of Eq.~(\ref{eq:G2Revolution}) relies on the particular form of the subtraction series given on the RHS of Eq.~(\ref{eq:GCIRsubtracted}), so that the expression on the RHS of Eq.~(\ref{eq:G2Revolution}) is in general scheme-dependent as well. However, since we can modify the scheme only by adding an additional convergent power series in $\bar a_R$ on the RHS of Eq.~(\ref{eq:GCIRsubtracted}), the property that the $R$-evolution equation has a convergent power expansion remains valid in all sensible schemes.

Still, the $R$-dependence of $\langle G^2\rangle(R^2)$ is not very convenient from the practical perspective, and we therefore take one additional step to define a scale-invariant renormalon-free GC matrix element. What we need is a closed function that obeys the same $R$-evolution equation.
Interestingly, such a function can be obtained from the Borel sum of the subtraction series in Eq.~(\ref{eq:GCIRsubtracted}) defined by
\begin{eqnarray}
\label{eq:subtractclosed}
\bar c_0(R^2) & \equiv & R^4\,\,\,{\rm PV}\,\int_0^\infty \!
\frac{ {\rm d} u \,\,e^{-\frac{u}{\bar a_R}}}{(2-u)^{1+4 \hat b_1}} \\ \nonumber
& = & \left\{
\begin{array}{ll}
- \frac{R^4 \,e^{-\frac{2}{\bar a_R}}}{(-\bar a_R)^{4\hat b_1}}\,
\Gamma\Big(\!-4\hat b_1,-\frac{2}{\bar a_R}\,\Big) -
{\rm sig}[{\rm Im}[a_R]]\,\frac{i\pi\,R^4 \,e^{-\frac{2}{\bar a_R}}}{\Gamma(1+4\hat b_1)\,\bar a_R^{4\hat b_1}}
& \mbox{\quad for\,\,}{\rm Im}[
R^2
]\neq 0
\\
- \frac{R^4\,e^{-\frac{2}{\bar a_R}}}{(\bar a_R)^{4\hat b_1}}\,
{\rm Re}\left[\, e^{4\pi \hat b_1 i}\, \Gamma\Big(\!-4\hat b_1,-\frac{2}{\bar a_R}\,\Big)\,\right]
&  \mbox{\quad for\,\,}{\rm Im}[
R^2
] = 0
\end{array}
\right.
\,,
\end{eqnarray}
where for completeness we have also displayed the result for complex $R^2$. Note that the expression for the general
complex-valued strong coupling reduces to the expressions of the real-valued one in the limit ${\rm Im}[\bar a_R]\to 0$. The $R^2$-derivative of $\bar c_0(R^2)$ gives exactly the expression on the RHS of Eq.~(\ref{eq:G2Revolution}) divided by $N_g$ for any complex $R^2$ for which the strong coupling $\alpha_s(R^2)$ is analytic. With this definition it is in principle possible to even consider complex values for $R^2$. However, in the following we only discuss real-valued $R^2$. As a consequence, the subtraction series in Eq.~(\ref{eq:GCIRsubtracted}) is intrinsically real-valued as well and should   in principle be strongly suppressed for the GCS spectral function moment, in the same way as the GC OPE correction. This is an essential aspect in the analysis carried out in Sec.~\ref{sec:toymodel}.

The Borel sum in Eq.~(\ref{eq:subtractclosed}) is a priori not unique due to the cut along the positive real axis, and we have adopted the common principal value prescription (PV), which is the average of deforming the contour above and below the real $u$-axis. The prescription is, however, not in any way essential since the choice of the function $\bar c_0(R^2)$ is simply defining the scheme of our scale-invariant and renormalon-free GC. In fact, any other choice for $\bar c_0(R^2)$ (related to adding a constant on the RHS of Eq.~(\ref{eq:subtractclosed})) would be equally feasible, as long as it satisfies the same $R$-evolution equation. We define our final scale-invariant renormalon-free GC matrix element $\langle G^2\rangle^{\rm RF}$ by the relation
\begin{eqnarray}
\label{eq:GCIRsubtracted2}
\langle G^2\rangle(R^2)
& \equiv &
\langle G^2\rangle^{\rm RF} + N_g \, \bar c_0(R^2)\,.
\end{eqnarray}
Our particular choice for the function $\bar c_0(R^2)$ has the nice feature that it implements the renormalon-free Borel sum scheme  as we show explicitly in Sec.~\ref{sec:momentsnewscheme}.
This means that $\langle G^2\rangle^{\rm RF}$ is closely related to the
scheme definitions implemented in Refs.~\cite{Caprini:2009vf,Lee:2010hd,Caprini:2011ya,Bali:2014sja,Ayala:2020pxq,Hayashi:2021vdq}.

We stress again that neither the exact form of the subtraction series in Eq.~(\ref{eq:GCIRsubtracted}) nor the function $\bar c_0(R^2)$ are in principle unique. The subtraction series merely needs to have the same asymptotic large order behavior as the one shown in Eq.~(\ref{eq:AdlerseriesOPEterm}) but may have additional convergent contributions.\footnote{Here we use the naming `convergent series' to signify that the series has a finite radius of convergence. We use naming 'a series converges to a value' to signify that the series converges to the value when the expansion parameter is smaller than the radius of convergence.} The function $\bar c_0(R^2)$ has mainly been introduced for practical convenience.
We have adopted a choice for $\bar c_0(R^2)$ such that it agrees with the Borel sum of the subtraction series as defined in Eq.~(\ref{eq:subtractclosed}) for any value of $N_g$. As a consequence the difference between our scale-invariant and renormalon-free GC $\langle G^2\rangle^{\rm RF}$ and the original
order-dependent $\overline{\rm MS}$ condensate $\langle \bar G^2\rangle^{(n)}$ is formally $\mathcal{O}\left(\bar a_R^{n+1} \right)$. In this sense our scheme can be considered minimal. (See also the comment after Eq.~(\ref{eq:invBorelDR}).) In this context the Borel sum's ambiguity due to the renormalon cut (which is $R$-independent) amounts to the freedom of defining an alternative scheme choice for $\bar c_0(R^2)$. The essential point is that our scheme is unambiguous and well-defined based on concrete expressions given in Eqs.~(\ref{eq:GCIRsubtracted}), (\ref{eq:subtractclosed}) and  (\ref{eq:GCIRsubtracted2}) and that it is straightforward to compute the perturbative relation to any other sensible renormalon-free GC scheme in a renormalon-free way.
A convenient practical property of our GC scheme is that for GCS spectral function moments the perturbation series are numerically close to the original unsubtracted series at intermediate orders.

We stress that our scheme is already fully specified from Eqs.~(\ref{eq:GCIRsubtracted}), (\ref{eq:subtractclosed}) and  (\ref{eq:GCIRsubtracted2}) and can in principle be implemented for any observable where the GC OPE correction plays an important role. As a consequence the observable independence of our scheme is manifest, an issue that is somewhat more involved in the approaches of Refs.~\cite{Caprini:2009vf,Lee:2010hd,Caprini:2011ya,Bali:2014sja,Ayala:2020pxq,Hayashi:2021vdq}
where renormalon-free schemes have been implemented directly at the level of the observable-dependent perturbation series.

We finally note that Eq.~(\ref{eq:GCIRsubtracted2}) implies the relation
\begin{equation}
\label{eq:G2scalerelation}
\langle G^2\rangle(R^2) - \langle  G^2\rangle(R^{\prime 2})
\,= \,
N_g\,[ \, \bar c_0(R^2) - \bar c_0(R^{\prime 2})\,]\,.
\end{equation}
This is compatible with Eq.~(\ref{eq:GCIRsubtracted}): the difference of the subtraction series for $R$ and $R^\prime$, which is renormalon- and ambiguity-free, indeed sums up to $\bar c_0(R^2) - \bar c_0(R^{\prime 2})$. To evaluate the difference correctly perturbatively, it is essential to expand the difference series with a common renormalization scale for the strong coupling such that the renormalon contained in both individual series can properly cancel order-by-order. Equality~(\ref{eq:G2scalerelation}) also shows that a possible alternative to our scale-invariant $\langle G^2\rangle^{\rm RF}$ would be to consider the $\langle G^2\rangle(R_0^2)$ at some reference scale $R_0^2$, which would yield
instead of Eq.~(\ref{eq:GCIRsubtracted2}) the equality $\langle G^2\rangle(R^2) = \langle G^2\rangle(R_0^2)
+ N_g\tilde c_0(R^2,R_0^2)$ with $\tilde c_0(R^2,R_0^2)=\bar c_0(R^2) - \bar c_0(R^{\prime 2})$.
Such a scale-dependent gluon condensate $\langle G^2\rangle(R_0^2)$ would be in analogy to the RS (renormalon subtracted) scheme advocated in Refs.~\cite{Pineda:2001zq,Bali:2003jq,Campanario:2005np}.
We also note that for large hierarchies between $R$ and $R^\prime$ the difference of the functions $\bar c_0(R^2)$ and $\bar c_0(R^{\prime 2})$ also provides a summation of large logarithms of $R/R^\prime$~\cite{Bali:2003jq,Campanario:2005np,Hoang:2008yj,Hoang:2009yr}.

\subsection{Spectral Function Moments in the New Scheme}
\label{sec:momentsnewscheme}

The GC scheme we have defined in the previous section subtracts the GC renormalon from the real-valued Euclidean Adler function. Due to the universality of the renormalons, the same scheme (for real-valued $R$) can also be applied for the complex-valued Adler function $D(s)$ and subsequently the spectral function moments. The scheme can of course be applied for any quantity for which the GC matrix element appears as an OPE correction.
In our scheme it is convenient to treat the function $\bar c_0(R^2)$, which is conceptually part of the GC OPE correction, like a {\it tree-level contribution that is not supposed to be expanded any more in powers of the strong coupling.}\footnote{This implies that $\delta^{(0)}_{W}(s_0)$ in general contains this tree-level contribution in our renormalon subtracted GC scheme.}
Parametrizing the subtraction series generated by Eq.~(\ref{eq:GCIRsubtracted}) back into a Borel function, the perturbation series for the Adler function $\hat D^{\rm RF}(s)$ in the renormalon-free GC scheme can be concisely written as
\begin{eqnarray}
\label{eq:invBorelDR}
\hat D^{\rm RF}(s,R^2) & = &
\frac{1}{s^2}\,\Big[ 1 + \bar c_{4,0}^{(1)} \bar a(-s) \Big]\, N_{4,0}\,\bar c_0(R^2)
\, + \,
\int_0^\infty \!\! {\rm d} u \,\, \bigg[\,B[\hat D(s)](u)\,\bigg]_{\rm Taylor}\,e^{-\frac{u}{\bar a(-s)}}
\nonumber \\
&&
\, - \,
\Big[ 1 + \bar c_{4,0}^{(1)}\, \bar a(-s)  \Big]\,N_{4,0}\,\frac{R^4}{s^2} \int_0^\infty \!\! {\rm d} u \,\, \bigg[\,
\frac{1}{(2-u)^{1+4 \hat b_1}}\,\bigg]_{\rm Taylor}\,e^{-\frac{u}{\bar a_R}}
\nonumber \\
& = &
\frac{1}{s^2}\,\Big[ 1 + \bar c_{4,0}^{(1)} \bar a(-s) \Big]\,N_{4,0}\, \bar c_0(R^2)
\, + \,
\sum_{\ell=1}^\infty \, \bar c_{\ell} \,\bar a^\ell(-s)\,
\nonumber \\
&&
\, - \,
\Big[ 1 + \bar c_{4,0}^{(1)}\, \bar a(-s)  \Big]\, N_{4,0}\,\frac{R^4}{s^2}
\sum\limits_{\ell=1}^\infty
\Big(\frac{1}{2}\Big)^{\ell+4 \hat b_1}\,\frac{\Gamma(\ell+4 \hat b_1)}{\Gamma(1+4 \hat b_1)} \,\bar a^\ell_R\,,
\end{eqnarray}
where $B[\hat D(s)](u)$ is the Borel function for the (original) Adler function in the $\overline{\rm MS}$ GC scheme with respect to the expansion in powers of $\bar\alpha_s(-s)$.
The GC OPE correction adopts the standard form
\begin{eqnarray}
\label{eq:AdlerOPEGCBS}
\delta D^{\rm OPE,RF}_{4,0}(s) & = &
\frac{1}{s^2}\frac{2\pi^2}{3}\,\Big[ 1 + \bar c_{4,0}^{(1)} \, \bar a(-s) \Big]\, \langle G^2\rangle^{\rm RF} \,.
\end{eqnarray}
We stress again, that for the correct perturbative evaluation of $\hat D^{\rm RF}(s,R^2)$ it is essential to consistently expand and truncate the series terms using the strong coupling at a common renormalization scale such that the GC renormalon cancels. We also remind the reader that the expressions in the above equations are written in the $C$-scheme for the strong coupling $\bar\alpha_s$ and that  for our numerical analyses below we have applied an additional reexpansion in terms of the $\overline{\rm MS}$ coupling $\alpha_s$.  The scheme change back to the $\overline{\rm MS}$ coupling is, however, not at all mandatory. For the determination of the perturbation series for the spectral function moments in the CIPT approach the expansion is in powers of $\alpha_s(-s)$; for the FOPT approach the expansion is in powers of $\alpha_s(s_0)$.\footnote{\label{ftn:scalevariation} For the scale variations considered in Secs.~\ref{sec:largeb0} and \ref{sec:MultiRenormalonModel} actually expand in powers of $\alpha_s(-\xi s)$ and $\alpha_s(\xi s_0)$, respectively, for $1/2\le\xi\le2$.}
We also remind the reader that the relation between the renormalon norm of the GC, $N_g$, and the norm of the GC renormalon in the Adler function, $N_{4,0}$, is given by $N_g=\frac{3}{2\pi^2}N_{4,0}$.

From the expressions in the first equality in Eq.~(\ref{eq:invBorelDR}) it is straightforward to prove that our renormalon-free GC scheme ensures that the Borel sum of  $\hat D^{\rm RF}(s,R^2)$ based on the PV prescription as defined in Eq.~(\ref{eq:subtractclosed})
is the same as the one for the original unsubtracted Adler function $\hat D(s)$ {\it irrespective of the value adopted for the norm parameter $N_g$}. Defining the Borel function of $\hat D^{\rm RF}(s,R^2)$ with respect to the series expansion in powers of $\bar \alpha_s(-s)$ as $B[\hat D^{\rm RF}(s)](u)$ we obtain the identity
\begin{eqnarray}
\label{eq:invBorelDRv2}
\lefteqn{\frac{1}{s^2}\,\Big[ 1 + \bar c_{4,0}^{(1)} \bar a(-s) \Big]\, N_{4,0}\,\bar c_0(R^2)
\, + \,
{\rm PV} \int_0^\infty \!\! {\rm d} u \,\, \bigg[\,B[\hat D^{\rm RF}(s)](u)\,\bigg]\,e^{-\frac{u}{\bar a(-s)}}
}
\nonumber \\
& = &
\frac{1}{s^2}\,\Big[ 1 + \bar c_{4,0}^{(1)} \bar a(-s) \Big]\,N_{4,0}\,\bar c_0(R^2)
\, + \,
{\rm PV}\int_0^\infty \!\! {\rm d} u \,\, \bigg[\,B[\hat D(s)](u)\,\bigg]\,e^{-\frac{u}{\bar a(-s)}}
\nonumber \\
&&
\, - \,
\Big[ 1 + \bar c_{4,0}^{(1)}\, \bar a(-s)  \Big]\, N_{4,0}\,\frac{R^4}{s^2}\, {\rm PV}\int_0^\infty \!\! {\rm d} u \,\, \bigg[\,
\frac{1}{(2-u)^{1+4 \hat b_1}}\,\bigg]\,e^{-\frac{u}{\bar a_R}}
\nonumber \\
& = &
{\rm PV}\int_0^\infty \!\! {\rm d} u \,\, \bigg[\,B[\hat D(s)](u)\,\bigg]\,e^{-\frac{u}{\bar a(-s)}}\,,
\end{eqnarray}
where the `tree-level' term involving the function $\bar c_0$ appears explicitly since the Borel function only accounts for terms with positive powers of the coupling.
The first equality follows from the fact that the Borel sum based on the PV prescription of any given perturbation series is independent of the choice of the renormalization scale for the strong coupling from which one defines the Borel function of the series. The second equality follows from the definition for $\bar c_0$ given in Eq.~(\ref{eq:subtractclosed}).
The fact that the Borel sum based on the PV prescription is also independent of the choice of the renormalization scheme for the strong coupling\footnote{
The renormalization scale and renormalization scheme invariance of the Borel sum based on the PV prescription is a well-known accepted fact in the renormalon calculus. It is trivial to see in the large-$\beta_0$ approximation, where essentially only renormalization scale variations can be considered. Here the Borel function $B_\mu(u)$ with respect to an expansion in $\alpha_s(\mu^2)$ is related to the Borel function with respect $\alpha_s(\tilde\mu^2)$ through the relation $B_{\tilde\mu}(u)=B_\mu(u)e^{-u\ln(\tilde\mu^2/\mu^2)}$
and the invariance can be seen trivially using $e^{-4\pi u/\beta_0\alpha_s(\mu^2)} = e^{-4\pi u/\beta_0\alpha_s(\tilde\mu^2)}e^{+u\ln(\tilde\mu^2/\mu^2)}$. An explicit examination in the context of full QCD has been discussed in Ref.~\cite{Ayala:2019uaw}.}
then allows us to conclude that the Borel sum of  $\hat D^{\rm RF}(s,R^2)$, based on the principal value prescription as defined in Eq.~(\ref{eq:subtractclosed}),
is the same as the one for the original unsubtracted Adler function $\hat D(s)$ in any renormalization scheme of the strong coupling, which includes of course also the $\overline{\rm MS}$ scheme.  It is also obvious that all these statements apply equally well to any other quantity for which the GC appears as an OPE correction.
This shows that the renormalon-free GC scheme we have defined is indeed a realization of the Borel sum scheme based on the PV prescription to define the Borel sum. In the remainder of this paper we refer to our renormalon-free GC scheme simply as the `RF GC scheme' or the `RF scheme'.

Even though the PV Borel sums of the original  $\overline{\rm MS}$ and the RF scheme Adler functions are identical and  $\langle G^2\rangle^{\rm RF}$ is a scale invariant quantity,
which shows that the dependence of $\hat D^{\rm RF}(s,R^2)$  on the IR factorization scale $R$ formally vanishes in the limit of large orders, it is clear from the form of Eq.~(\ref{eq:invBorelDR}) that finite truncations depend on the value for $R$, regardless which renormalization scale prescription for the strong coupling is used.
We can view variations in $R$ as a different ways to sum logarithms involving the scale ratio $R^2/s_0$. So, while it is natural consider
$R^2<s_0$, since $R$ constitutes an IR factorization scale, the existence of logarithms of $R^2/s_0$ implies that the strong hierarchy $R^2\ll s_0$ should be avoided. This means that $R^2$ should be taken smaller, but still of order $s_0$, i.e.\ $R^2\lesssim s_0$. If widely different hierarchical choices were possible for $s_0$, this prescription could take care of the proper resummation of all potentially large logarithms related to the RF scheme.
For the $\tau$ hadronic spectral functions such hierarchical choices do, however, not exist.
Because $R^2$ should remain in the realm of QCD perturbation theory, and $s_0$ is bounded from above by the square of the $\tau$ mass, the freedom in the choice of $R$ is quite limited  for the spectral function moments and the problem of large logarithms does never arise in practice. Still, variations of $R$ constitute an important practical (and in our view very welcome) diagnostic tool for the reliability and effectiveness of the renormalon subtraction scheme at finite truncation orders. Variations of $R$ should therefore play an important new role in the estimate of the perturbative uncertainties when the RF  scheme is used in phenomenological applications.

\subsection{Toy Model Analysis and the Breakdown of the Standard OPE for CIPT}
\label{sec:toymodel}

Following Refs.~\cite{Hoang:2020mkw,Hoang:2021nlz}, the cancellation of the GC renormalon in $\hat D^{\rm RF}(s,R^2)$ entails that the major source of the disparity between the FOPT and CIPT series for the hadronic $\tau$ decay spectral function moments at intermediate orders is eliminated. So using $\hat D^{\rm RF}(s,R^2)$ instead of the original unsubtracted series in Eq.~(\ref{eq:Dresum}) for the parton-level Adler function, the perturbation series for the moments obtained from both methods should provide compatible perturbative descriptions.

It is well worth to explore this using a very simplistic toy model for the Adler function, which contains only the GC renormalon contribution for $N_{4,0}=\frac{2\pi^2}{3} N_g =1$ and $\bar c_{4,0}^{(1)}=0$.
The Borel function of the complex-valued parton-level Adler function for this toy model with respect to the expansion in powers of $\bar\alpha_s(-s)$ reads
\begin{eqnarray}
\label{eq:AdlerBorelToy}
B[\hat D_{\rm toy}(s)](u) & = & \frac{1}{(2-u)^{1+4 \hat b_1}}\,.
\end{eqnarray}
In this model the Adler function's perturbation series in the $\msb$ GC scheme has the form
\begin{eqnarray}
\label{eq:AdlerGCmodel}
\hat D_{\rm toy}(s) & = &
\sum\limits_{\ell=1}^\infty r_\ell^{(4,0)}\,\bar a^\ell(-s)\,,
\end{eqnarray}
with $r_\ell^{(4,0)}$ given in Eq.~(\ref{eq:rn4zero}),  while the corresponding series in the RF scheme reads
\begin{eqnarray}
\label{eq:AdlerGCmodelsub}
\hat D_{\rm toy}^{\rm RF}(s,R^2) & = &
\frac{\bar c_0(R^2)}{s^2}
\, + \,
\sum\limits_{\ell=1}^\infty \,
r_\ell^{(4,0)} \,\bar a^\ell(-s) -
\frac{R^4}{s^2}
\sum\limits_{\ell=1}^\infty
r_\ell^{(4,0)} \,\bar a^\ell_R\,.
\end{eqnarray}
This toy model does not have any direct phenomenological relevance, but it illustrates the impact the GC renormalon has concerning the disparity between the CIPT and FOPT moment perturbation series at intermediate orders using $\hat D_{\rm toy}(s)$ in the $\overline{\rm MS}$ GC scheme,
and how this discrepancy is eliminated when the RF scheme is applied in $\hat D_{\rm toy}^{\rm RF}(s,R^2)$. We can also use the moments obtained from $\hat D_{\rm toy}(s)$ to test the numerical size of the asymptotic separation determined in Ref.~\cite{Hoang:2020mkw} and to dwell further on the apparent contradiction we pointed out at the end of Sec.~\ref{sec:renormalon}. The important phenomenological aspect of this toy model is that its series shown in Eq.~(\ref{eq:AdlerGCmodel}) constitutes an important contribution of the true Adler function series coefficients, if the GC renormalon norm $N_{4,0}$ of the Adler function in full QCD is sizeable (and not strongly suppressed).

To be definite we consider the perturbative spectral function moments
\begin{eqnarray}
\label{eq:deltatoy}
\delta^{(0)}_{W}(s_0)
& = &
\frac{1}{2\pi i}\,\,
\ointctrclockwise_{|s|=s_0}  \frac{{\rm d}s}{s}\,W({\textstyle \frac{s}{s_0}})\, \hat D_{\rm toy}(s)
\\
\delta^{(0)}_{W}(s_0,R^2)
& = &
\frac{1}{2\pi i}\,\,
\ointctrclockwise_{|s|=s_0}  \frac{{\rm d}s}{s}\,W({\textstyle \frac{s}{s_0}})\, \hat D^{\rm RF}_{\rm toy}(s,R^2)
\end{eqnarray}
for the weight functions $W_m(x)=2(1-x^{m})/m$ with $m=1,3,4,5$. They arise from the monomial spectral weight functions $w_{m-1}(x)=x^{m-1}$
and are relevant for the total hadronic decay width, $W_\tau(x) = W_1(x)-3W_3(x)+2W_4(x)$.
They all lead to GCS moments where the strong divergent behavior of GC renormalon series is damped.  Our Adler function toy model has the particular feature (because $\bar c_{4,0}^{(1)}=0$) that for these spectral functions moments the GC OPE correction as well as the scheme compensating term $\bar c_0(R^2)/s^2$
vanish identically.
Following the standard logic concerning the association of the OPE corrections and IR renormalons, the results based on expansion methods that are consistent with this standard association, have to satisfy the following conditions:
\begin{itemize}
\item [(i)] The perturbation series for the spectral function moments of the original unsubtracted series $\delta^{(0)}_{W}(s_0)$  in the $\overline{\rm MS}$ GC scheme is convergent.
\item [(ii)] The subtractions generated by the change from the $\msb$ GC scheme to the RF scheme, i.e.\ the series for  $\delta^{(0)}_{W}(s_0,R^2)-\delta^{(0)}_{W}(s_0)$ either vanish or at least constitute a series that converges to zero.
\end{itemize}
Any expansion method for which the unsubtracted and subtracted series for $\delta^{(0)}_{W}(s_0)$ and $\delta^{(0)}_{W}(s_0,R^2)$, respectively, do not satisfy conditions (i) and (ii) are not compatible with the standard association of the OPE corrections and IR renormalons.
In the following we show that (i) and (ii) are realized for the FOPT expansion method, but they are not for the CIPT expansion method. For constant $R^2$, property (ii) is true for the FOPT moment series by construction:
\begin{eqnarray}
\label{eq:FOPTexp}
\frac{1}{2\pi i}\,\,
\ointctrclockwise_{|s|=s_0}  \frac{{\rm d}s}{s}\,W_{m\neq 2}({\textstyle \frac{s}{s_0}})\,\bigg[\!-\frac{R^4}{s^2}
\sum\limits_{\ell=1}^\infty
r_\ell^{(4,0)} \,\bar a^\ell_R \,\bigg]^{\mbox{FOPT}} \, = \, 0\,.
\end{eqnarray}
This is because the FOPT expansion method is based on an expansion in terms of the real valued coupling $\alpha_s(s_0)$ (or $\alpha_s(\xi s_0)$) such that the subtraction series (generated by the third term in Eq.~(\ref{eq:AdlerGCmodelsub})) is constant and real-valued up to the overall factor $1/s^2$. So the subtraction series contributions at any order (and the GC OPE correction in Eq.~(\ref{eq:AdlerOPEGC}) as well as the term $\bar c_0(R^2)/s^2$) vanish identically due to the contour residue relation
\begin{eqnarray}
\label{eq:contour}
\frac{1}{2\pi i}\,\,
\ointctrclockwise_{|s|=s_0}  \frac{{\rm d}s}{s}\,\left(\!-\frac{s}{s_0}\right)^k\, \frac{1}{s^2}
\, = \,
\left\{
\begin{array}{ll}
0\,,  & \,\,\mbox{for}\,\,k\neq 2 \\
\frac{1}{s_0^2}  & \,\,\mbox{for}\,\,k = 2
\end{array}
\right.
\,.
\end{eqnarray}

\begin{figure}
	\centering
	\begin{subfigure}[b]{0.44\textwidth}
		\includegraphics[width=\textwidth]{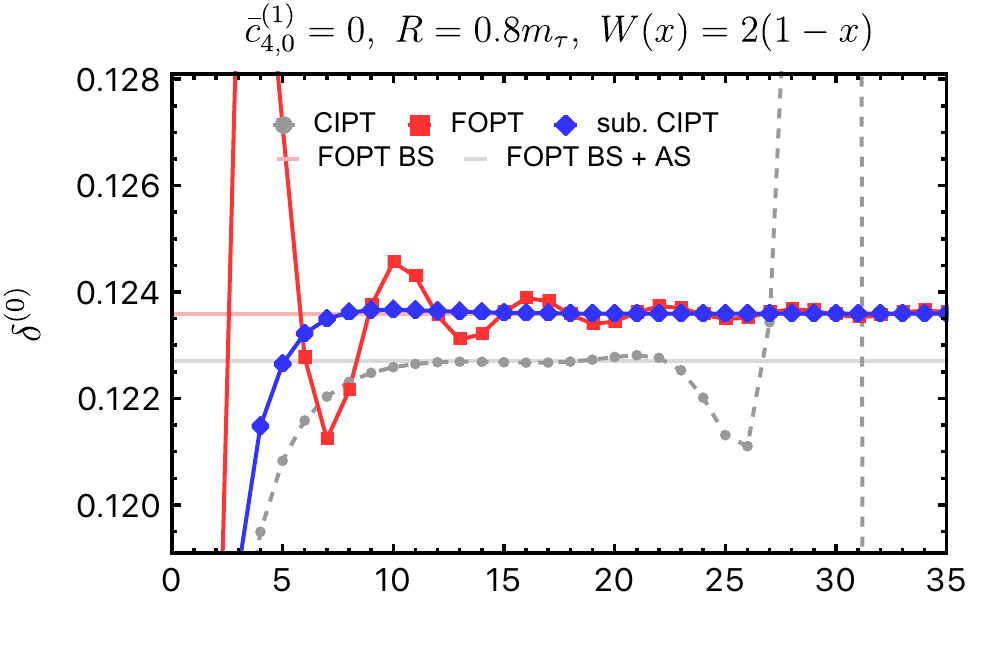}
	\end{subfigure}
	~ 
	\begin{subfigure}[b]{0.44\textwidth}
		\includegraphics[width=\textwidth]{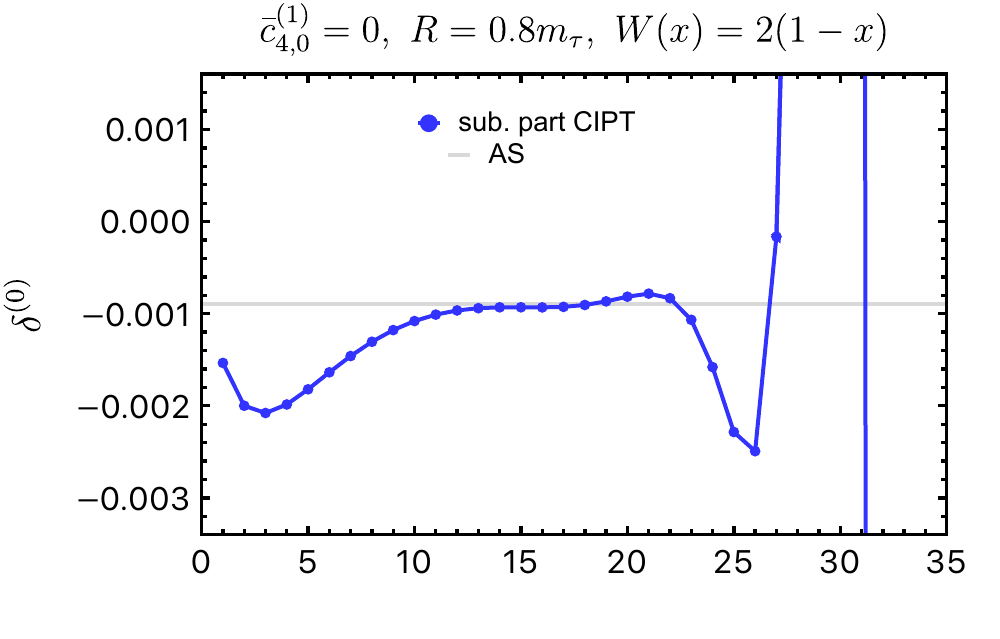}
	\end{subfigure}

	\begin{subfigure}[b]{0.44\textwidth}
		\includegraphics[width=\textwidth]{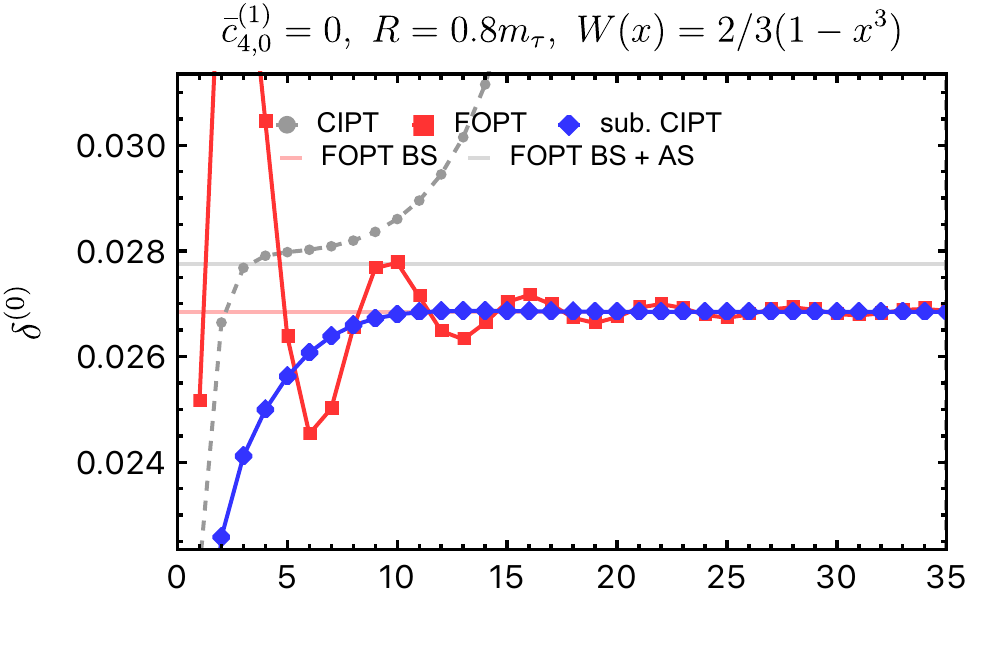}
	\end{subfigure}
	~
	\begin{subfigure}[b]{0.44\textwidth}
		\includegraphics[width=\textwidth]{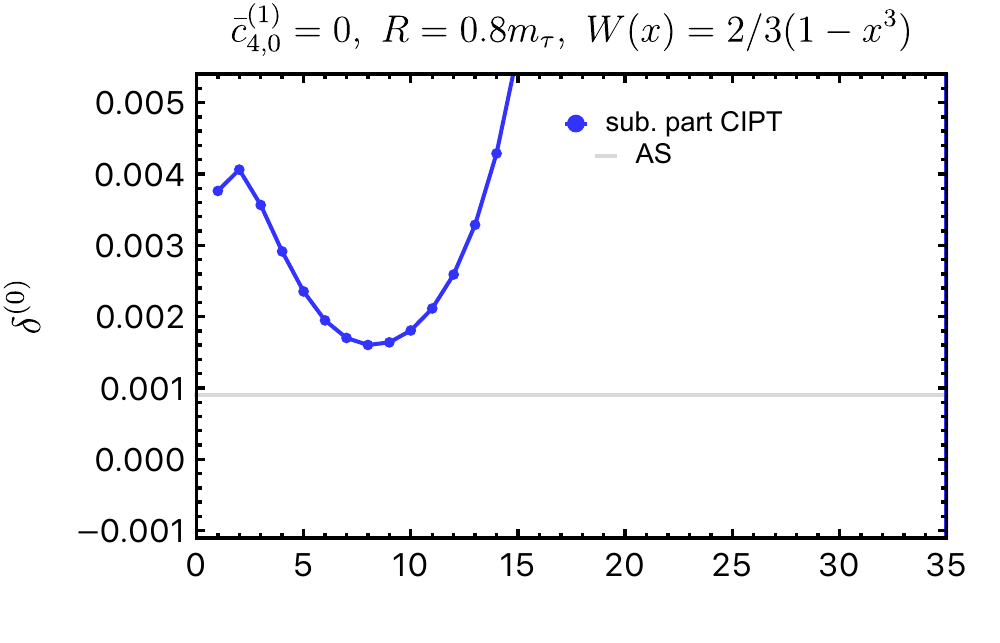}
	\end{subfigure}

	\begin{subfigure}[b]{0.44\textwidth}
		\includegraphics[width=\textwidth]{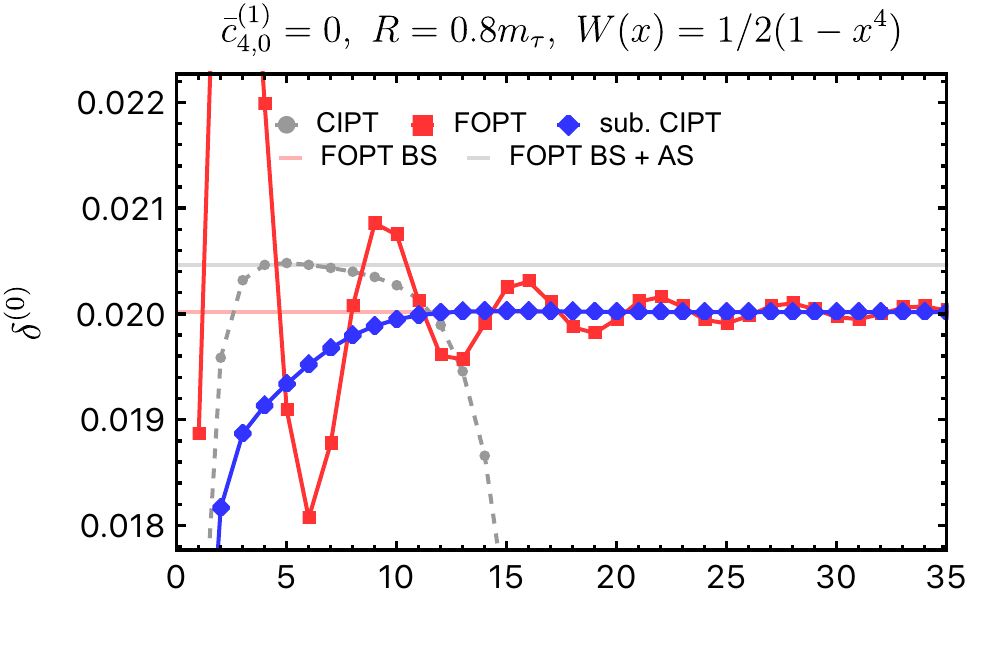}
	\end{subfigure}
	~
	\begin{subfigure}[b]{0.44\textwidth}
		\includegraphics[width=\textwidth]{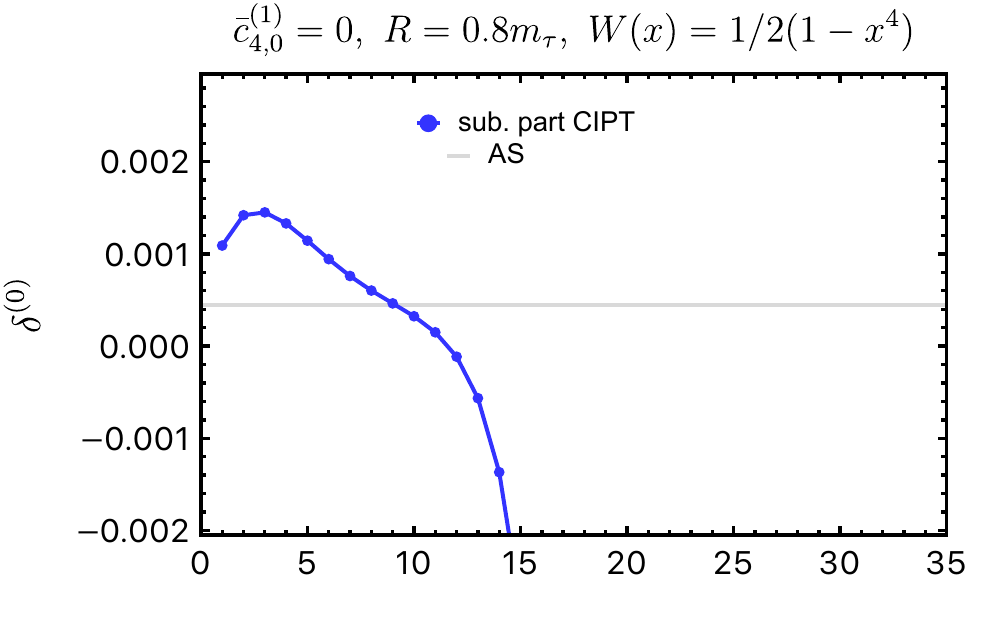}
	\end{subfigure}

	\begin{subfigure}[b]{0.44\textwidth}
		\includegraphics[width=\textwidth]{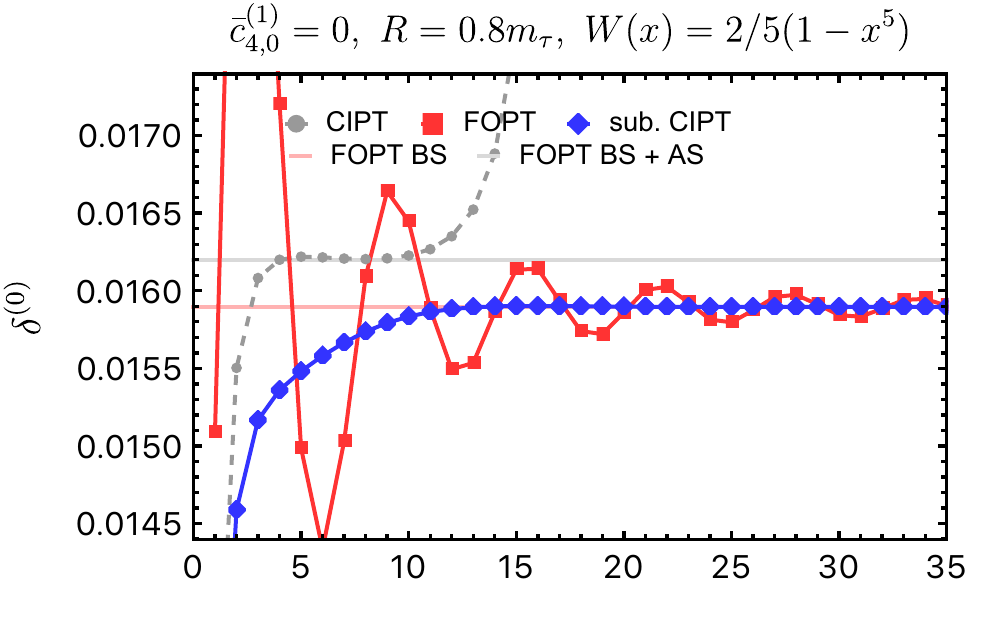}
	\end{subfigure}
	~
	\begin{subfigure}[b]{0.44\textwidth}
		\includegraphics[width=\textwidth]{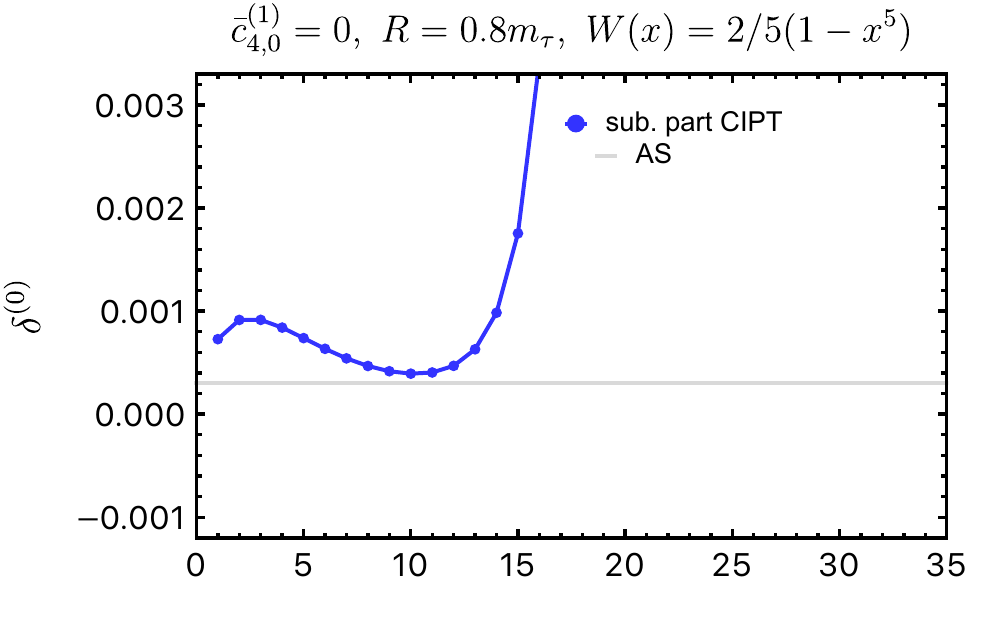}
	\end{subfigure}
	\caption{\label{fig:niceplots}
	Left panels: Series for  $\delta^{(0)}_{W}(m_\tau^2)$ and $\delta^{(0)}_{W}(m_\tau^2,R^2)$ for $W(x)=2(1-x^{m})/m$ with $m=1,3,4,5$ and the FOPT and CIPT expansion method for $\alpha_s(m_\tau^2)=0.315$ ($\bar\alpha_s(m_\tau^2)=0.30827$)
	Righ panels: The CIPT expansion for subtraction contribution $\delta^{(0)}_{W}(m_\tau^2,R^2)-\delta^{(0)}_{W}(m_\tau^2)$ for the CIPT series in the corresponding left panels.}
\end{figure}

In  the left panels of Fig.~\ref{fig:niceplots} we show $\delta^{(0)}_{W}(s_0)$ and  $\delta^{(0)}_{W}(s_0,R^2)$ for $W_m(x)=2(1-x^{m})/m$ with $m=1,3,4,5$ for $s_0=m_\tau^2=(1.77686~\mbox{GeV})^2$ and $\alpha_s(m_\tau^2)=0.315$ ($\bar\alpha_s(m_\tau^2)=0.30827$) as a function of order up to order 35 (in the $\overline{\rm MS}$ scheme for the strong coupling). For the subtraction scale we have adopted $R=0.8\,m_\tau$. The results for other $R$ values look very similar as long as $R$ is larger than $1.2$~GeV. Values for $R$ smaller than $0.7\,m_\tau$ should not be considered since the subtraction series becomes unstable due to the size of the strong coupling.

The red dots are the FOPT series where the original unsubtracted and the subtracted series terms agree identically, i.e.
$\delta^{(0)}_{W}(s_0)=\delta^{(0)}_{W}(s_0,R^2)$ at any order in the FOPT expansion. The FOPT series oscillate and eventually converge\footnote{We note that the FOPT series has a finite radius of convergence which is not precisely known and which could actually be slightly smaller than the value $\alpha_s(m_\tau^2)=0.315$~\cite{LeDiberder:1992jjr}.}
to the values of the Borel sum integral
\begin{equation}
\label{eq:BorelFOPT}
	\delta_{W_m,{\rm Borel}}^{(0),{\rm FOPT}}(s_0) = {\rm PV} \int_0^\infty \!\! {\rm d} u \,\,
	\frac{1}{2\pi i}\,\ointctrclockwise_{|s|=s_0}  \frac{{\rm d}s}{s}\,W_m({\textstyle \frac{s}{s_0}}) \,
	\,
	\frac{e^{-\frac{u}{\bar a(-s)}}}{(2-u)^{1+4 \hat b_1}}\,,
\end{equation}
indicated by the red horizontal line.  The oscillations observed in the FOPT  series are a particular feature of the simplistic toy model~(\ref{eq:AdlerBorelToy}) and get washed out for Borel models with multiple renormalons or when pinched moments are considered.
There is no ambiguity in the Borel sum integral of Eq.~(\ref{eq:BorelFOPT}) for $m\neq 2$, as we show in Eq.~(\ref{eq:IRBorelIntFOPTambi}) in the Appendix since the effects of the cut in the toy model are eliminated exactly through the contour integration.\footnote{ \label{ftn:BSambiguity}
In general the PV definition of the Borel sum integral is ambiguous and the size of this ambiguity is commonly defined to be the size of the imaginary part of the $u$-contour integral either above or below the real axis multiplied by $1/\pi$.
}}
We have checked that the FOPT series are converging to $\delta_{W_m,{\rm Borel}}^{(0),{\rm FOPT}}(s_0)$ for any integer $m\neq 2$ that may be realistically employed.
So the results for the FOPT expansions of $\delta^{(0)}_{W}(s_0)$ satisfy the conditions (i) and (ii) mentioned above and are thus consistent with the standard properties of the OPE and its association to IR renormalons.
The PV Borel sum integral of the form of Eq.~(\ref{eq:BorelFOPT}) is the Borel sum that is commonly assigned as the `true' value of spectral function moment series in the literature.
Given that the FOPT series expansion fulfills conditions (i) and (ii) and converges to $\delta_{W_m,{\rm Borel}}^{(0),{\rm FOPT}}(s_0)$ this assignment is consistent.

Let us now discuss the results for the CIPT perturbation series for $\delta^{(0)}_{W}(s_0)$ (gray dots) and $\delta^{(0)}_{W}(s_0,R^2)$ (blue dots).
We remind the reader that for the CIPT method the series for the original unsubtracted and the subtracted Adler function are coherently expanded and truncated with respect to power of $\alpha_s(-s)$ prior to carrying out the contour $x$-integration.
We clearly see that none of the unsubtracted CIPT moment series $\delta^{(0)}_{W}(s_0)$ converges. We have checked there is also no convergence for any other integer $m$ that may be realistically employed.
This behavior already contradicts condition (i) and shows that the CIPT method does not follow the standard logic concerning the one-to-one association of the OPE corrections and IR renormalons. Rather, despite the fact that the GC OPE correction is vanishing, the CIPT series for $\delta^{(0)}_{W}(s_0)$ still appears to contain some kind of renormalon divergence.
Interestingly, for $m=1,4,5$ the unsubtracted moment series approach an asymptotic value at intermediate orders (of around 10 to 20 for $m=1$ and orders 5 to 10 for $m=4,5$), but this value clearly disagrees with the value of $\delta_{W_m,{\rm Borel}}^{(0),{\rm FOPT}}(s_0)$ defined in Eq.~(\ref{eq:BorelFOPT}).
This asymptotic value is in agreement with the value of
$\delta_{W_m,{\rm Borel}}^{(0),{\rm FOPT}}(s_0)$ plus the asymptotic separation~\cite{Hoang:2020mkw}, which reads
$2/m[\Delta (0,2,1+4 \hat b_1,s_0)-(-1)^m \Delta (m,2,1+4 \hat b_1,s_0)]$  for the
Borel function $B[\hat D_{\rm Toy}(s)](u)$ in Eq.~(\ref{eq:AdlerBorelToy}) and the weight functions $W_m(x)=2(1-x^{m})/m$.
The sum of $\delta_{W_m,{\rm Borel}}^{(0),{\rm FOPT}}(s_0)$ and the asymptotic separation is indicated by the gray horizontal line. For the convenience of the reader we have given the formula for the functions $\Delta$, that are given for a general scheme of the strong coupling in Ref.~\cite{Hoang:2020mkw}, in the $C$-scheme in Eq.~(\ref{eq:Deltafct}) in the appendix.

It is straightforward to check (either from the analytic expression for the $\Delta$ functions or from numerical studies) that the difference between this asymptotic value and $\delta_{W_m,{\rm Borel}}^{(0),{\rm FOPT}}(s_0)$,  scales with $1/s_0^2$, clearly indicating that it is arising from a quartic IR sensitivity that can only be attributed to the GC renormalon.
For $m=3$ the unsubtracted CIPT moment series behaves quite badly and does not show a clear range of orders where an asymptotic value is approached. But at order $3$ to $5$ the series shows a flattening behavior at a value that is much better described by $\delta_{W_3,{\rm Borel}}^{(0),{\rm FOPT}}(s_0)$ plus the value of the asymptotic separation $2/3[\Delta (0,2,1+4 \hat b_1,s_0)+\Delta (3,2,1+4 \hat b_1,s_0)]$ than by $\delta_{W_3,{\rm Borel}}^{(0),{\rm FOPT}}(s_0)$ alone.
Overall, since the unsubtracted CIPT series for $\delta^{(0)}_{W}(s_0)$ does not satisfy expectation (i), we must conclude that for the CIPT expansion method  the renormalon is not eliminated through the contour integral in the original $\overline{\rm MS}$ GC scheme.\footnote{This was suggested prior to Refs.~\cite{Hoang:2020mkw,Hoang:2021nlz} in Refs.~\cite{Beneke:2008ad,Beneke:2012vb}.}

The results for the subtracted CIPT moment series $\delta^{(0)}_{W}(s_0,R^2)$ are shown as the blue dots. We see that they are fully compatible with the FOPT series, and no discrepancy arises at any intermediate order. Indeed, the subtracted CIPT moment series converge to $\delta_{W_m,{\rm Borel}}^{(0),{\rm FOPT}}(s_0)$, and they converge even much faster than the FOPT moment series without exhibiting any oscillating behavior. We thus see that the CIPT subtraction series generated through switching to the RF GC scheme does not converge to zero. Rather, they eliminate the disparity between the unsubtracted CIPT moment series and the FOPT series:
\begin{eqnarray}
\label{eq:CIPTexp}
\frac{1}{2\pi i}\,\,
\ointctrclockwise_{|s|=s_0}  \frac{{\rm d}s}{s}\,W_{m\neq 2}({\textstyle \frac{s}{s_0}})\,\bigg[\!-\frac{R^4}{s^2}
\sum\limits_{\ell=1}^\infty
r_\ell^{(4,0)} \,\bar a^\ell_R \,\bigg]^{\mbox{CIPT}} \, \neq \, 0 \,\,\mbox{and divergent}\,.
\end{eqnarray}
For illustration we have displayed the CIPT subtraction series, i.e. the CIPT expansion of $\delta^{(0)}_{W_m}(s_0,R^2)-\delta^{(0)}_{W_m}(s_0)$ in the right panels of  Fig.~\ref{fig:niceplots}.
This shows that the CIPT expansion contradicts also condition (ii).
It is a remarkable fact that, when the CIPT method is used to determine the spectral function moment series, the  change induced by switching from the $\overline{\rm MS}$ to the RF GC scheme --- which is supposed to vanish in the contour integration within the standard approach to the OPE as in Eq.~(\ref{eq:FOPTexp}) --- can generate a subtraction series that makes the CIPT series to be fully compatible with the FOPT series.

These observations confirm the suggestion made in Ref.~\cite{Hoang:2020mkw} that the unsubtracted CIPT approach leads to a perturbation series that is inconsistent with the
standard OPE corrections as already mentioned in Sec.~\ref{sec:renormalon}.
At this point the following important conclusions can be drawn:
\begin{enumerate}
\item The asymptotic value that the unsubtracted CIPT expansion series approach at intermediate orders in the $\overline{\rm MS}$ GC scheme should not be considered as the `true' value of the moment series in the context of using standard OPE terms in the Adler function of the form in Eqs.~(\ref{eq:DOPE}) or (\ref{eq:AdlerOPEGCv2}) to parametrize non-perturbative OPE corrections.
\item Strong coupling determinations based on the original unsubtracted CIPT expansion in the $\overline{\rm MS}$ GC scheme combined with OPE corrections obtained from
standard OPE terms in the Adler function of the form in Eqs.~(\ref{eq:DOPE}) or (\ref{eq:AdlerOPEGCv2}) are physically inconsistent.
\item The assignment of the commonly used PV Borel sum integral
\begin{equation}
\label{eq:BorelFOPTgen}
\delta_{W,{\rm Borel}}^{(0),{\rm FOPT}}(s_0) = {\rm PV} \int_0^\infty \!\! {\rm d} u \,\,
\frac{1}{2\pi i}\,\ointctrclockwise_{|s|=s_0}  \frac{{\rm d}s}{s}\,W({\textstyle \frac{s}{s_0}}) \,
\,B[\hat D(s)](u) e^{-\frac{u}{\bar a(-s)}}\,,
\end{equation}
with $B[\hat D(s)](u)$ being the Borel function of the Adler function $\hat D(s)$ with respect to the expansion in powers of $\bar\alpha_s(-s)$, as the `true' value for the perturbative spectral function moments (modulo its ambiguity as defined in footnote~\ref{ftn:BSambiguity}), is consistent with the standard OPE and the common association of OPE corrections and IR renormalons.
\end{enumerate}
The results shown in Fig.~\ref{fig:niceplots} furthermore confirm the main idea of this work that --- if the Adler function perturbation series has a sizeable contribution from the GC renormalons so that $N_{4,0}$ is not strongly suppressed --- the observed disparity between the unsubtracted FOPT and CIPT moment series may be removed or at least reduced when using the RF GC scheme. As a result, at least for the GC the same OPE correction can be applied for FOPT and CIPT and both expansions may lead to physically consistent results. The fact that the CIPT expansion may lead to much faster converging series than the FOPT expansion, indicates that the renormalon-free CIPT expansion may still remain a viable method for strong coupling determinations.
These conclusions are further supported by the numerical studies in the following two sections.

\section{The  \boldmath Large-\texorpdfstring{$\beta_0$}{beta0} Approximation}
\label{sec:largeb0}

Let us now apply the RF GC scheme to the spectral function moments in the large-$\beta_0$ limit, where the perturbation series for the Adler function and the analytic form of its Borel function are known exactly and to all orders~\cite{Broadhurst:1992si} (see also Ref.~\cite{Beneke:1992ch} for the calculation of the perturbative coefficients to any order). This analysis tests the RF scheme in the context of a more realistic case where infinitely many other IR and UV renormalons are present. Even though it is well-known that the large-$\beta_0$ approximation does not provide a quantitatively accurate approximation to full QCD it exhibits many important qualitative features of full QCD that can be conveniently studied in an explicit way at all orders in perturbation theory.

Within our convention for the renormalon calculus the exact Borel function of the Adler function $\hat D(s)$ with respect to the expansion in powers of $\alpha_s(-s)$ has the form
\begin{eqnarray}
\label{eq:AdlerBorelb0}
\lefteqn{ B[\hat D(s)](u)
\, = \, \frac{128}{3\beta_0}\,\frac{e^{5u/3}}{2-u}\,\sum_{k=2}^\infty \, \frac{(-1)^k \,k}{[k^2-(1-u)^2]^2} } \\ \nonumber
& = &
\frac{128}{3\beta_0}\,e^{5u/3}\Big\{
\frac{3}{16(2-u)} +\sum\limits_{p=3}^\infty \Big[ \frac{d_2(p)}{(p-u)^2} - \frac{d_1(p)}{p-u} \Big]
-\sum\limits_{p=-1}^{-\infty} \Big[ \frac{d_2(p)}{(u-p)^2} + \frac{d_1(p)}{u-p} \Big]
\Big\}\,,
\end{eqnarray}
where
\begin{eqnarray}
d_2(p) &=& \frac{(-1)^p}{4(p-1)(p-2)} \\
d_1(p) &=& \frac{(-1)^p(3-2p)}{4(p-1)^2(p-2)^2}\,.
\end{eqnarray}
Here the normalization of the GC renormalon pole reads $N_{4,0}=\frac{2\pi^2}{3} N_g =8 e^{10/3}/9=24.92$, so that
the construction of our renormalon-free scheme is straightforward. Some of the results have already been discussed briefly in Ref.~\cite{Benitez-Rathgeb:2021gvw}.

In the large-$\beta_0$ approximation, the evolution equation of $\alpha_s$ is already exact at ${\cal O}(\alpha_s^2)$  and the $C$ and $\overline{\rm MS}$ schemes for the strong coupling are identical. The GC renormalon singularity becomes a simple pole, since the coefficient $\hat b_1$ vanishes. Furthermore, the higher order corrections to the Wilson coefficients of all OPE corrections vanish such that the coefficient $\bar c_{4,0}^{(1)}=0$, see Ref.~\cite{Hoang:2021nlz}. In order to set up the RF scheme in the large-$\beta_0$ approximation we can conveniently adopt the full QCD formulae in the $C$-scheme for the strong coupling simply setting the coefficients  $\hat b_1$ and $\bar c_{4,0}^{(1)}=0$ to zero.
The Adler function perturbative coefficients $\bar c_\ell$ can be obtained using Eqs.~(\ref{eq:BTaylor}) and  (\ref{eq:AdlerBorelb0}). Together with Eq.~(\ref{eq:invBorelDR})
this sets up the RF scheme for the Adler function to all orders which can then be used to determine spectral function moments in the FOPT and CIPT expansion.

\begin{figure}
	\centering
	\begin{subfigure}[b]{0.48\textwidth}
		\includegraphics[width=\textwidth]{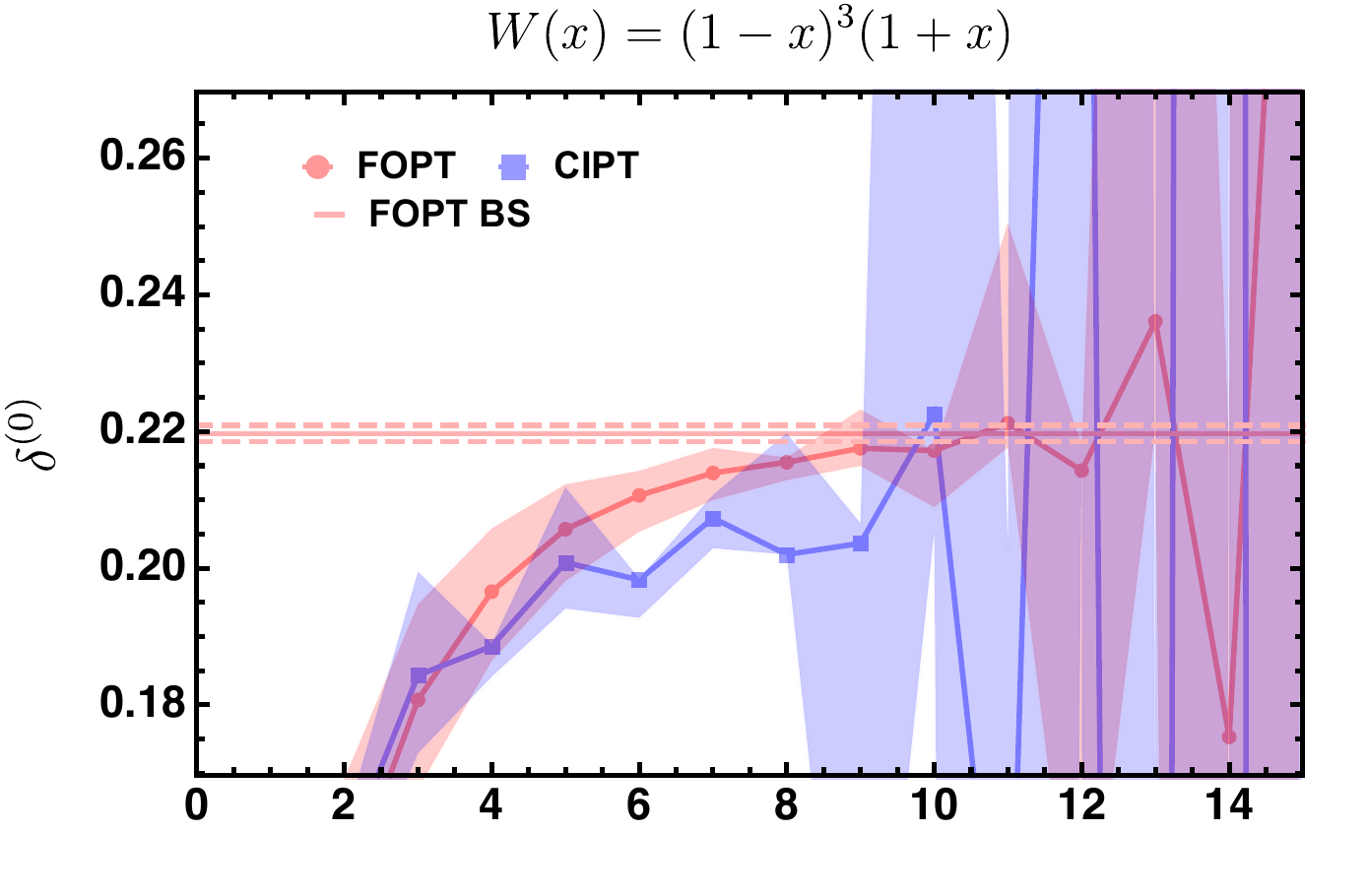}
	\end{subfigure}
	~ 
	\begin{subfigure}[b]{0.48\textwidth}
		\includegraphics[width=\textwidth]{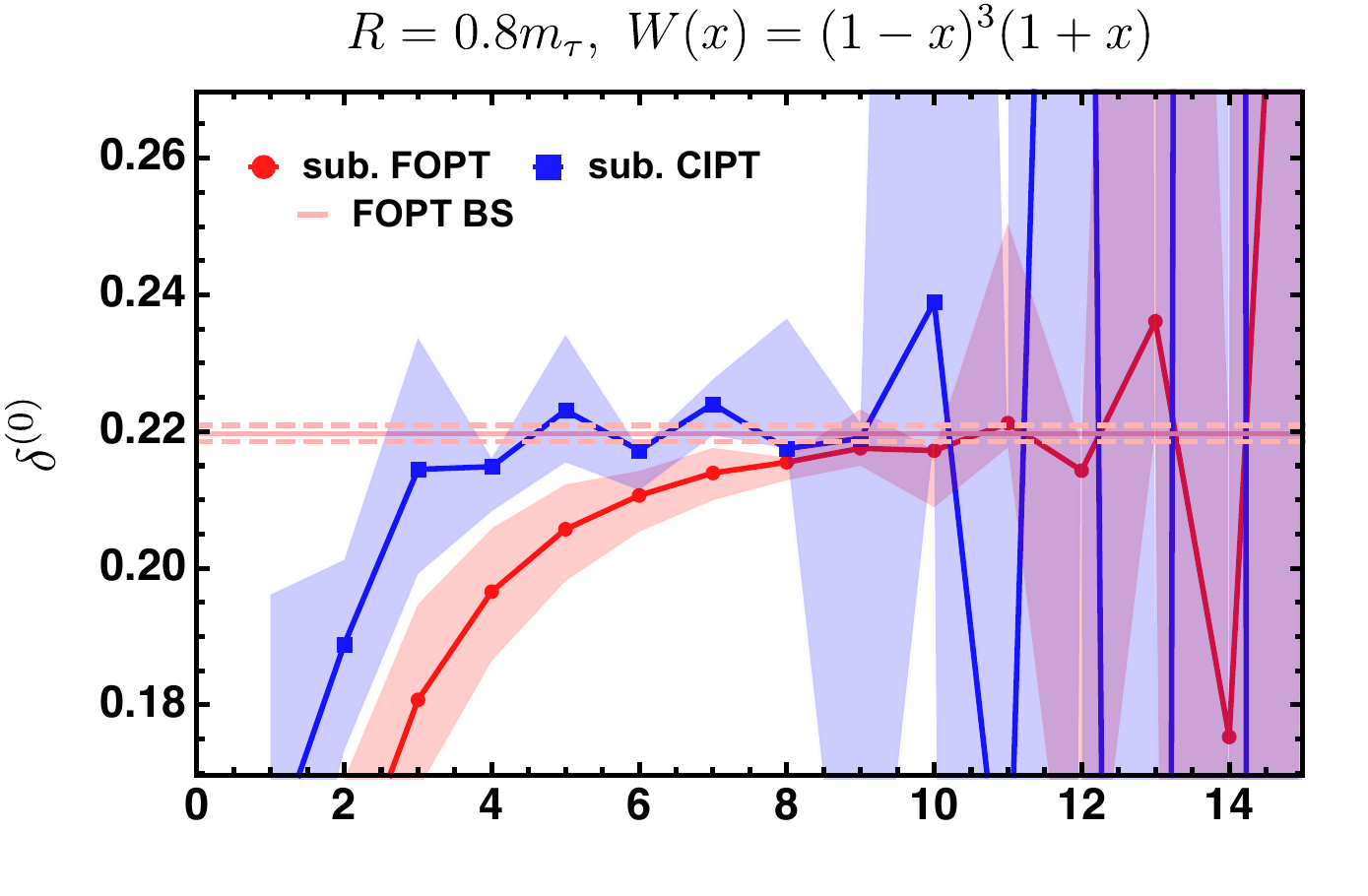}
	\end{subfigure}

	\begin{subfigure}[b]{0.48\textwidth}
		\includegraphics[width=\textwidth]{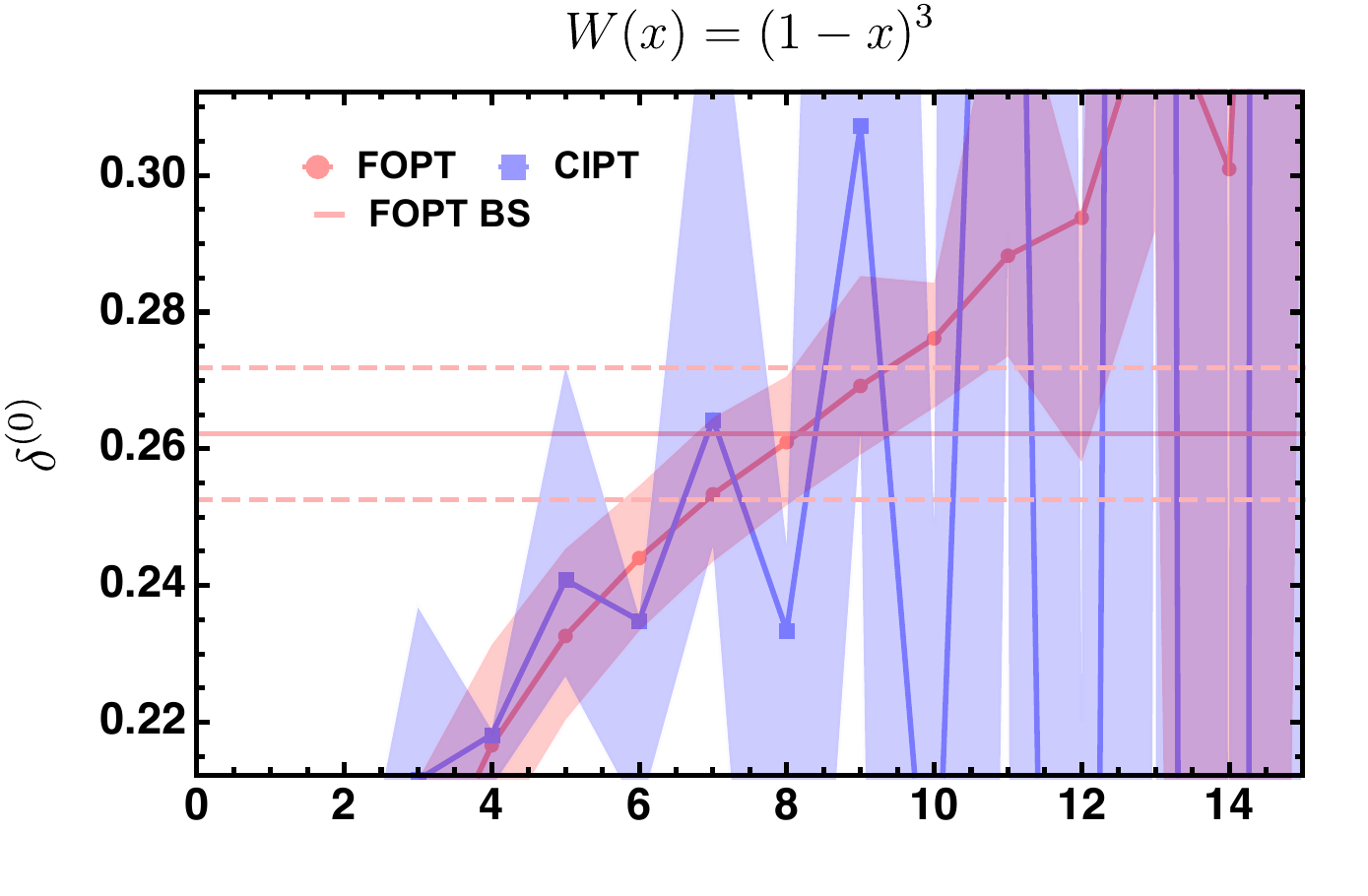}
	\end{subfigure}
	~
	\begin{subfigure}[b]{0.48\textwidth}
		\includegraphics[width=\textwidth]{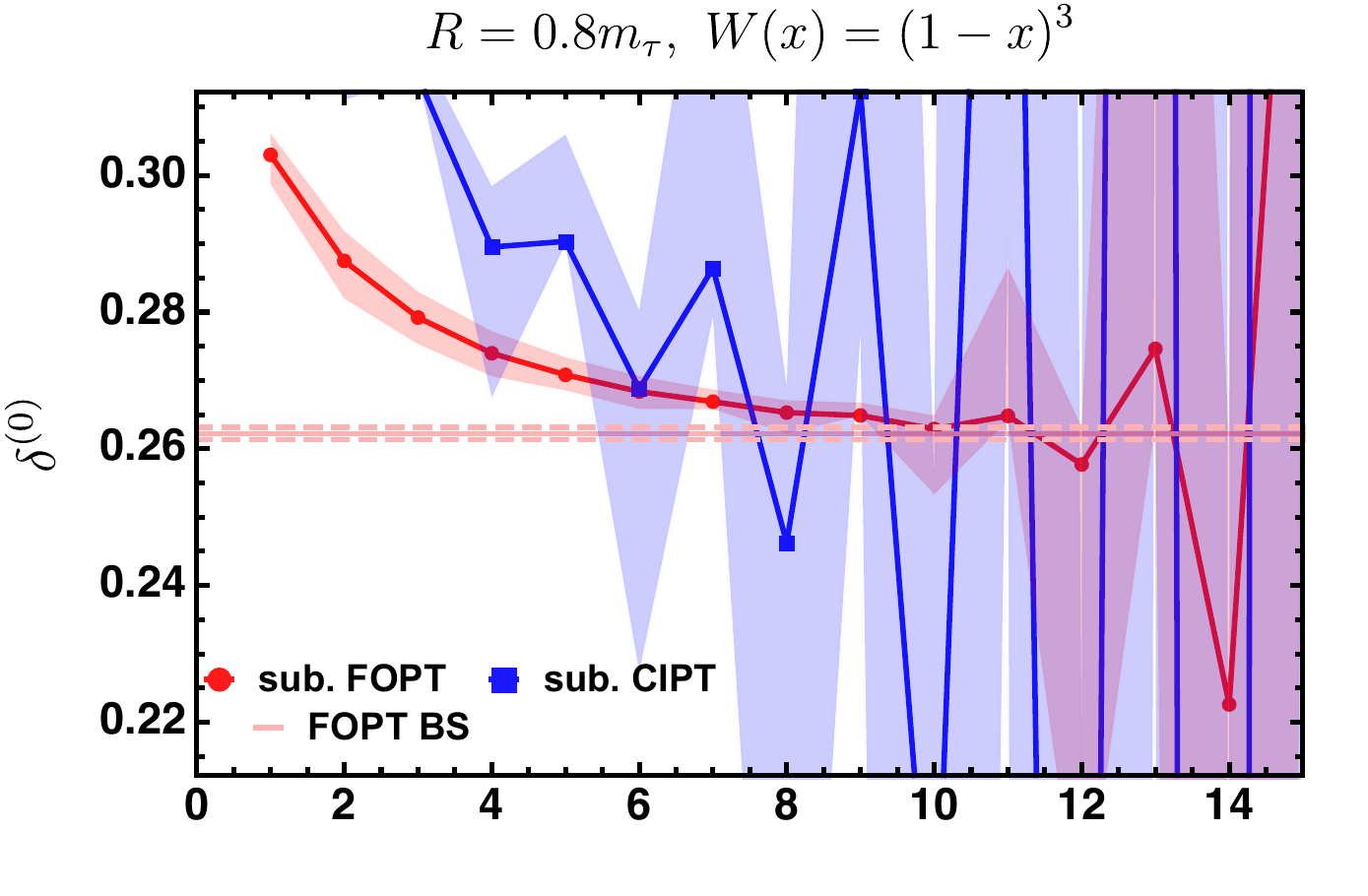}
	\end{subfigure}

	\begin{subfigure}[b]{0.48\textwidth}
		\includegraphics[width=\textwidth]{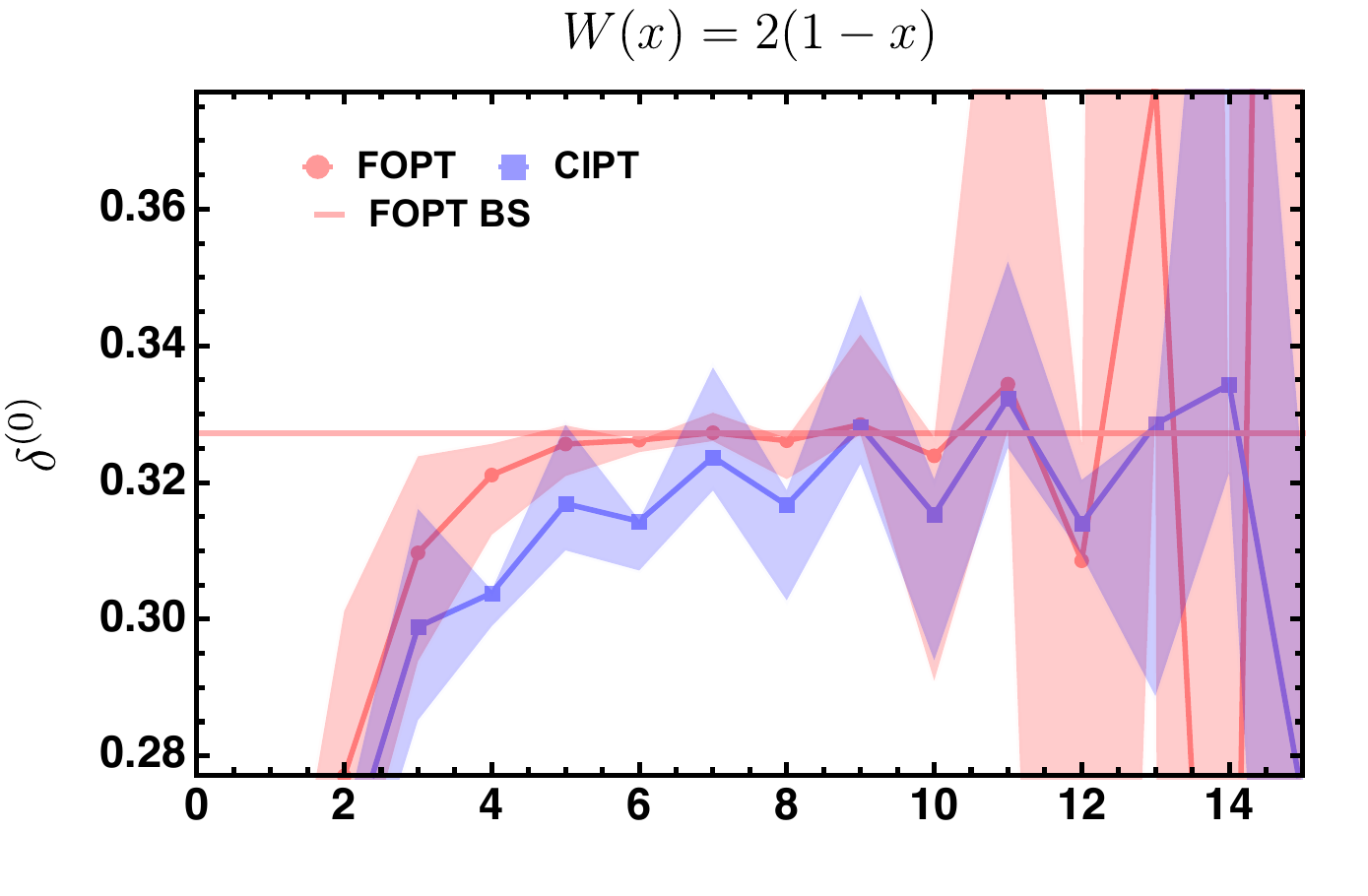}
	\end{subfigure}
	~
	\begin{subfigure}[b]{0.48\textwidth}
		\includegraphics[width=\textwidth]{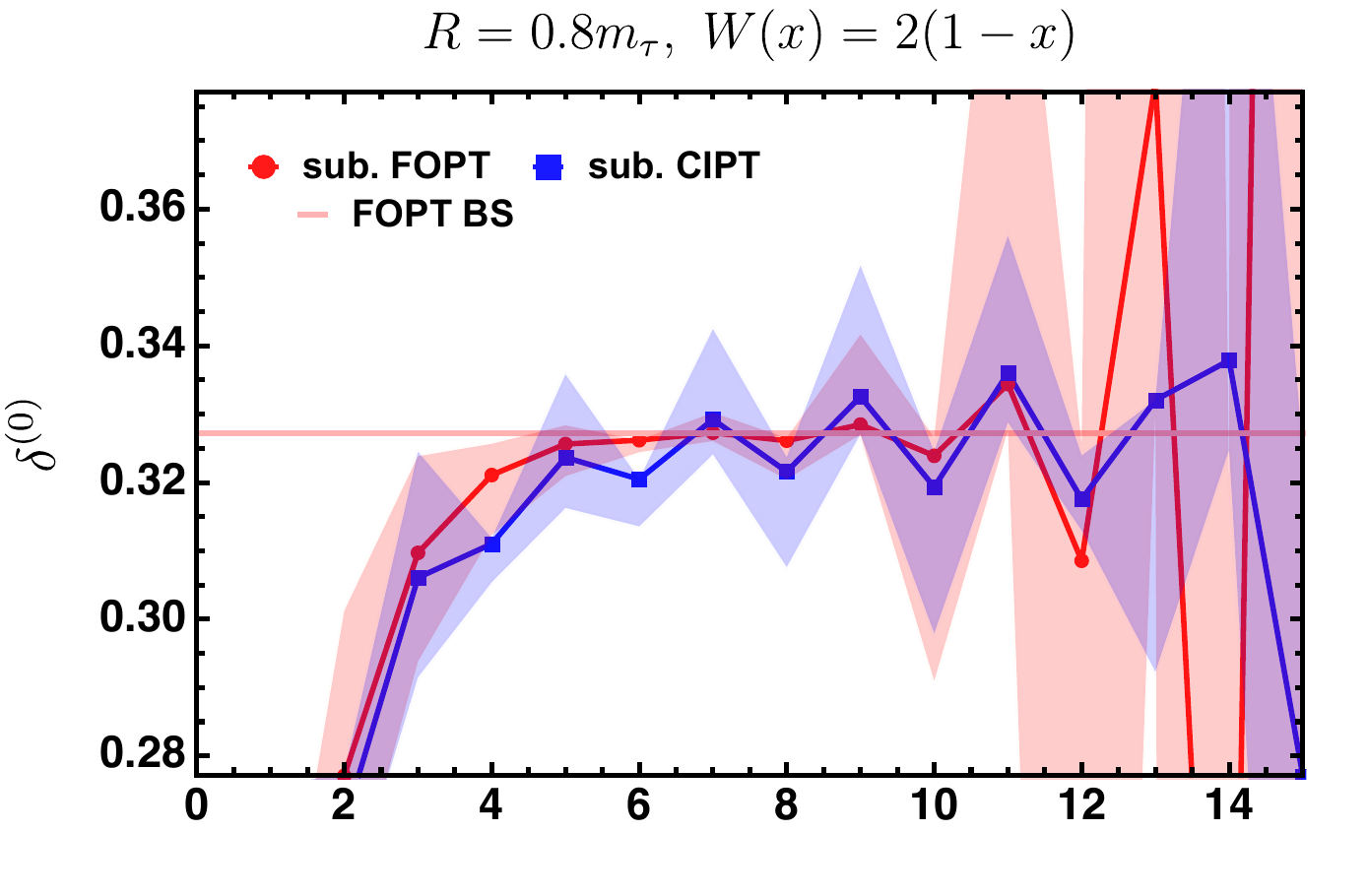}
	\end{subfigure}
	\caption{\label{fig:serieslb0}
		Left panels: Series for FOPT and CIPT expansions for $\delta^{(0)}_{W}(m_\tau^2)$ as a function of order in the large-$\beta_0$ approximation in the $\overline{\rm MS}$ GC scheme for different weight functions $W(x)$ for $\alpha_s(m_\tau^2)=0.315$. Renormalization scale variations are indicated by the colored bands.
		Right panels: Corresponding series for FOPT and CIPT expansions for $\delta^{(0)}_{W}(m_\tau^2,R^2)$ in the RF GC scheme.
		}
\end{figure}

In Fig.~\ref{fig:serieslb0} we compare the FOPT (red color) and CIPT (blue color) series as a function of the order in perturbation theory in the large-$\beta_0$ limit for three representative spectral function moments for $s_0=m_\tau^2$ for $\alpha_s(m_\tau^2)=0.315$.
The left panels display the unsubtracted results in the original $\overline{\rm MS}$ GC scheme, and the right panels show the results in the RF scheme for $R=0.8\,m_\tau$. The connected colored dots represent the truncated series values for the default renormalization scale choice for $\xi=1$. The shaded colored bands show the respective renormalization scale variations arising from $1/2\le \xi\le 2$ (see footnote~\ref{ftn:scalevariation}).
It is conventional to adopt these bands as an estimate for the perturbative uncertainty associated to the series truncation at the respective order, but we remind the reader that the usual care should be applied not to overinterpret this association.
The red horizontal line in each of the plots represents the PV Borel sum $\delta_{W(x),{\rm Borel}}^{(0),{\rm FOPT}}(s_0)$ defined in Eq.~(\ref{eq:BorelFOPTgen}) and
the area between the dashed red horizontal lines its ambiguity, see footnote~\ref{ftn:BSambiguity}.

In the upper two panels of Fig.~\ref{fig:serieslb0} we show the results for the doubly-pinched kinematic moment based on the weight function $W(x)=1-2x+2x^3-x^4$. Due to the absence of a quadratic $x^2$ term this is the prototypical example for a GCS spectral function moment. But this moment still has unsuppressed sensitivity to dimension-6 and dimension-8 OPE corrections and the associated IR renormalons. For the original unsubtracted series (upper left panel) the well-known discrepancy between FOPT and CIPT emerging at orders $5$ to $8$ is clearly visible. The unsubtracted CIPT series (upper left panel) approaches a value at intermediate orders significantly below the value of the FOPT series, and later runs into the asymptotic regime and diverges. The unsubtracted FOPT series, on the other hand, slowly approaches the `true' Borel sum of Eq.~(\ref{eq:BorelFOPTgen}), before it starts diverging at orders 11 and higher. The discrepancy between the unsubtracted FOPT and CIPT series is substantially larger than the ambiguity of the `true' Borel sum. Using the RF GC scheme (upper right panel) we clearly see that the subtracted CIPT series now approaches the same value as the FOPT\footnote{
We remind the reader that in the large-$\beta_0$ approximation the GCS moment series in the unsubtracted and subtracted FOPT expansions are identical.
For GCE moments, this is not the case since their GC corrections do not vanish.
}
series, for which the subtracted series is identical to the unsubtracted one. The subtracted FOPT and CIPT series now both approach the `true' Borel sum of Eq.~(\ref{eq:BorelFOPTgen}). Interestingly, the subtracted CIPT series approaches the true value much faster than the FOPT expansion exhibiting again the viability of the subtracted CIPT expansion. Starting from order 6, the two series are in good agreement within renormalization scale variations, before they diverge at orders beyond. This result shows that in the subtracted scheme the FOPT and CIPT expansion series clearly approach equivalent values and the fundamental discrepancy between them in the original unsubtracted scheme is eliminated. The results show that the GC renormalon is responsible for the largest contribution to the discrepancy between the two unsubtracted expansion prescriptions, and that the discrepancy arising from higher power suppressed IR renormalons (which still exist in the RF  scheme) can be neglected from the practical perspective.
This corroborates the conclusions of Refs.~\cite{Hoang:2020mkw,Hoang:2021nlz}.

Next, in the second line of panels of Fig.~\ref{fig:serieslb0} we show the moment series results for the doubly-pinched weight function $W(x)=(1-x)^3=1-3x+3x^2-x^3$, corresponding to the weight function $w(x)=3/2(1-x)^2$ for the spectral function integral.
Due to the presence of the quadratic $x^2$ term this is an example for a GCE spectral function moment, where the divergent asymptotic series behavior related to the GC is not suppressed. As a consequence the series obtained in the unsubtracted FOPT and CIPT expansions (middle left panel) immediately show the divergent run-away behaviour discussed before in Ref.~\cite{Beneke:2012vb} coming from the GC renormalon.
This is also reflected in the huge renormalon ambiguity indicated by the band between the dashed red horizontal lines.
In contrast to the FOPT series, the CIPT series shows a significant sign alternation at all orders. Neither of the two series show any sign of stabilization before they run into the divergent asymptotic regime.  In the RF scheme (middle right panel)  we see that the subtracted FOPT series is very well behaved and slowly approaches the `true' Borel sum of Eq.~(\ref{eq:BorelFOPTgen}). The run-away behaviour is eliminated by the GC renormalon subtraction, as expected. The subtracted CIPT series also approaches the same value, albeit with a sign alternation which is a manifestation of the strong leading UV renormalon in the large-$\beta_0$ approximation.
 The renormalon ambiguity has become tiny and reduced to a practically negligible size.
We note that the previously mentioned sign alternation is not observed in the first few orders in full QCD, which suggests that the UV renormalon is weaker when compared with large-$\beta_0$~\cite{Beneke:2008ad,Beneke:2012vb,Boito:2018rwt}.
This view is further supported by our model study in Sec.~\ref{sec:MultiRenormalonModel} in the context of full QCD.

Finally, in the lowest two panels of Fig.~\ref{fig:serieslb0}, we show the results for the  moment $W(x)=2(1-x)$, which corresponds, in terms of the spectral function integrals, to the weight function $w(x)=1$. This GCS moment is not pinched, but it has the additional interesting property that for it, the dominant contributions from all OPE corrections beyond the dimension-4 GC are strongly suppressed.\footnote{Due to the double pole structures of these higher dimensional IR renormalons in the large-$\beta_0$ approximation (see Eq.~(\ref{eq:AdlerBorelb0}) and Ref.~\cite{Hoang:2021nlz}), they are not entirely eliminated in contrast to the GC renormalon which is related to a simple pole.} It is also important phenomenologically, since the main results for the $\alpha_s$ determinations in Refs.~\cite{Boito:2020xli,Boito:2014sta,Boito:2012cr} are obtained from sum rules based on this moment. The unsubtracted FOPT expansion (middle left panel) is well behaved and approaches relatively fast to the `true' Borel sum of Eq.~(\ref{eq:BorelFOPTgen}). The unsubtracted CIPT series (middle left panel)  is systematically below the FOPT results at intermediate orders, which again is a manifestation of the fundamental discrepancy between the two expansion prescriptions and the inconsistency of unsubtracted CIPT with the standard OPE expansion. Turning to the results in the RF scheme (middle right panel) we see that, again, the discrepancy between the two prescription is, for all practical purposes, eliminated thanks to the subtraction of the divergence associated with the GC renormalon.

We have repeated this analysis for many other moments and we found that the results are fully consistent with the results shown Fig.~\ref{fig:serieslb0} which can thus be taken as representative for all spectral function moments. Overall our findings can be summarized as follows:
\begin{itemize}
\item[(1)] In the RF scheme, the discrepancy between FOPT and CIPT for GCS moments is strongly suppressed and becomes numerically negligible. Furthermore, the subtracted FOPT and CIPT expansions both approach the `true'  Borel sum of Eq.~(\ref{eq:BorelFOPTgen}). The results suggest that $\alpha_s$ extractions using the CIPT and FOPT expansion in the RF scheme should lead to results in much better agreement in comparison to the original unsubtracted FOPT and CIPT expansions in the $\overline{\rm MS}$ scheme for the GC.
\item[(2)] In the RF scheme, GCE moments display a much better perturbative behaviour since the subtraction eliminates the run-away behaviour discussed in Ref.~\cite{Beneke:2012vb} coming from the GC renormalon. An important benefit is that these moments, which are nowadays excluded in most phenomenological analyses due their bad perturbative behaviour, can now in principle be reemployed in high-precision determinations of the strong coupling, with the advantage of providing information about the GC matrix element $\langle G^2\rangle^{\rm RF}$ as defined through Eqs.~(\ref{eq:GCIRsubtracted}) and (\ref{eq:GCIRsubtracted2}).
\end{itemize}
In the next section we will see that these findings also apply in the context of full QCD assuming a multi-renormalon Borel function model
for the Adler function following Ref.~\cite{Beneke:2008ad} where the GC renormalon norm $N_{4,0}=\frac{2\pi^2}{3}N_g$ is close to $4$. What is more important, for this value of the GC renormalon norm, the improvements that arise in the RF scheme are consistently visible already at ${\cal O}(\alpha_s^4)$ and ${\cal O}(\alpha_s^5)$.

\section{Full QCD and a Natural Renormalon Model for Higher Orders}
\label{sec:MultiRenormalonModel}

\subsection{Comments on the Multi-Renormalon Model}
\label{sec:comments}

We now apply the RF GC scheme to the spectral function moments in full QCD using the coefficients $\bar c_{n,1}$ up to ${\cal O}(\alpha_s^5)$ given in Eqs.~(\ref{cn1}) and (\ref{eq:c51MSb}) (and Eqs.~(\ref{cb3cb4}) and (\ref{eq:barc51}) for the $C$-scheme strong coupling). We furthermore analyze higher order corrections beyond ${\cal O}(\alpha_s^5)$ based on a model for the Borel function of the Euclidean Adler function motivated by the multi-renormalon Borel model suggested in Ref.~\cite{Beneke:2008ad} (see also Ref.~\cite{Jamin:2021qxb}). This Borel model,  which is in accordance with proposition (a) mentioned in Sec.~\ref{sec:intro}, is based on the assumption that the effects of IR and UV renormalons related to renormalon singularities located closer to the origin of the complex $u$ plane are enhanced compared to those located further away from the origin already in the known perturbative coefficients up to ${\cal O}(\alpha_s^4)$ or ${\cal O}(\alpha_s^5)$. This assumption is analogous (but not equivalent) to the canonical generic assumption that OPE corrections from condensates with lower dimension $d$ constitute larger non-perturbative corrections than those with larger $d$.
In this sense this Borel model is `natural'. This naturalness assumption also implies that the imprint of the first few renormalons close to the origin is already visible in the order-dependence and size of the known perturbative Adler function coefficients and that there are no  fine-tuned cancellations or particular enhancements or suppressions of the norms of the renormalon singularities that are accounted for in the model. In other words, the naturalness assumption implies that the perturbative corrections up to ${\cal O}(\alpha_s^4)$ or ${\cal O}(\alpha_s^5)$ allow at least some approximate determination of the renormalon norms that enter the model. As was shown in Ref.~\cite{Beneke:2012vb}, the behavior of the known perturbative corrections is fully consistent with the naturalness assumption, and that within this context the Borel model suggested in Ref.~\cite{Beneke:2008ad} can provide an adequate quantitative prediction for the true higher order corrections and also provide a reasonable estimate for the GC renormalon norm $N_{4,0}=\frac{2\pi^2}{3} N_g$, which turns out to be close to $4$.
This value for $N_{4,0}$ is sizeable\footnote{We remind the reader that the actual value of the renormalon norm depends on our conventions for the renormalon calculus given in Sec.~\ref{sec:renormalon}.} and, as we show below, when employed in the RF GC scheme consistently improves the behavior of the spectral function moments FOPT and CIPT expansions in a way very similar to the large-$\beta_0$ study in Sec.~\ref{sec:largeb0}.
The model, however, does not provide any estimate concerning the theoretical uncertainty on the value of $N_{4,0}$.

In the context of the naturalness assumption the renormalon structure of the perturbative coefficients can be modeled by the first few renormalons located close to the origin (because the effects of the renormalons further away from the origin are small) and a regular function that quantifies the potentially sizeable non-asymptotic contributions present at lower orders. The parameters of this model are fixed from the condition that it reproduces the perturbative coefficients up to ${\cal O}(\alpha_s^5)$,  where we use the central value for $\bar c_{5,1}$ given in Eq.~(\ref{eq:barc51}).
To be concrete, the multi-renormalon Borel model proposed in Ref.~\cite{Beneke:2008ad} contains the GC renormalon term $B_{4,0}(u)$ given in Eq.~(\ref{eq:AdlerBorelGC}) (generalized for complex-valued $\alpha_s(-s)$) and one generic dimension-6 renormalon with a vanishing anomalous dimension and vanishing higher order Wilson coefficient corrections, $B_{6,0}(u)$, see Eq.~(\ref{eq:AdlerOPEBorel}). Furthermore it accounts for one $2p=d=-2$ UV renormalon with $\alpha=1$, motivated by the double pole structure known from the large-$\beta_0$ approximation and a linear polynomial to account for sizeable  pre-asymptotic contributions in the one- and two-loop coefficients unrelated to renormalons. The model's five parameters are the norms $N_{4,0}$, $N_{6,0}$ and $N_{-2}$ and the two coefficients of the linear polynomial, which are fixed from the requirement that the model reproduces the Adler function coefficients $\bar c_{n,1}$ for $n=1,\ldots 5$.

We have rederived the model in the $C$-scheme for the strong coupling using the convention of Eq.~(\ref{eq:AdlerOPEBorel})
and included the effects of the $5$-loop correction to the QCD $\beta$-function that were not yet available in Ref.~\cite{Beneke:2008ad}; see also Ref.~\cite{Hoang:2020mkw} for the analogous model rederived in the $\overline{\rm MS}$ scheme for the strong coupling. The explicit expression for the Borel model $B[\hat D(s)]_{\rm mr}(u)$ is given in Eq.~(\ref{eq:Bmodel}) in the appendix. Numerically, the differences in the norms  $N_{4,0}$, $N_{6,0}$ and $N_{-2}$ and the predictions of the higher order Adler function coefficients between the forms of the model given here and in Refs.~\cite{Beneke:2008ad,Hoang:2020mkw} are tiny and insignificant for the purpose of our analysis.
Since the known perturbative Adler function coefficients are consistent with a sizeable contribution coming from the GC renormalon~\cite{Beneke:2012vb}\footnote{We stress that the model construction does not rely on the assumption that the GC renormalon dominates the first five coefficients.}, the value $N_{4,0}=4.2$ contained in the model is sizeable as well.
 The model also has (in contrast to the large-$\beta_0$ approximation, see Eqs.~(\ref{eq:AdlerBorelb0}) and (\ref{eq:Bmodel})) a quite small $d=-2$ UV renormalon norm $N_{-2}=-0.03$, because the known full QCD series coefficients do not show a significant sign alternation behavior (in contrast to the coefficients in the large-$\beta_0$ approximation, compare Figs.~\ref{fig:serieslb0} and \ref{fig:seriesqcd}), see also Refs.~\cite{Beneke:2012vb,Boito:2018rwt}.

We emphasize that the naturalness assumption explained above is well-known in the literature and also applied in the context of the common practice of using short-distance heavy-quark mass renormalization schemes in favor of the pole mass scheme in order to improve the convergence behavior of perturbative calculations for quark mass dependent observables~\cite{ParticleDataGroup:2020ssz,Hoang:2020iah,Beneke:2021lkq}. The now common practice in current state-of-the-art high-precision strong coupling determinations from $\tau$ hadronic spectral function moments to exclusively employ GCS spectral function moments, where the perturbative uncertainties are in general much smaller than for GCE moments (see the end of Sec.~\ref{sec:renormalon} for references), is taking advantage of the naturalness assumption as well --- regardless of whether this has been done intentionally or not --- because otherwise the resulting improvement in theoretical precision would have to be considered as accidental and not as systematic.
We stress that the naturalness assumption (and that the resulting form of the Borel model in Eq.~(\ref{eq:Bmodel}) provides a good approximation to the true Borel function of the Euclidean Adler function) can currently not be proven from first principles in QCD. If it were not true, the construction of the Borel function would require further unknown input that may not even be gained when additional higher order coefficients become available and no concrete statement on the value of $N_{4,0}$ could be made. In other words, the corrections coming from beyond ${\cal O}(\alpha_s^4)$ may be very large and completely change the character and the order-dependence of the available series terms.  The results from Ref.~\cite{Hoang:2020mkw} and the RF GC scheme would still be valid  at the purely conceptual level, but it would not be clear to which extent their implications would be relevant in practice. However, we consider such an unnatural scenario as rather unlikely. Additional quantitative plausibility arguments supporting this view have been given in Refs.~\cite{Beneke:2008ad,Beneke:2012vb}.

\subsection{Numerical Analysis}
\label{sec:numerics}

\begin{figure}
	\centering
	\begin{subfigure}[b]{0.48\textwidth}
		\includegraphics[width=\textwidth]{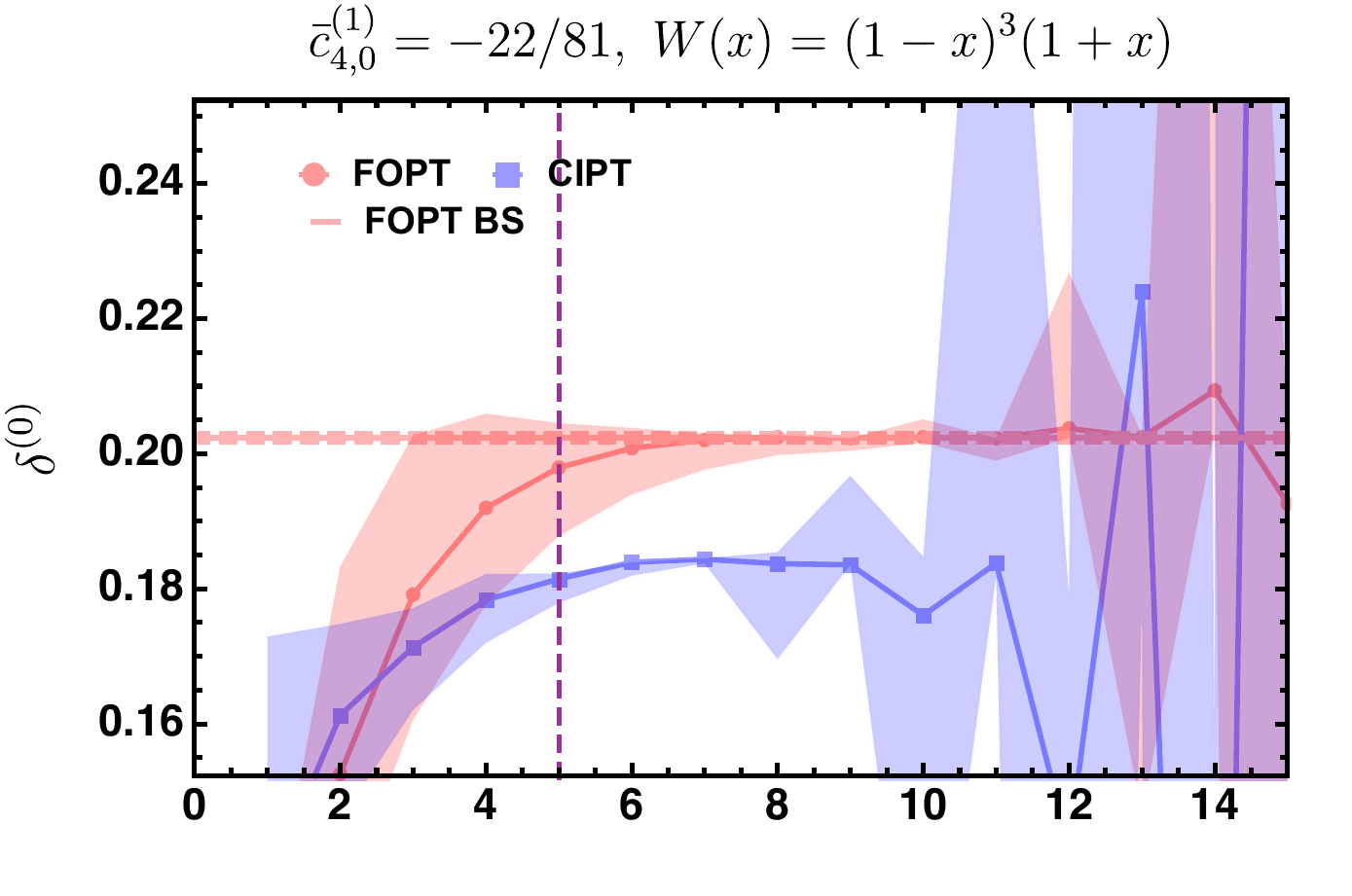}
	\end{subfigure}
	~ 
	\begin{subfigure}[b]{0.48\textwidth}
		\includegraphics[width=\textwidth]{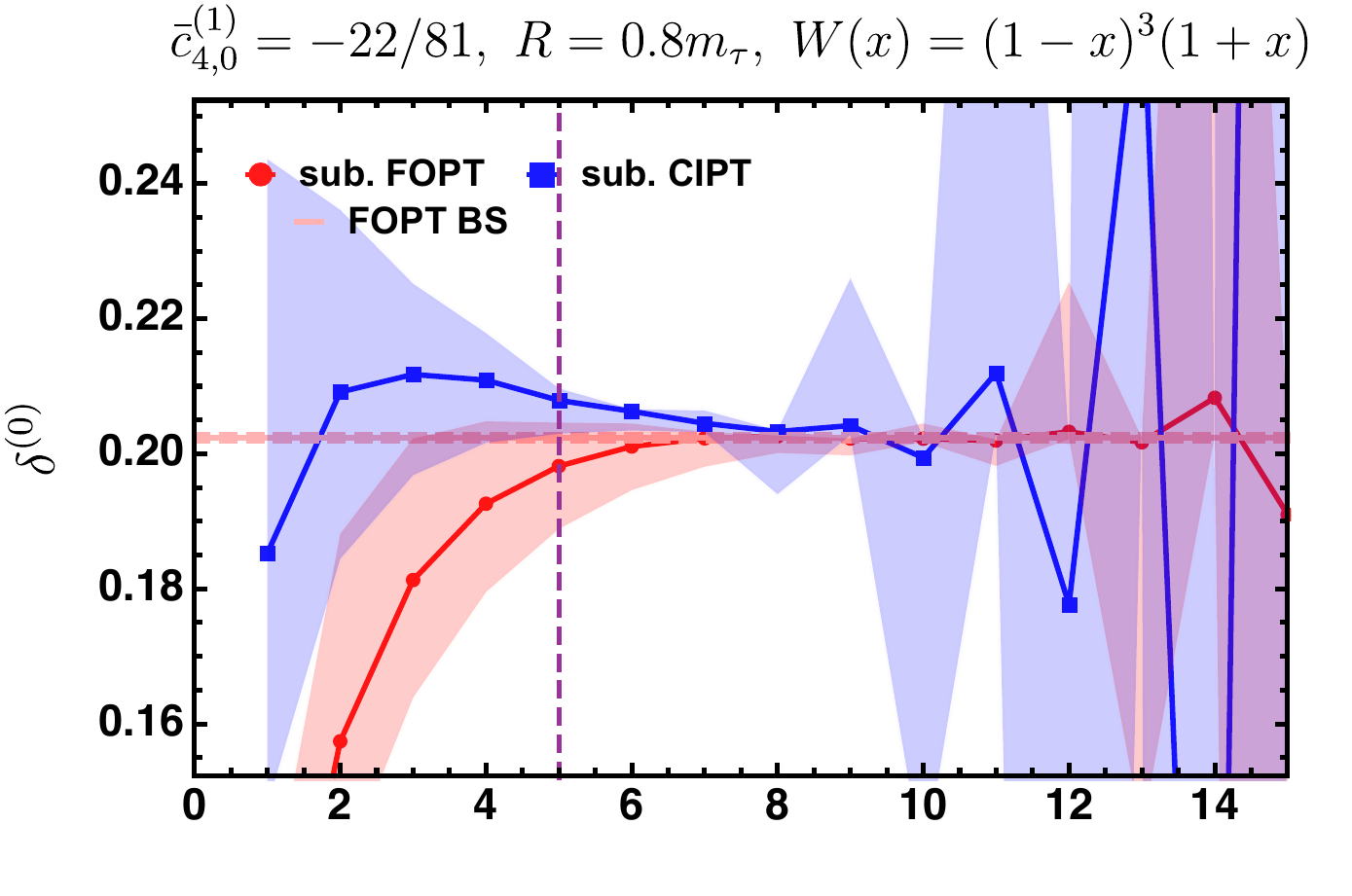}
	\end{subfigure}

	\begin{subfigure}[b]{0.48\textwidth}
		\includegraphics[width=\textwidth]{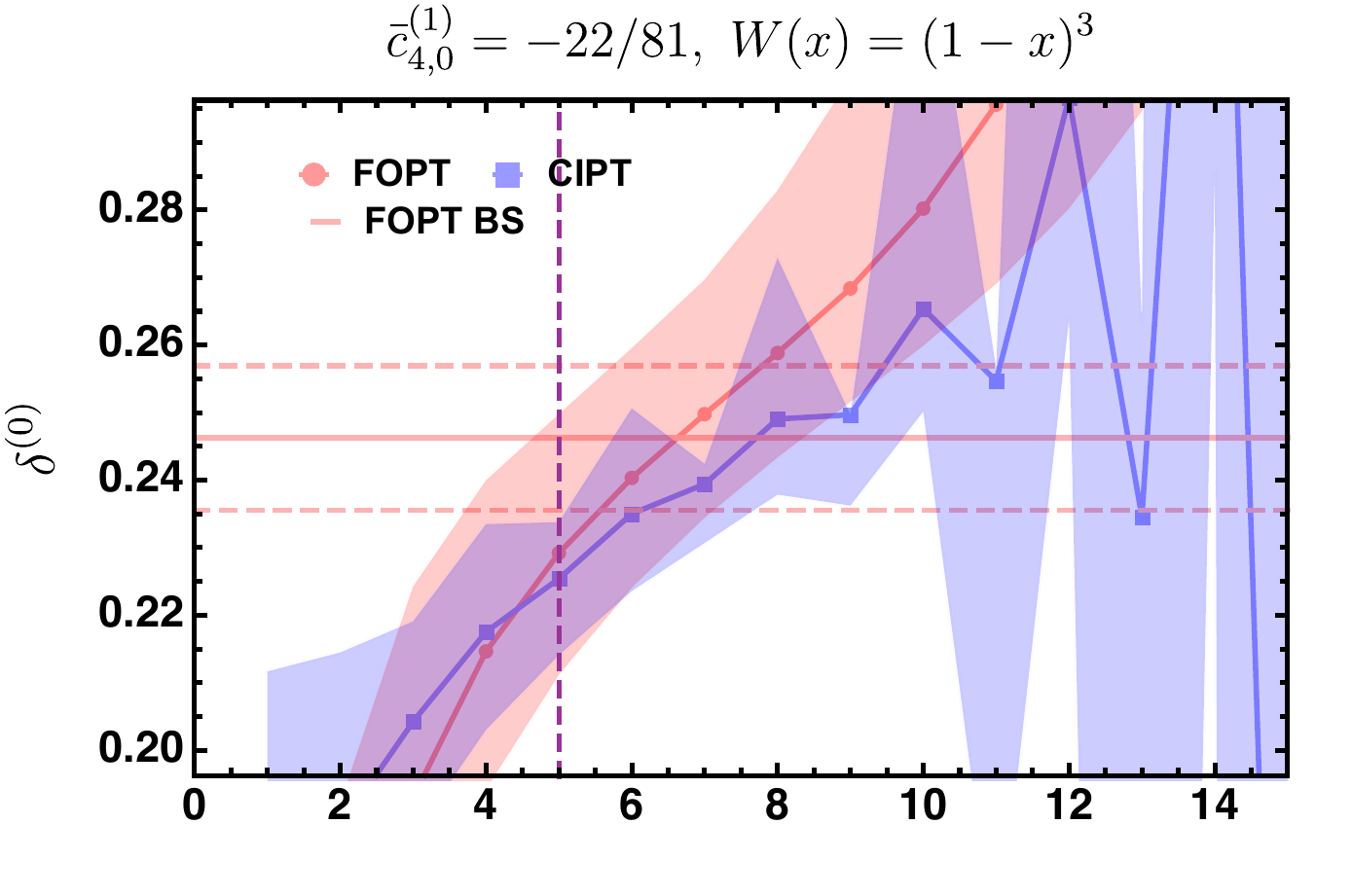}
	\end{subfigure}
	~
	\begin{subfigure}[b]{0.48\textwidth}
		\includegraphics[width=\textwidth]{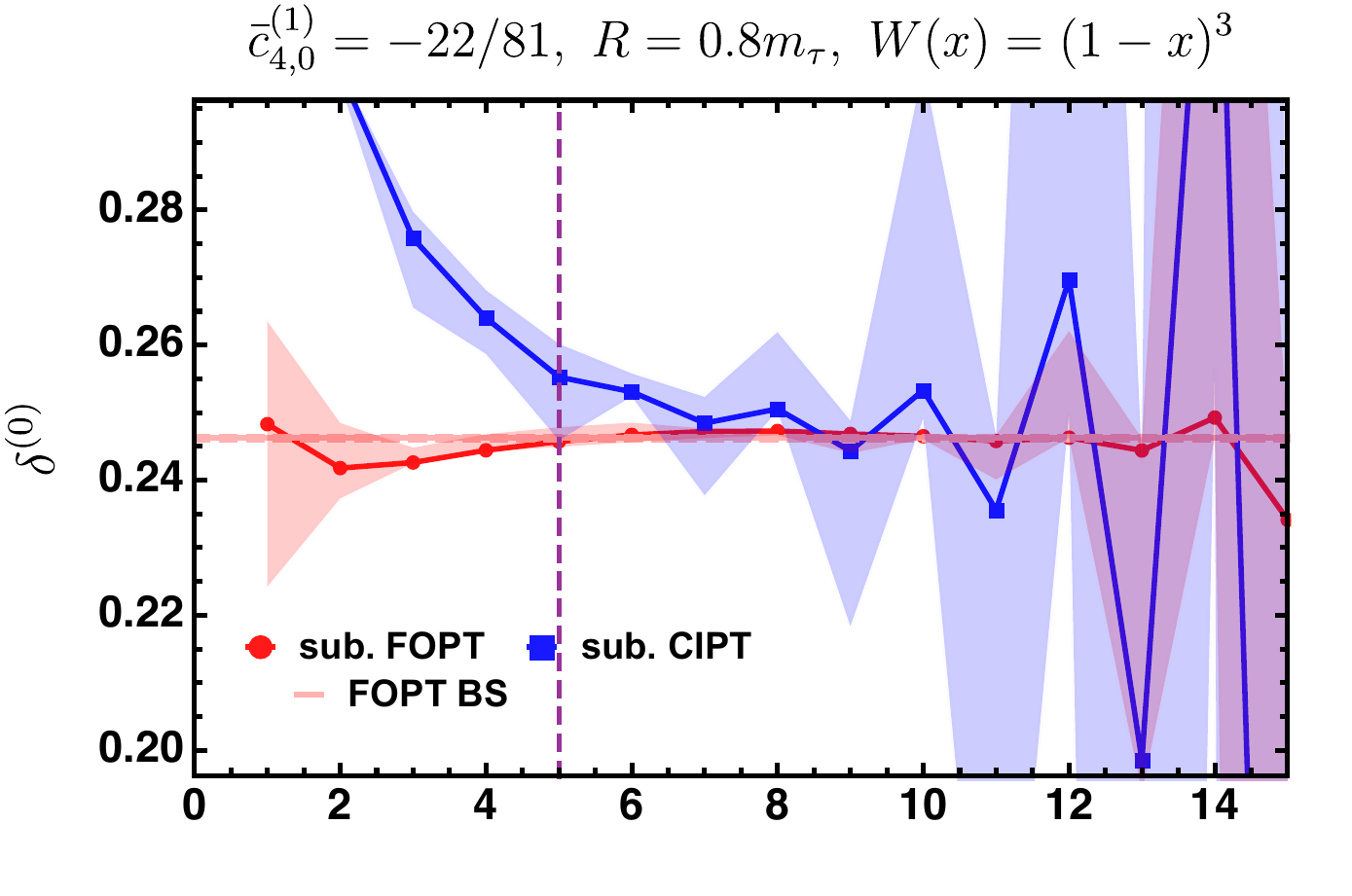}
	\end{subfigure}

	\begin{subfigure}[b]{0.48\textwidth}
		\includegraphics[width=\textwidth]{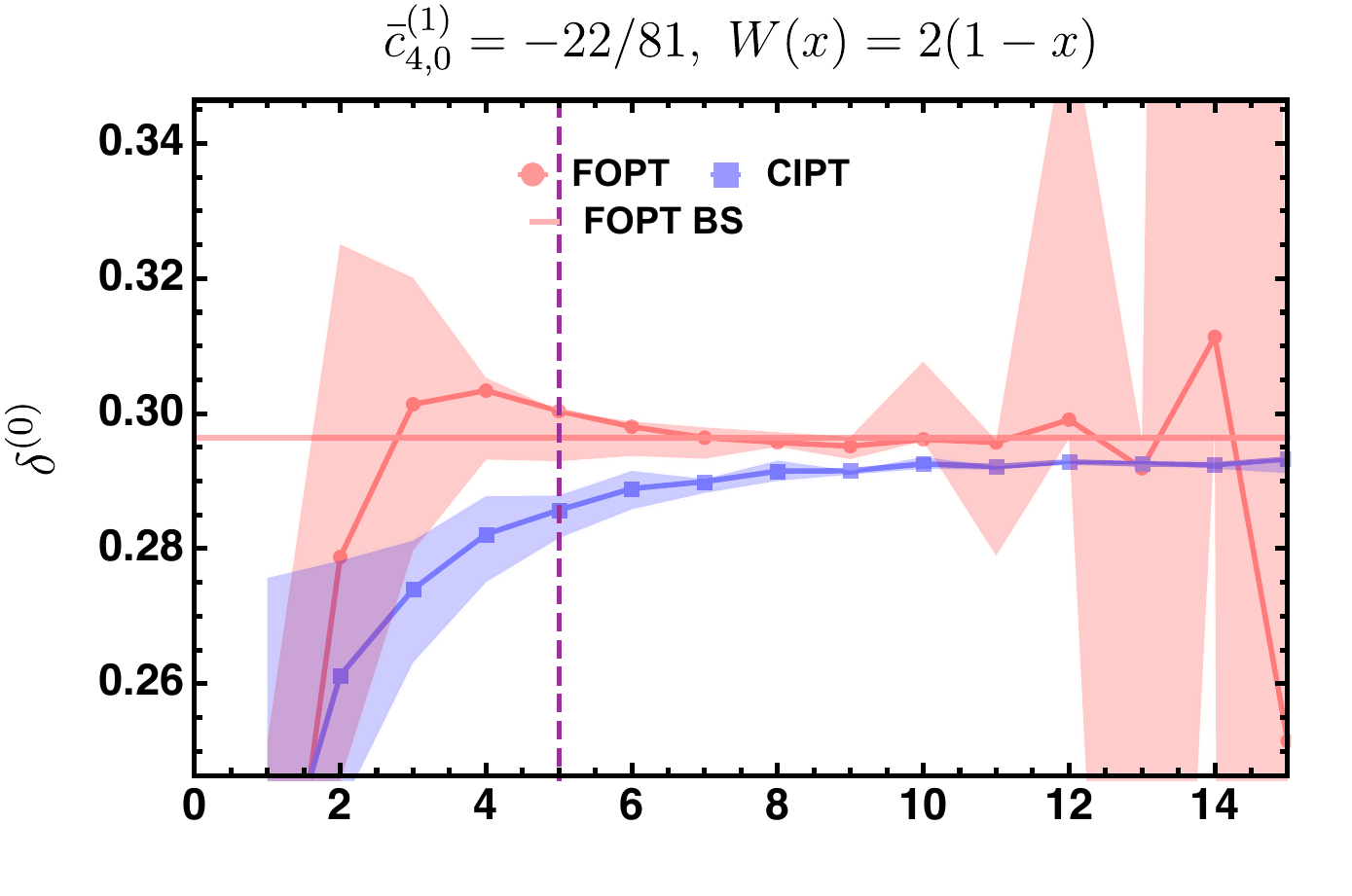}
	\end{subfigure}
	~
	\begin{subfigure}[b]{0.48\textwidth}
		\includegraphics[width=\textwidth]{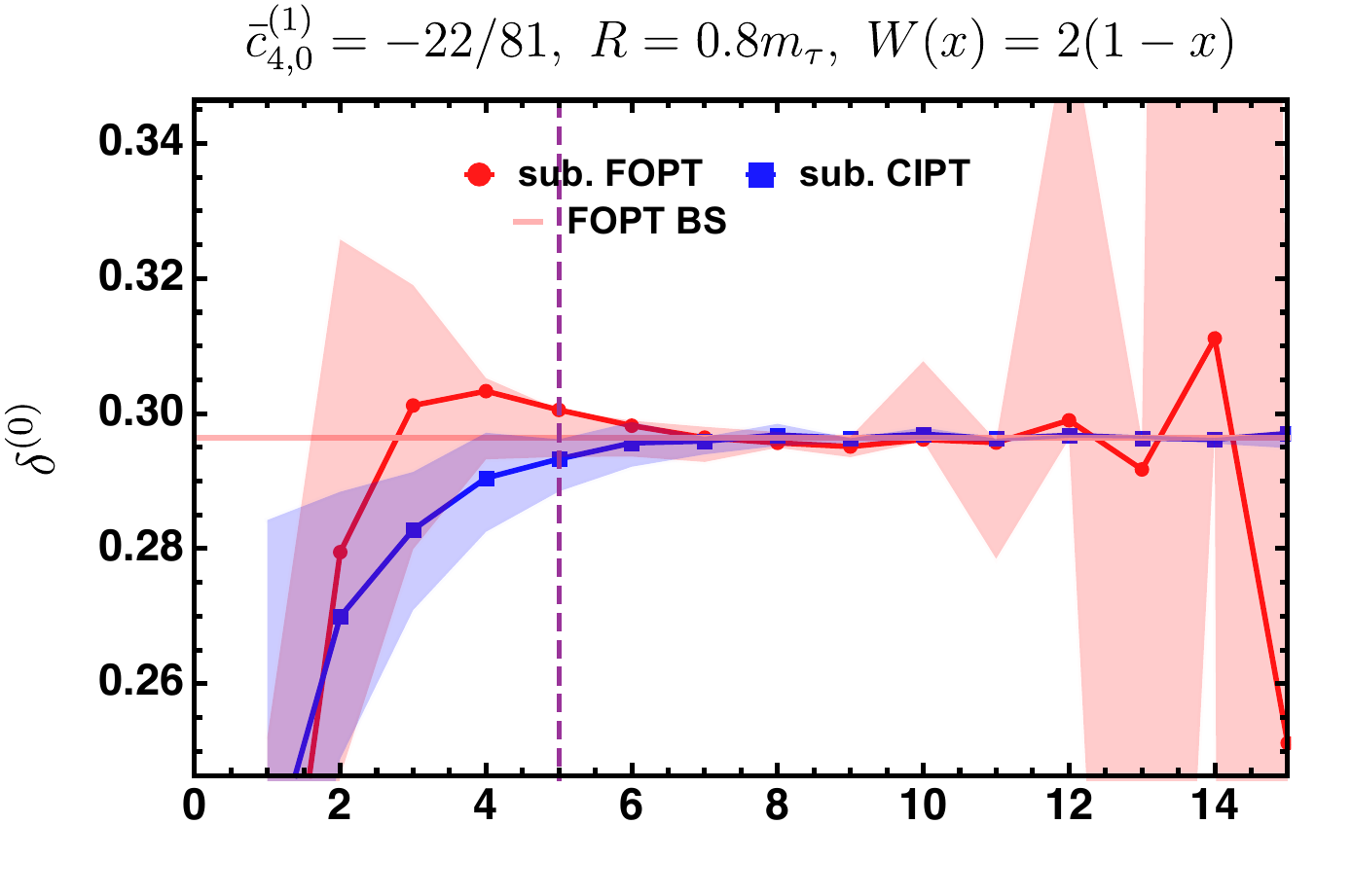}
	\end{subfigure}
	\caption{\label{fig:seriesqcd}
	Left panels: Series for FOPT and CIPT expansions for $\delta^{(0)}_{W}(m_\tau^2)$ as a function of order in full QCD in the $\overline{\rm MS}$ GC scheme for different weight functions $W(x)$ and $\alpha_s(m_\tau^2)=0.315$. The orders beyond 5 are obtained from the multi-renormalon Borel model. Renormalization scale variations are indicated by the colored bands.
	Right panels: Corresponding series for FOPT and CIPT expansions for $\delta^{(0)}_{W}(m_\tau^2,R^2)$ in the RF GC scheme.
	}
\end{figure}

In Fig.~\ref{fig:seriesqcd}, we compare the FOPT (red color) and CIPT (blue color) expansion series as a function of the order for the three spectral function moments with
$W(x)=(1-x)^3(1+x)$, $W(x)=(1-x)^3$ and $W(x)=2(1-x)$, already considered  in Sec.~\ref{sec:largeb0}, for $s_0=m_\tau^2$ and  $\alpha_s(m_\tau^2)=0.315$.
The left panels display the unsubtracted results in the $\overline{\rm MS}$ GC scheme, and the right panels show the subtracted results in the RF scheme for $R=0.8\,m_\tau$.  The connected colored dots represent the truncated series values for the default renormalization scale choice for $\xi=1$. The shaded bands again show the respective renormalization scale variations arising from $1/2\le \xi\le 2$ (see footnote~\ref{ftn:scalevariation}). The red horizontal line in each of the plots represents the PV Borel sum from Eq.~(\ref{eq:BorelFOPTgen}) (based on the expression for the multi-renormalon model function $B[\hat D(s)]_{\rm mr}(u)$ in Eq.~(\ref{eq:Bmodel})) and the area between the dashed red horizontal lines its ambiguity, see footnote~\ref{ftn:BSambiguity}. We remind the reader that the first five orders of the unsubtracted moments series (indicated by the magenta vertical line) are obtained from the coefficients in Eqs.~(\ref{cb3cb4}) and (\ref{eq:barc51}) and not predicted by the model. For the subtracted series all series terms are affected by the model since the subtraction depends on the model's value for the GC renormalon norm $N_{4,0}=4.2$.

In the upper two panels of Fig.~\ref{fig:seriesqcd}, we show the results for the GCS kinematic moment with $W_\tau(x)=(1-x)^3(1+x)$. The first five terms of the unsubtracted FOPT and CIPT series expansion (upper left panel) clearly show the well-known discrepancy between FOPT and CIPT and that the unsubtracted FOPT series closely approaches the `true' Borel sum of Eq.~(\ref{eq:BorelFOPTgen}). From the results for the first five orders we can see the well-known fact that the discrepancy {\it increases} with the inclusion of the ${\cal O}(\alpha_s^3)$, ${\cal O}(\alpha_s^4)$, and the ${\cal O}(\alpha_s^5)$ terms. The results beyond ${\cal O}(\alpha_s^5)$, which are predictions of the Borel model, show that both series stabilize and that the discrepancy persists at intermediate orders until both series become unstable beyond order $10$. The discrepancy is substantially larger than the ambiguity of the `true' Borel sum, which is hardly visible in the figure. This is a consequence of the sizeable GC renormalon norm $N_{4,0}$. Using the RF GC scheme (upper right panel) the discrepancy is now reduced, systematically, order by order in perturbation theory and at ${\cal O}(\alpha_s^5)$ it is already significantly smaller than for the unsubtracted series. While the FOPT expansion has only been slightly modified by the subtraction and again approaches the `true' Borel sum, the CIPT expansion has received sizeable modifications. At orders 7 and 8 the results from the subtracted FOPT and CIPT series are essentially equivalent. At order 10 and beyond the sign alternation associated with the UV renormalon becomes visible and both series diverge.
These results again show that employing the RF scheme eliminates, for all practical purposes, the discrepancy between the FOPT and CIPT expansion methods that was persistent in the $\overline{\rm MS}$ GC scheme. To the extent that the assumption that go into the Borel model are correct, we can expect that predictions from the FOPT and CIPT expansion series become practically equivalent if one or two more perturbative coefficients (ever) become available.

The significant reduction in the difference of the FOPT and CIPT series in the RF scheme that is exhibited for the terms up to ${\cal O}(\alpha_s^5)$
is not a purely academic issue. Already at ${\cal O}(\alpha_s^5)$ when extracting $\alpha_s$ from experimental data,
the reduced difference will translate into results for $\alpha_s$ values in much better agreement. This, again, corroborates the conclusions of Refs.~\cite{Hoang:2020mkw,Hoang:2021nlz} since the subtraction of the GC renormalon may be sufficient to eliminate almost completely the discrepancy between the two series. We also observe that the main qualitative features of the large-$\beta_0$ results are corroborated by this model for QCD. The only exception is the size of the contribution of the UV renormalon, which appears to be less prominent in QCD, as suggested by the exactly known coefficients.

The second line of panels in Fig.~\ref{fig:seriesqcd} shows results for the GCE moment obtained from $W(x)=(1-x)^3$. For the unsubtracted FOPT and CIPT series (middle left panel), the picture is qualitatively similar to the results in large-$\beta_0$ approximation with both series displaying the run-away behaviour, with a CIPT series that shows the sign alternation that is substantially weaker than in the large-$\beta_0$ case. This run-away behaviour is clearly visible at the first 5 orders and motivated the suggestion of Ref.~\cite{Beneke:2012vb} that  GCE moments are not suitable for high-precision extractions of $\alpha_s$ from hadronic $\tau$ decay data.
In the RF scheme (middle right panel) the FOPT and CIPT series, are significantly improved and the run-away behaviour is eliminated. For the subtracted FOPT series the improvement is quite dramatic. It approaches the `true' Borel sum already in the first few series terms and stabilizes around this value with tiny renormalization scale variations, until the UV renormalon becomes visible at orders $10$ and beyond. The subtracted CIPT results overshoots the Borel sum for the first few orders but approaches the subtracted FOPT series with large steps and become compatible within the renormalization scale variation at order $5$.
In the subtracted CIPT series, the sign alternating effects of the UV renormalon become visible already at order $6$, but at orders 6 and 7 its value is already very close to the FOPT series.
These results show that in the RF scheme the perturbative behaviour of the GCE moments is tamed and these moments may indeed become eligible for high-precision analyses. We note that the rather small renormalization scale variation of the subtracted CIPT series at low orders is caused by the phase cancellations of scale variation effects in the powers of the strong coupling and their coefficients entering the contour integration. Here, renormalization scale variation does not provide an adequate estimate of the perturbative error, and alternative methods such as the size of the last included series term (or the $R$ variation discussed below) have to be used.

 In the second line of panels we can also see the dramatic difference concerning the Borel sum ambiguity. For the unsubtracted series (middle left panel) the ambiguity is huge, while it has been reduced to practically negligible size for the subtracted series in the GC renormalon-free RF scheme (middle right panel), as we have already observed in the large-$\beta_0$ approximation. The actual value of the Borel sum ambiguity relies, of course, on the particular form of our Borel model, because one cannot specify the ambiguity exactly in the context of truncated perturbation theory. However, in the context of the naturalness assumption the displayed ambiguities should provide an adequate estimate for the actual Borel sum ambiguity in full QCD.

Finally, in the lowest two panels of Fig.~\ref{fig:seriesqcd} we show results for the unpinched moment $W(x)=2(1-x)$. In the $\overline{\rm MS}$ scheme for the GC (lower left panel) both series behave as typical asymptotic series, showing a plateau at intermediate orders, but the discrepancy between the values approached by FOPT and CIPT is clearly visible up to ${\cal O}(\alpha_s^5)$ and persists to higher orders within the Borel model. The FOPT expansion approaches the `true' Borel sum of Eq.~(\ref{eq:BorelFOPTgen}), while the CIPT expansion does not. In the RF scheme we see that already at ${\cal O}(\alpha_s^5)$ the discrepancy is substantially reduced and for orders 7 to 11 the two expansion series have stabilized around the `true' Borel sum. Again, the discrepancy between FOPT and CIPT exhibited in the $\overline{\rm MS}$ scheme for the
GC is eliminated in the RF scheme will lead, in actual extractions of $\alpha_s$ from experimental data to a substantially better agreement between the FOPT and CIPT expansion methods.

\begin{figure}
	\centering
	\begin{subfigure}[b]{0.48\textwidth}
		\includegraphics[width=\textwidth]{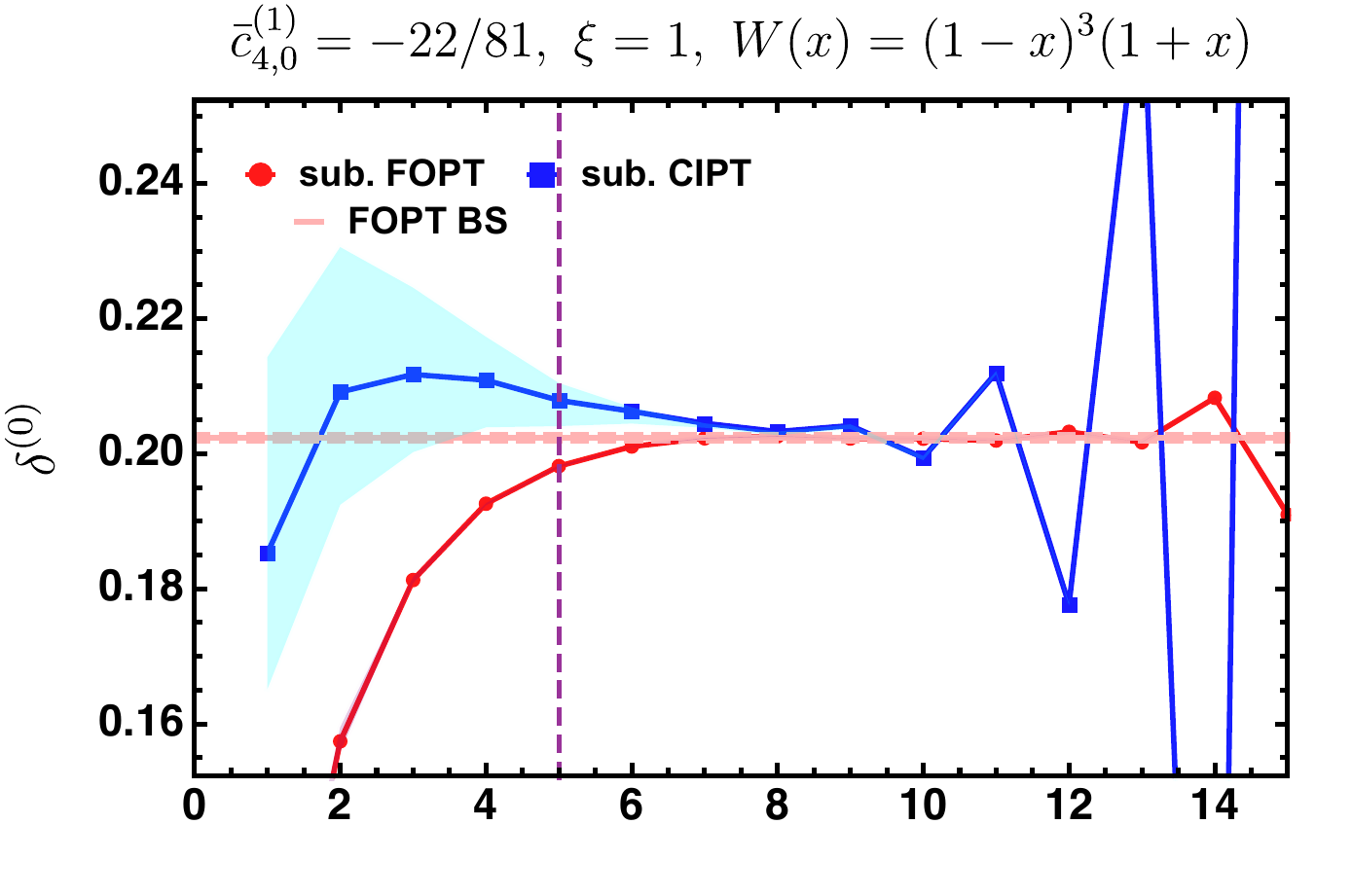}
	\end{subfigure}
	~ 
	\begin{subfigure}[b]{0.48\textwidth}
		\includegraphics[width=\textwidth]{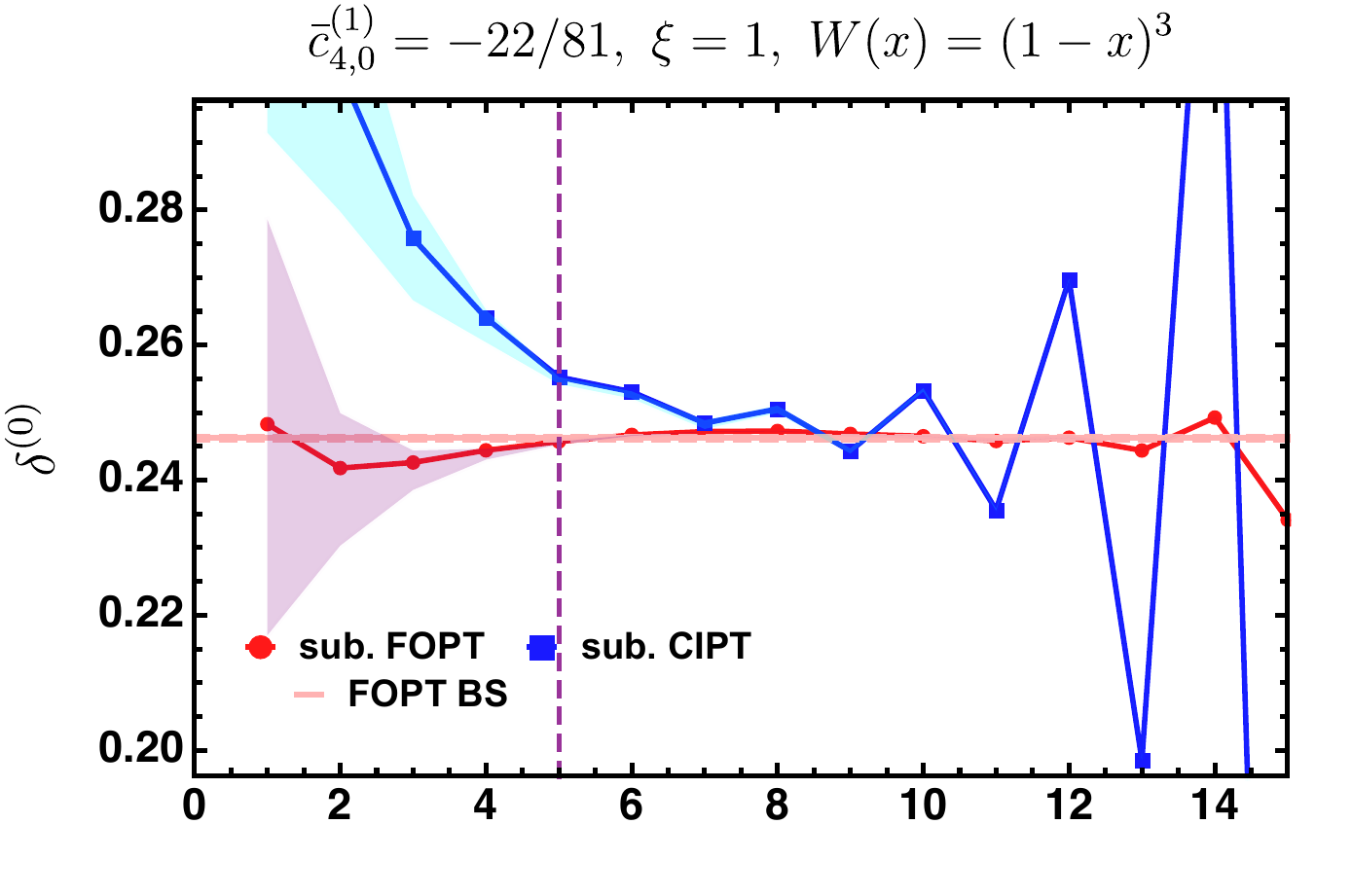}
	\end{subfigure}

	\begin{subfigure}[b]{0.48\textwidth}
		\includegraphics[width=\textwidth]{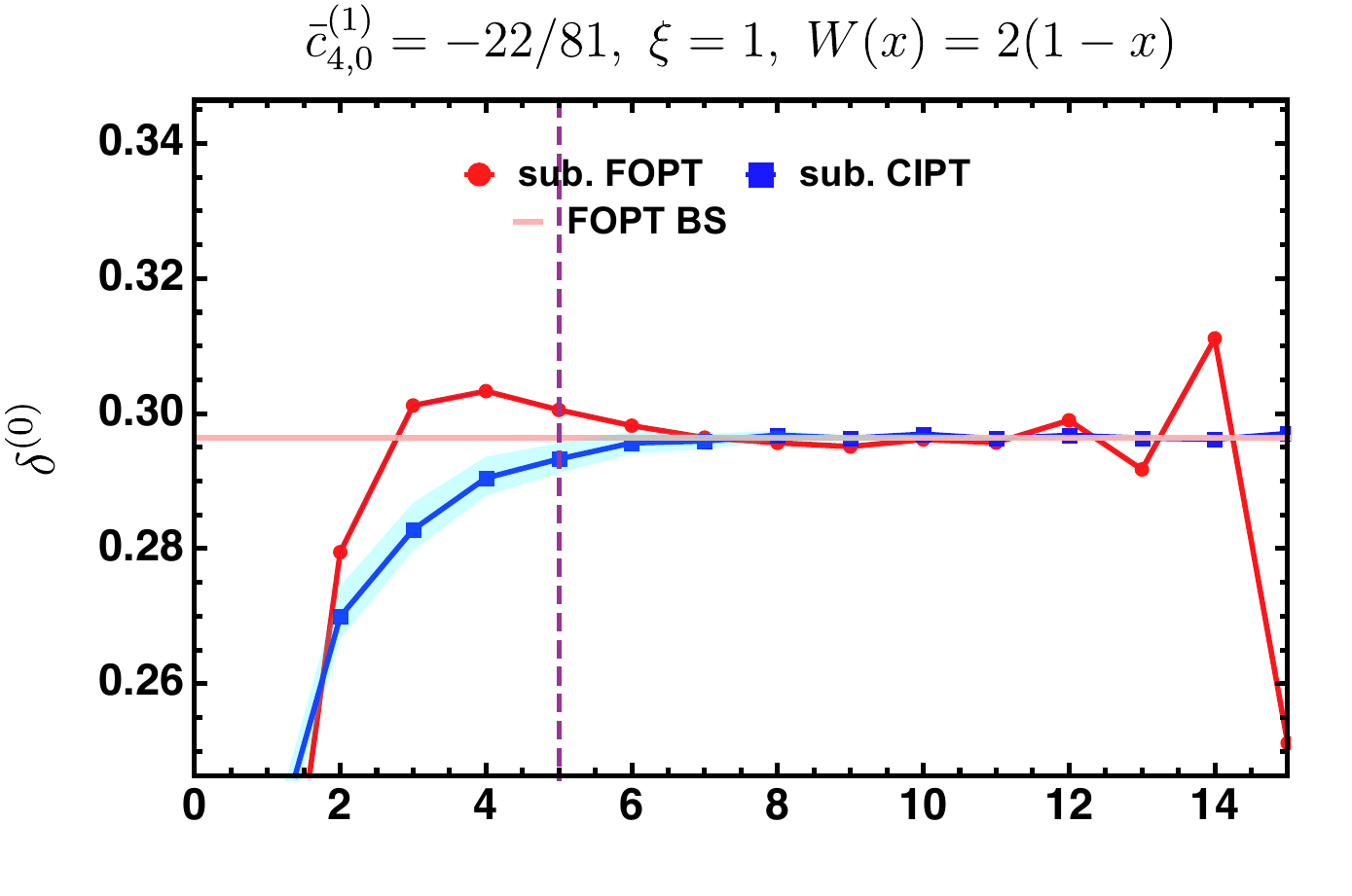}
	\end{subfigure}
	\caption{\label{fig:seriesqcdeta}
	Dependence FOPT and CIPT expansions for $\delta^{(0)}_{W}(m_\tau^2,R^2)$ in the RF GC scheme for the default renormalization scale choice $\xi=1$ with respect to changes of the IR subtraction scale in the range $0.7\,m_\tau\le R\le 0.9\,m_\tau$ for different weight functions $W(x)$ and $\alpha_s(m_\tau^2)=0.315$. The orders beyond 5 are obtained from the multi-renormalon Borel model.}
\end{figure}

An important aspect of the RF GC scheme we have not illuminated numerically so far is the dependence on the IR subtraction scale $R$.
As mentioned at the end of Sec.~\ref{sec:momentsnewscheme}, the variations of the scale $R$ constitute an important diagnostic tool concerning the
reliability and perturbative convergence of the RF GC scheme and should be accounted for in estimations of the perturbative uncertainty in the context of phenomenological analyses. While it is formally clear from the construction that the $R$-dependence decreases when the truncation order is increased, it is phenomenologically interesting to assess, how the remaining $R$-scale variation compares to the common renormalization scale variations of the strong coupling.
In Fig.~\ref{fig:seriesqcdeta} we compare the FOPT (red color) and CIPT (blue color) expansion series in the RF scheme for $\xi=1$ and for the three spectral function moments with $W_\tau(x)=(1-x)^3(1+x)$, $W(x)=(1-x)^3$ and $W(x)=2(1-x)$ already discussed in Fig.~\ref{fig:seriesqcd}, again for  $s_0=m_\tau^2$ for $\alpha_s(m_\tau^2)=0.315$.
The connected colored dots represent the truncated series values for the default IR subtraction scale $R=0.8\,m_\tau$ and are identical to those in Fig.~\ref{fig:seriesqcd}.
The shaded colored bands show the order-dependent variations arising from $0.7\, m_\tau\le R \le 0.9 \,m_\tau$.
We see that for the GCS moments (upper left and lower panels) the $R$ scale variation for the FOPT expansion is almost invisible. This is because for the FOPT expansion the subtractions arising in the renormalon-free scheme are strongly suppressed. The GCS moments in the CIPT expansion, on the other hand, exhibit a visible $R$ dependence, which, however, decreases strongly with order as it should. In comparison with the renormalization scale variations displayed in the right panels of Fig.~\ref{fig:seriesqcd}, the $R$ variation is, however, in general smaller. For the GCE moment (upper right panel) the $R$ scale variation is larger than the renormalization scale variation for both expansion methods for orders up to 4, but smaller for orders above, confirming the renormalization scale variation alone does  in general not provide an adequate estimate of the perturbative uncertainty. Overall, we see that the $R$ variation is under good perturbative control and in general smaller than the renormalization scale variation already at order $5$. Its size is, however, in general not negligible, so that the $R$ variations should be an integral part of the assessment of the perturbative uncertainties within the RF  scheme.

We have again checked  thoroughly that the results exemplified by the three spectral function moments shown in  Figs.~\ref{fig:seriesqcd} and \ref{fig:seriesqcdeta} are by no means specific for the weight functions we have picked. The investigation of many different GCS moments using $N_{4.0}=4.2$ shows that the strong reduction of the discrepancy between CIPT and FOPT that arises from switching from the $\overline{\rm MS}$  to the RF scheme and our observations concerning the size of the remaining $R$ variations are generally true and systematic. The investigation of many different GCE moments, for which the GC OPE corrections is unsuppressed and the FOPT and CIPT moment series exhibit a runaway behavior already at very low orders with very large renormalization scale variations in the  $\overline{\rm MS}$ GC scheme, shows that switching to the RF scheme systematically leads to very well-behaved series expansions with small renormalization scale variations. Our observations concerning the $R$ variation of the GCE moments are general as well. This systematic improvement for GCS as well as GCE spectral function moments in the RF scheme is already visible at ${\cal O(\alpha}_s^4)$ and  ${\cal O}(\alpha_s^5)$ and not a speculative feature of the higher order predictions generated by our Borel model. This strongly supports the view that original FOPT-CIPT discrepancy indeed originates from a sizeable GC renormalon contribution contained in the already known QCD corrections to the Adler function  and is not an entirely accidental characteristics of the low truncation order. This in turn supports the view that the GC norm $N_{4.0}$ is indeed sizeable and close to the value $4.2$.

\subsection[Impact on  \texorpdfstring{$\alpha_s$}{alphas} determinations]{Impact on \boldmath \texorpdfstring{$\alpha_s$}{alphas} determinations}
\label{sec:numex}

It is now interesting to assess the impact of the use of the RF GC scheme in concrete $\alpha_s$ extractions from $\tau$ decay data. The improvement in the behaviour of the spectral function moments and the suppression of the asymptotic separation is visible in Fig.~\ref{fig:seriesqcd}, but it remains to be shown that this is sufficient to lead to a significant reduction on the discrepancy of $\alpha_s$ values obtained by using FOPT and CIPT.  This will be addressed in our follow-up paper in the context of a full phenomenological analysis.

For the time being, we perform the following exercise, which nicely illustrates the improvement due to the use of the RF GC scheme. We take the Borel sum of the multi-renormalon model for each of the three moments  of Fig.~\ref{fig:seriesqcd}, shown in the  corresponding panels as the horizontal red solid lines (calculated for $\alpha_s(m_\tau^2)=0.315$), as a proxy for the perturbative part of the ``experimental" moment values. We  then extract $\alpha_s(m_\tau^2)$  for each moment using their perturbative contributions at $\mathcal{O}(\alpha_s^5)$  in FOPT and CIPT both in the $\overline{\rm MS}$ as well as in the RF GC scheme.
 Since non-perturbative contributions do not play a role here, this analysis mimics the perturbative aspects of realistic analyses of $\tau$ decay data.
The focus  is on the discrepancy of the perturbative FOPT and CIPT series, and the respective values of $\alpha_s$. We estimate the pertubative error in each of the $\alpha_s$ extractions varying $\xi$ as described above as well as varying $c_{5,1}$ by $50\%$ (see Eq.~(\ref{eq:c51MSb})). The intrinsic uncertainty associated with the implemenation of the RF scheme is estimated varying the scale $R$ as in Fig.~\ref{fig:seriesqcdeta}.  We again adopt $N_{4,0}=\frac{2\pi^2}{3}\,N_g=4.2$, which is the GC renormalon norm of the multi-renormalon model.

\begin{table}
	\begin{center}
		\begin{tabular}{|p{1.204cm}|p{0.45cm}|p{4.65cm}|p{7.1cm}|}
			\hline
			$W(x)$& &$\hspace{1.2cm}\text{$\overline{\rm MS}$ GC scheme}$&$\hspace{2.4cm}\text{RF GC scheme}$\\
			\hline
			kin&$\begin{matrix}\text{FO} \\\text{CI} \end{matrix}$&$\begin{matrix}0.319 \pm (0.008)_{\xi} \pm (0.001)_{c_{5,1}} \\ 0.341 \pm (0.003)_{\xi} \pm (0.002)_{c_{5,1}} \end{matrix}$&$\begin{matrix} 0.319 \pm (0.007)_{\xi}  \pm (0.0003)_{R}  \pm (0.001)_{c_{5,1}} \\ \hspace{-0.15cm}0.311 \pm (0.002)_{\xi}  \pm (0.003)_{R}  \pm (0.002)_{c_{5,1}} \end{matrix}$\\
			\hline
			$(1-x)^3$&$\begin{matrix}\text{FO} \\\text{CI} \end{matrix}$&$\begin{matrix}0.326 \pm (0.015)_{\xi} \pm (0.001)_{c_{5,1}} \\ 0.329 \pm (0.009)_{\xi} \pm (0.003)_{c_{5,1}}\end{matrix}$&$\begin{matrix} 0.315 \pm (0.0009)_{\xi}  \pm (0.0002)_{R}  \pm (0.0009)_{c_{5,1}} \\ \hspace{-0.35cm}0.310 \pm (0.003)_{\xi}  \pm (0.0009)_{R}  \pm (0.002)_{c_{5,1}}\end{matrix}$\\
			\hline
			$2(1-x)$&$\begin{matrix}\text{FO} \\\text{CI} \end{matrix}$&$\begin{matrix}0.312 \pm (0.004)_{\xi} \pm (0.002)_{c_{5,1}} \\ 0.325 \pm (0.003)_{\xi} \pm (0.001)_{c_{5,1}} \end{matrix}$&$\begin{matrix} 0.312 \pm (0.004)_{\xi}  \pm (0.0001)_{R}  \pm (0.002)_{c_{5,1}} \\ \hspace{-0.175cm}0.318 \pm (0.004)_{\xi}  \pm (0.003)_{R}  \pm (0.004)_{c_{5,1}} \end{matrix}$\\
			\hline
		\end{tabular}
		\caption{Results for $\alpha_s(m_\tau^2)$ from the toy analysis described in Sec.~\ref{sec:numex} for the input value $\alpha_s(m_\tau^2)=0.315$ in the FOPT (FO) and CIPT (CI) expansions.  The central values are obtained using $c_{5,1}=280$, $\xi=1$ and $R=0.8m_{\tau}$.}
		\label{tab:numex}
	\end{center}
\end{table}

\begin{figure}[!t]
	\centering
		\includegraphics[width=\textwidth]{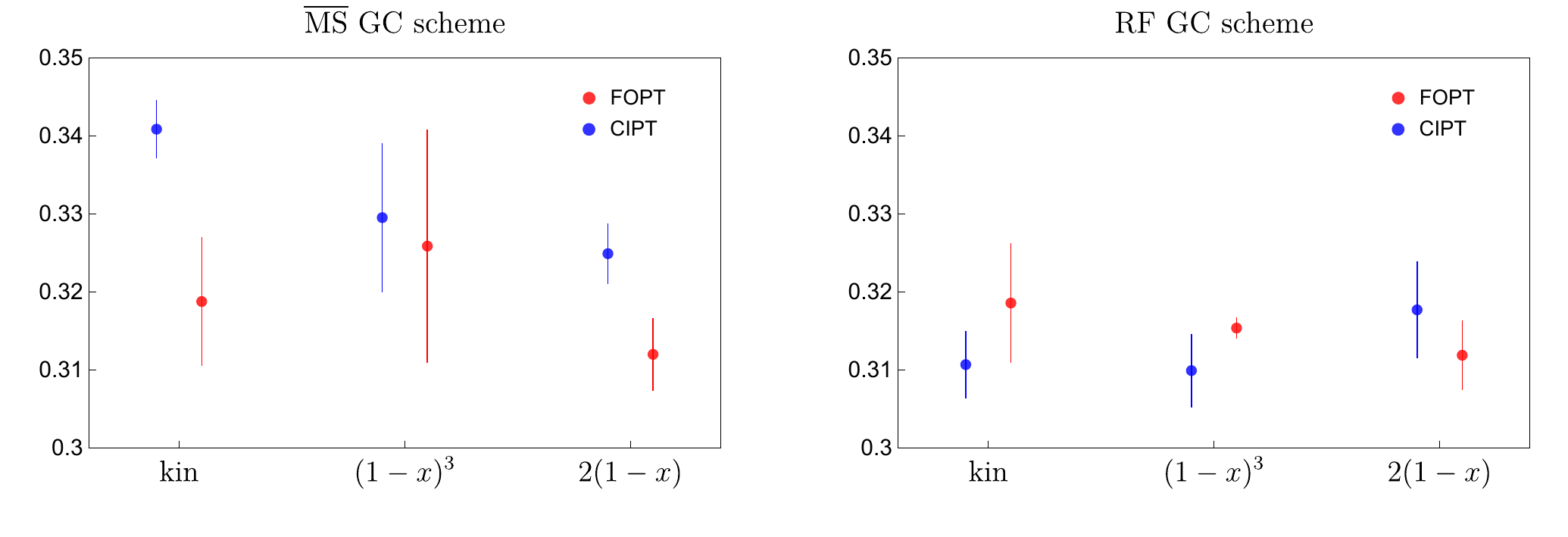}
		\caption{Central values und total uncertainties for $\alpha_s(m_\tau^2)$ from the toy analysis described in Sec.~\ref{sec:numex} for the input value $\alpha_s(m_\tau^2)=0.315$ from the results in Tab.~\ref{tab:numex} in the $\overline{\rm MS}$ GC scheme (left panel) and the RF GC scheme (right panel). }
		\label{fig:numex}
\end{figure}

Results for this exercise are shown in Table~\ref{tab:numex}. The case of the kinematic moment, labeled `kin', which enters many of the recent $\alpha_s$ analyses from $\tau$ decays, is emblematic. In the  unsubtracted $\overline{\rm MS}$ GC scheme  we find a discrepancy in the $\alpha_s(m_\tau^2)$ values extracted with FOPT and CIPT of $0.022$, which is significantly larger than the theoretical uncertainties arising from $\xi$ and $c_{5,1}$ variations.  This agrees very well with the discrepancy that was found in the recent $V+A$ fits of Ref.~\cite{Pich:2016bdg} based on simultaneous fits of several moments (including the kinematical moment) where, for example, the discrepancy between FOPT and CIPT was quoted as $0.019$, see their Table 8. On the other hand, carrying out the $\alpha_s$ determination in the RF GC scheme we now find a discrepancy of  $0.008$, which  renders the two results compatible within the theoretical uncertainties. The  outcome is similar for the moment $W(x)=2(1-x)$ shown in the last row of Table~\ref{tab:numex},  which enters the recent analysis of Ref.~\cite{Boito:2020xli}. The discrepancy in $\alpha_s(m_\tau^2)$ is reduced from $0.013$ to $0.006$ when the RF GC scheme is used. For the GCE moment $W(x)=(1-x)^3$, which has not been employed in recent $\alpha_s$ determinations due to its badly behaved series, we see that the very large error in $\alpha_s(m_\tau^2)$ arising from the $\xi$ variation in the $\overline{\rm MS}$ GC scheme, is  reduced by at least an order of magnitude and thus completely tamed in the RF GC scheme. Finally, one can also see that the results for $\alpha_s(m_\tau^2)$ obtained in the CIPT expansion are much closer to the input value $\alpha_s(m_\tau)=0.315$ when the RF GC scheme employed. A visualization of these results is depicted in Fig.~\ref{fig:numex}.

Overall, we find that the conclusions (1) and (2) drawn from our analysis in the large-$\beta_0$ approximation in Sec.~\ref{sec:largeb0} also apply in the context of full QCD. The RF GC scheme we have set up in this work eliminates in a systematic way the FOPT-CIPT discrepancy for GCS spectral function moments observed in the common $\overline{\rm MS}$ GC scheme and, in addition, improves significantly the perturbative behavior of the FOPT as well as CIPT for GCE spectral function moments. This makes GCE spectral function moments eligible for high-precision phenomenological applications. These improvements take place already at the ${\cal O}(\alpha_s^5)$ level, which is the order used for state-of-the-art strong coupling determinations, and do not rely on extrapolations to larger orders based on Borel function models. From the fact that the improvements take place for any type of spectral function moment we conclude that it is highly unlikely and implausible that the known perturbative coefficient of the Adler function only accidentally mimic a behavior consistent  with
$N_{4.0}$ being close to that value. In the context of proposition (a), which we stated at the very beginning of this paper, we therefore consider the FOPT-CIPT discrepancy problem, that has permeated the literature for a very long time, as resolved, and we corroborate the findings of Refs.~\cite{Hoang:2020mkw,Hoang:2021nlz}.
Within the RF GC scheme, the CIPT expansion method for spectral function moments remains a viable theoretical expansion method for phenomenological applications.

Of course, the issues of whether the true value of GC renormalon norm $N_{4.0}=\frac{2\pi^2}{3} N_g$ is indeed close to $4$ and accounting for a reasonable estimate of its uncertainty remain important subjects of further theoretical studies. These  issues will be addressed in detail in our follow-up paper , where we will also carry out a more thorough phenomenological $\alpha_s$ determination.

\section{Summary and Conclusions}
\label{sec:conclusions}

In this work we propose a simple and easy-to-implement renormalon-free gluon condensate (GC) scheme, dubbed the `RF scheme'. In the literature compatible schemes for the GC matrix element have been discussed before, but the corresponding implementations primarily focused on modifications of the original perturbation series (either at the level of the loop integrals or through high-order calculations or all-order rearrangements). This makes these implementations technically complicated and issues such as observable independence somewhat involved. In contrast to these previous approaches, the RF scheme we propose starts from a perturbative redefinition of the GC matrix element itself. The approach is analogous to the well-established scheme change procedure where the heavy-quark pole mass scheme is replaced by short-distance mass schemes. In our approach the GC matrix element in the original scheme, which contains an ${\cal O}(\Lambda_{\rm QCD}^4)$ renormalon and which we call $\overline{\rm MS}$, is treated in analogy to the heavy-quark pole mass. It is replaced by a renormalon-free GC matrix element plus a perturbative subtraction series that encodes the renormalon.
The resulting subtraction series can then be combined with the original perturbation series such that the diverging asymptotic behavior of the
GC renormalon is being eliminated order-by-order in a controlled and transparent way. The subtraction series, which quantifies the difference between the
original $\overline{\rm MS}$ GC and the one in the RF scheme, depends on the IR factorization scale $R$ that can be adapted such that the subtraction is suitable for the relevant dynamical scale of the observable where the scheme change is applied. Since the $R$-dependence of the subtraction series is known in exact closed form to all orders, the  GC matrix element in the RF scheme is scale invariant while maintaining the possibility to adapt the $R$ in order to sum potentially large logarithms related to the observables dynamical scale. This allows us to define the RF scheme GC matrix element such that it implements the same Borel sum scheme that has been the scheme of choice for most of the previously literature on renormalon-free OPE schemes.
The RF scheme depends on the norm $N_g$ of the GC renormalon, which has to be supplemented independently.

The primary motivation of the  RF GC scheme  in the context of this work is to apply it to the QCD perturbation series for $\tau$ hadronic spectral function moments determined with the FOPT and CIPT  expansion methods {\it in their original form}. Our work constitutes the first study of this kind in the literature.
For spectral function moments, where the GC OPE correction is strongly suppressed, and which are used for most state-of-the-art high-precision strong coupling extractions, the FOPT and CIPT series expansions show good convergence in the $\overline{\rm MS}$ GC scheme. However, there has been the problem that both series approach perturbatively incompatible values at ${\cal O}(\alpha_s^4)$ and ${\cal O}(\alpha_s^5)$. This FOPT-CIPT discrepancy problem has permeated the literature for many years. It has been suggested recently~\cite{Hoang:2020mkw,Hoang:2021nlz} that this discrepancy arises from a remaining strong sensitivity to the GC renormalon that is inherent to the CIPT expansion prescription despite the fact that the GC OPE correction is strongly suppressed. In Refs.~\cite{Hoang:2020mkw,Hoang:2021nlz} the discrepancy was calculated quantitatively and it was concluded that the CIPT expansion provides inconsistent physical results in the $\overline{\rm MS}$ GC scheme.

Due to its simplicity, the RF GC scheme allows for the application of the FOPT and CIPT expansion methods such that we can study the impact of the renormalon subtraction on both expansion methods. We find in the large-$\beta_0$ approximation for the important spectral function moments, where the GC OPE correction is strongly suppressed, that using the RF scheme has little impact on the FOPT series expansion, but significantly modifies the CIPT series expansion such that it becomes compatible with the FOPT expansion. For spectral function moments where the GC OPE correction is unsuppressed, and which have not been considered in most recent state-of-the-art strong coupling determinations, the quite bad convergence behavior of the FOPT and CIPT expansions present in the $\overline{\rm MS}$ GC scheme becomes substantially improved. Applying the RF scheme to the spectral function moments in full QCD at ${\cal O}(\alpha_s^4)$ or ${\cal O}(\alpha_s^5)$ we make the same observations as in the large-$\beta_0$ approximation, when we adopt the value $N_{4,0}=4.2$ for the GC renormalon norm of the Adler function. Our results show that the known perturbative coefficients of the Adler function are fully consistent with the view that the GC renormalon already has a significant impact on the moments' perturbation series at ${\cal O}(\alpha_s^5)$ and the true value for $N_{4,0}$ is indeed close to $4$.
The RF scheme with $N_{4,0}=4.2$ is capable of resolving the long-standing FOPT-CIPT discrepancy problem for moments with suppressed GC OPE correction and improves significantly the perturbative behavior for moments with unsuppressed GC OPE corrections.  Given that both improvements take place simultaneously and systematically for all choices of spectral function moments, we believe that it is highly unlikely that the actual value of $N_{4,0}$ is substantially different from $4$ and the known perturbative coefficients of the Adler function only accidentally mimic a behavior consistent with a GC norm close to this value. To the extent that this view is correct, we can consider the long-standing FOPT-CIPT discrepancy problem for moments which suppressed GC OPE corrections as resolved.
Our results fully corroborate the findings in Refs.~\cite{Hoang:2020mkw,Hoang:2021nlz}, but also show that CIPT remains a viable theoretical expansion method for the computation of the spectral function moments in the context of renormalon-free GC schemes such as the RF scheme.

Of course, the issues of whether the true value of GC renormalon norm $N_{4.0}=\frac{2\pi^2}{3}N_g$ is indeed close to $4$ and accounting for a reasonable estimate of its uncertainty remain important subjects of further theoretical investigations. These are addressed in more detail in our follow-up paper, where we also discuss how employing the RF scheme affects the outcome of strong coupling determinations from $\tau$ hadronic spectral function moments using experimental data.

\section*{Acknowledgments} DB and MJ would like to thank the Particle Physics Group of the University of Vienna for hospitality.
We acknowledge partial support by the FWF Austrian Science Fund under the Doctoral Program ``Particles and Interactions'' No.\ W1252-N27 and under the Project No. P32383-N27.  We also thank the Erwin-Schr\"odinger International Institute for Mathematics and Physics for partial support. DB's work was supported by  the Coordena\c c\~ao de Aperfei\c coamento de Pessoal de N\'ivel Superior -- Brasil (CAPES) -- Finance Code 001 and by the S\~ao Paulo Research Foundation (FAPESP) Grant No.~2021/06756-6.
\vspace*{0.3cm}

\begin{appendix}

\section{Borel Model, \boldmath \texorpdfstring{$\alpha_s$}{alphas} conversion and Asymptotic separation}
\label{app:modelsC}

The multi-renormalon Model in the $C$-scheme for the Borel function of the Adler function $\hat D(s)$ with respect to the series expansion in powers of $\bar \alpha_s(-s)$ reads
\begin{align}
\label{eq:Bmodel}
		\begin{split}
		B[\hat{D}(s)]_{\rm mr}(u) &= b^{(0)} + b^{(1)}u + \frac{N_{4,0}
			\left[1+\bar{a}(-s) \bar c_{4,0}^{(1)}\right]}{(2-u)^{\gamma_4}} + \frac{N_{6,0}}{(3-u)^{\gamma_6}} + \frac{N_{-2}}{(1+u)^{\gamma_{(-2)}}},
		\end{split}
\end{align}
\begin{align*}
N_{4,0}&= \frac{2\pi^2}{3}N_g=4.20757\,,   &  N_{6,0} &= -15.64640\,,    & N_{-2}&= -0.02714\,,\\
 \gamma_4 &= \frac{209}{81}\,, &\gamma_6 &= \frac{91}{27}\,,  & \gamma_{(-2)} &= \frac{98}{81}\,, \\
b^{(0)} &= 0.15381\,,   &  b^{(1)} &=0.00790\,, &
\bar c_{4,0}^{(1)} &= -\frac{22}{81}\,.
\end{align*}
The parameter $\bar c_{4,0}^{(1)}$ denotes the $\mathcal{O}(\alpha_s)$ Wilson coefficient correction of the gluon condensate, and $N_g$ is the renormalon norm of the gluon condensate.

The perturbative relation between the coupling in the $C$-scheme and in the $\overline{\text{MS}}$-scheme at any common renormalization scale and for $n_f=3$ dynamic quark flavors reads
\begin{align}
\label{eq:asCtoMSbar}
\begin{split}
\frac{\bar\alpha_s}{\alpha_s} &= 1 -0.132789 ~ \alpha_s^2-0.247879
~ \alpha_s^3-0.103368 ~ \alpha_s^4 +0.137953 ~ \alpha_s^5+0.146252 ~ \alpha_s^6
\\
&+0.0729472 ~ \alpha_s^7-0.081650 ~ \alpha_s^8-0.122995
~ \alpha_s^9-0.042241 ~ \alpha_s^{10}+0.050279 ~ \alpha_s^{11}
\\
&+0.091138 ~ \alpha_s^{12} +0.033505 ~ \alpha_s^{13}-0.038225 ~ \alpha_s^{14} + \ldots.
\end{split}
\end{align}
For the relation the truncated 5-loop $\msb$ QCD $\beta$-function with the known coefficients $\beta_{0,1,2,3,4}$ is treated as exact, i.e.\ the $\beta$-function coefficients $\beta_n\le 5$ are set to zero. This means that the coefficients at ${\cal O}(\alpha_s^5)$ and beyond receive additional contributions once additional $\msb$ QCD $\beta$-function coefficients become available.

The asymptotic separation for a spectral function moments is the difference in the Borel sums for the CIPT expansion series and that of the FOPT expansion series based on the PV prescription. In the $C$-scheme for a generic weight function $W(x)=(-x)^m$ and for a generic IR renormalon term in the Borel function of $\hat D(s)$ with respect to the series expansion in powers of $\bar \alpha_s(-s)$ of  the form
\begin{eqnarray}
\label{eq:AdlerBorelgen}
B_{\hat D,d,\gamma}^{\rm IR}(u) & = & \frac{1}{(p-u)^{\gamma}}\,,
\end{eqnarray}
it can be derived from the formulae given in Ref.~\cite{Hoang:2020mkw}
and has the form
\begin{eqnarray}
\label{eq:Deltafct}
\lefteqn{\Delta (m\ne p, p, \gamma, s_0) \, = \,
\delta_{(-x)^m,{\rm Borel}}^{(0),{\rm CIPT}}(s_0)-\delta_{(-x)^m,{\rm Borel}}^{(0),{\rm FOPT}}(s_0)
}\\   \nonumber
& = &  \bigg( \frac{\Lambda_{\text{QCD}}^2}{s_0} \bigg)^{m} \,\frac{2^{2m\hat{b}_1}}{\Gamma(\gamma)}\times \\   \nonumber
&&\Big\{\text{Re}\left[ (p-m+i0)^{2m\hat{b}_1-\gamma}\Gamma(-2m\hat{b}_1 + \gamma, -2(p-m)t_{-}) \right] -
 \\ \nonumber
&& -2\hat{b}_1 \text{Re}\left[ (p-m+i0)^{2m\hat{b}_1-\gamma+1}\Gamma(-2m\hat{b}_1 + \gamma-1, -2(p-m)t_{-}) \right]\Big\}\,,
\label{eq: AS in the C scheme leading contribution}
\end{eqnarray}
where $\Gamma(a,z)$ is the (upper) incomplete gamma function.
Note that the $t$-variable notation from Ref.~\cite{Hoang:2020mkw} was used for the formula above.

The QCD scale $\Lambda_{\rm QCD}$ that appears in Eq.~(\ref{eq:Deltafct}) is defined as
\begin{align}
\label{eq:LambdaQCDdef}
\Lambda_{\rm QCD}^2 \, \equiv \, Q^2\,\left(\frac{2\pi}{\beta_0\bar \alpha_s(Q^2)}\right)^{2\hat b_1}\,\exp\left(-\frac{4\pi}{\beta_0\bar \alpha_s(Q^2)}\right)
\, = \,
 Q^2\,(2 \bar a_Q)^{-2\hat b_1}\,\exp\left(-\frac{1}{\bar a_Q}\right)\,
\end{align}
and is independent of the value of $Q^2$ including real- as well as complex values.
The formula for $\Lambda_{\rm QCD}$ in terms of the strong coupling $\alpha_s$ in the $\overline{\rm MS}$ scheme cannot be written down in closed form and has been given in Sec.~3 of Ref.~\cite{Hoang:2020mkw}. The QCD scale $\Lambda_{\rm QCD}$ is
related to what is sometimes referred to as the `$\overline{\rm MS}$ definition of the QCD scale' from Ref.~\cite{DallaBrida:2016uha} (that is referred to in QCD review section of Ref.~\cite{ParticleDataGroup:2020ssz}) by $\Lambda_{\rm QCD}^{\overline{\rm MS}}=2^{\hat b_1} \Lambda_{\rm QCD}$.

The ambiguity of the  Borel sum integral in Eq.~(\ref{eq:BorelFOPT}) is obtained using
\begin{align}
\label{eq:IRBorelIntFOPTambi}
&
\frac{1}{2\pi i}\,\ointctrclockwise_{|s|=s_0}  \frac{{\rm d}s}{s}\,\Big(\!-\frac{s}{s_0}\Big)^k \,
\frac{i}{2\pi}
\int_0^\infty \!\! {\rm d} u \, \left[\, \frac{e^{-\frac{u}{\bar a(-s)}}}{(2+i 0 - u)^{1+4 \hat b_1}}
\, - \,
\frac{e^{-\frac{u}{\bar a(-s)}}}{(2-i 0 - u)^{1+4 \hat b_1}}
\,\right]  \\
& \, = \,
\frac{1}{2\pi i}\,\ointctrclockwise_{|s|=s_0}  \frac{{\rm d}s}{s}\,\Big(\!-\frac{s}{s_0}\Big)^k \,
\frac{(\bar a(-s))^{-4 \hat b_1}}{\Gamma(1+4 \hat b_1)}\,e^{-\frac{2}{\bar a(-s)}}\nonumber \\
& \, = \,
\frac{1}{2\pi i}\,\ointctrclockwise_{|s|=s_0}  \frac{{\rm d}s}{s}\,\Big(\!-\frac{s}{s_0}\Big)^k \,
\frac{1}{s^2}\frac{2^{4\hat b_1}\,\Lambda_{\rm QCD}^4}{\Gamma(1+4 \hat b_1)}
\, = \, \left\{
\begin{array}{ll}
0\,,  & \,\,\mbox{for}\,\,k\neq 2 \\
 \frac{2^{4\hat b_1}\,\Lambda_{\rm QCD}^4}{s_0^2\,\Gamma(1+4 \hat b_1)} & \,\,\mbox{for}\,\,k = 2 \nonumber
\end{array}
\right.\,.
\end{align}

\end{appendix}

\bibliography{./sources}
\bibliographystyle{JHEP}

\end{document}